\documentclass[fleqn]{mnras}
\usepackage{newtxtext,newtxmath}

\usepackage[T1]{fontenc}

\DeclareRobustCommand{\VAN}[3]{#2}
\let\VANthebibliography\thebibliography
\def\thebibliography{\DeclareRobustCommand{\VAN}[3]{##3}\VANthebibliography}

\usepackage{graphicx}	
\usepackage{amsmath}	

\newcommand{\Lstar}{$L_{\ast}$}
\newcommand{\Msunyr}{\hbox{M$_\odot\,$yr$^{-1}$}}
\newcommand{\Teff}{$T_{\rm eff}$}
\newcommand{\Rstar}{$R_{\ast}$}
\newcommand{\Rsun}{$R_{\odot}$}
\newcommand{\Msun}{${\rm M}_{\odot}$}
\newcommand{\Mdot}{${\dot M}$}
\newcommand{\Lsun}{$\rm{L_{\odot}}$}
\newcommand{\vinf}{$v_{\rm \infty}$}

\title[The Arches cluster revisited: IV]{The Arches cluster revisited: IV. Observational constraints on the binary properties of very massive stars\thanks{Based on observations made at the European Southern Observatory, 
Paranal, Chile under programmes ESO 087.D-0317, 091.D-0187, 093.D-0306, 099.D-0345 and
0101.D-0141.}}

\author[J.~S.~Clark et al.]{
J.~S.~Clark,$^{1}$\thanks{In Memoriam: This work is dedicated to the memory of our dearest friend and colleague Simon Clark. His great enthusiasm pushed all of us to work towards unveiling massive young clusters.  
The Arches cluster and Westerlund~1 will certainly miss one of their best ambassadors.}
M.~E.~Lohr,$^{1}$
F.~Najarro$^{2}$\thanks{E-mail: najarro@cab.inta-csic.es (FN)}
L.~R.~Patrick$^{2, 3, 1}$
and B.~W.~Ritchie$^{1}$
\\
$^{1}$School of Physical Sciences, The Open University, Walton Hall, Milton Keynes MK7~6AA, UK\\
$^{2}$Departamento de Astrof\'{\i}sica, Centro de Astrobiolog\'{\i}a,
(CSIC-INTA), Ctra. Torrej\'on a Ajalvir, km 4,  28850 Torrej\'on de Ardoz,
Madrid, Spain\\
$^{3}$Departamento de F\'{\i}sica Aplicada, Facultad de Ciencias, 
Universidad de Alicante, Carretera San Vicente  s/n,
E03690, San Vicente del Raspeig, Spain
}

\date{Accepted 2023 February 2}

\pubyear{2022}

\begin{document}
\label{firstpage}
\pagerange{\pageref{firstpage}--\pageref{lastpage}}
\maketitle

\begin{abstract}
Serving as the progenitors of electromagnetic and gravitational wave transients, massive stars have received renewed interest in recent years. However, many aspects of their birth and evolution remain opaque, particularly in the context of binary interactions. The centre of our galaxy hosts a rich cohort of very massive stars, which appear to play a prominent role in the ecology of the region. In this paper we investigate the binary properties of the Arches cluster, which is thought to host a large number of very massive stars. A combination of multi-epoch near-IR spectroscopy and photometry was utilised to identify binaries. 13 from 36 cluster members meet our criteria to be classed as RV variable. Combining the spectroscopic data with  archival radio and X-ray observations - to detect colliding wind systems - provides a lower limit to the binary fraction of $\sim43$\%; increasing to $\gtrsim50$\% for the O-type hypergiants and WNLha. Dynamical and evolutionary masses reveal the primaries to be uniformly massive ($\gtrsim50M_{\odot}$). Where available, orbital analysis reveals a number of short period, highly eccentric binaries, which appear to be pre-interaction systems. Such systems are X-ray luminous, with 80\% above an empirical bound of  $(L_{\rm x}/L_{\rm bol})\sim10^{-7}$ and  their orbital configurations suggest formation and evolution via a single star channel; however, we cannot exclude a binary formation channel for a subset. Qualitative comparison to surveys of lower mass OB-type stars confirms that the trend to an extreme binary fraction ($\geq60$\%) extends to the most massive stars currently forming in the local Universe.
\end{abstract}




\section{Introduction}

Despite their rarity, massive stars are a principal agent in galactic evolution, due to the
deposition of mechanical, chemical and radiative feedback. They - and their relativistic descendants
- dominate the electromagnetic emission of galaxies at both high (UV and X-ray) and low (IR and
radio) energies, the former directly and the latter via reradiation. In both life and death they are
thought to be an important source of cosmic rays, while the nature and demographics of gravitational
wave sources directly follow from the properties of their stellar antecedents.

Consequently, it is a matter of regret that significant facets of the lifecycle of massive stars
remain poorly understood, from their formation mechanism through to their anticipated deaths in
either (core-collapse or pair instability)  supernovae or direct collapse events. Moreover
spectroscopic and imaging surveys of both clustered and isolated OB stars over the past decade have
revealed a further complication: that many such objects are to be found in binaries or higher order
multiples (Sana et al. \cite{sana12}, \cite{sana13a}, \cite{sana14}, Kobulnicky et al.
\cite{kobulnicky}, Sota et al. \cite{sota}, Dunstall et al. \cite{dunstall}, Almeida et al.
\cite{almeida}). This finding implies  significant modification of the already unresolved
evolutionary pathways of single stars,  driven by tidal interaction, binary mass transfer and, in
extreme cases, merger (de Mink et al. \cite{demink09}, \cite{demink14}, Schneider  et al.
\cite{schneider}). 

Such uncertainty is particularly acute for very massive stars and gives rise to  several
fundamental and interwoven  questions. Is there  an upper limit to the mass  with which a star may
be born? How many such stars  are found   within hierarchical systems? Do the majority achieve their
final mass during formation or instead via binary-driven mass transfer or merger at a later stage
(cf. Schneider  et al. \cite{schneider})? And, at the end of their lives do massive stars evolving via
either single or binary channels retain sufficient mass to undergo pair instability supernova  and,
if not, how does radiative and binary driven mass loss affect the demographics of the resultant
neutron star (NS) and black hole (BH) populations?

One region where these uncertainties are manifest is the Central Molecular Zone (CMZ) of the Milky
Way. Observations suggest that the  properties of the  CMZ are particularly extreme, with the mean
temperature, density, pressure and velocity dispersion of molecular material, the magnetic field
strength and the cosmic ray density and ionisation rate significantly greater than those found in molecular clouds in
the Galactic disc, in some cases by orders of magnitude. An obvious question is therefore whether
(massive)  stars form in the same manner in this tract as they do in more quiescent regions of  the
Galaxy. This is of particular importance given that the  physical conditions present within  the CMZ
are anticipated to bear close resemblance to  those of  starburst galaxies (Kruijssen \& Longmore
\cite{dkl}); hence determining the nature of star formation in the CMZ will provide unique insights
into such activity at high redshift. 

Based on the numbers  of supernova  remnants and young neutron stars observed within  the CMZ, it
must host a population of massive stars of a size capable of yielding one core collapse event every
thousand years (Deneva et al. \cite{deneva}, Kennea et al. \cite{kennea}, Ponti et al.
\cite{ponti}). Near-IR imaging and spectroscopy reveal that massive stars are indeed present - albeit
not yet detected in the quantities required - and reside in three apparently co-eval massive
clusters as well as being  distributed, in apparent isolation, throughout  the region (e.g. Clark et
al. \cite{clark18c} and refs. therein).  

Of the former cohort, with an age of $\sim2-3$Myr (Clark et al. \cite{clark18a}) the Arches is the
youngest of the massive clusters  within the CMZ (Figer \cite{1995PhDT.........6F}, Nagata et al. \cite{nagata}, Cotera et al.
\cite{cotera96}). As a consequence one would not expect it to have lost any stars to core collapse
at this juncture; hence the observed population of  13 Wolf-Rayet and $\sim$100 O-type stars should
reflect the original yield of the star formation process that gave rise to it (Blum  et al.
\cite{blum01}, Figer et al. \cite{figer02}, Martins et al. \cite{martins08}, Clark et al.
\cite{clark18a}, \cite{clark19b}). Moreover, with an integrated mass of $\gtrsim10^4M_{\odot}$
(Clarkson et al. \cite{clarkson}) the upper reaches of the initial mass function (IMF) are richly
populated; as such, it provides an unique astrophysical laboratory for the study of the birth
channel(s) and physical properties of the most massive stars Nature currently permits to form in the
local Universe (Figer~\cite{figer05}).

In order to exploit this potentiality, between 2011 and 2018 we undertook a new multi-epoch
spectroscopic survey of the Arches, reported in an ongoing series of papers.  In Clark et al.
(\cite{clark18a}, \cite{clark19b}; Papers~I and~III, respectively) we combined multiple epochs to produce  spectra reaching stars as
faint as O9.5V for the first time in order to produce a cluster census. Quantitative
model-atmosphere analysis of these data will facilitate the construction of an HR diagram and the
calibration of the luminosity and (initial) mass functions for the cluster. A parallel goal of our
programme was to identify massive binaries via radial velocity (RV) variability across epochs in the
brightest cluster members, presenting the first example - the eclipsing WN8-9h + O5-6Ia$^+$ binary
F2 - in Lohr et al. (\cite{lohr}; Paper~II). In this paper we present the results for the remaining cluster
members bright enough to permit such analysis. This yields a robust sample size of 36 stars,
allowing us to address a number of the outstanding issues regarding the properties and formation of massive binaries ($\gtrsim40M_{\odot}$; Sects. 4 and 5) in an  extreme physical
environment. Moreover, the long observational baseline  available to us (Sect. 2) allows us to
investigate secular variability in this population.

The paper is therefore structured as follow. In Sect. 2 we define our stellar sample and discuss the
spectroscopic and photometric datasets employed in this work. Sect. 3 presents an overview of these
datasets and, where appropriate, detailed discussion of individual stars,  supplemented by extant
radio and X-ray observations. We provide quantitative model-atmosphere analysis of binary candidates
in Sect. 4, utilising these to infer current and initial masses for system components. We discuss
the implications of these results in Sect. 5 in terms of  the binary frequency and properties of
very massive stars and, subsequently, the nature  of their formation mechanism and evolutionary
pathways. Finally we summarise our conclusions in Sect. 6. We also provide three appendices; Sect
A.1 provides  a comprehensive observational log of our Arches spectroscopic observations; Sect. B
discusses the implications of the lack of secular variability evident in the spectroscopic dataset;
Sect. C presents a comprehensive  assessment of the binary fraction amongst WNLha stars within
stellar clusters and associations and includes a detailed reappraisal of the X-ray properties of
single and binary star examples.

\section{Observations}

\subsection{Spectroscopic data and reduction}

The SINFONI integral field spectrograph on the ESO/VLT (Eisenhauer et al. \cite{eisenhauer}, Bonnet
et al. \cite{bonnet}) was used in service mode to make repeated observations of overlapping fields
covering the central region of the Arches, and several distinct outlying fields, in the $K$ band, in
2011, 2013, 2017 and 2018\footnote{ESO programmes 087.D-0317, 091.D-0187, 099.D-0345 and
0101.D-0141.}. The long baseline imposed by these multi-semester observations has the advantage of
increasing sensitivity to long period systems. Unfortunately, however, the uneven temporal sampling
imposed by observations undertaken in service mode complicates future Monte Carlo analysis of the
complete dataset.

 A further epoch of SINFONI observations for outlying fields from 2005 was furnished by data cubes
used for Martins et al. (\cite{martins08})\footnote{ESO programme 075.D-0736.}.  For seven
Wolf-Rayet targets, observations were also made in 2014 with KMOS/VLT (Sharples et al.
\cite{sharples})\footnote{ESO programme 093.D-0306.}, as part of a wider survey of massive stars
across the central molecular zone (Clark et al. \cite{clark18c}).  Lohr et al. (\cite{lohr}) also
give details of additional archival SINFONI and Keck/NIRSPEC spectra used in the analysis of F2;
these are not considered here in the interests of survey consistency.

A detailed account of the acquisition and reduction approach taken for the SINFONI observations is
given in Clark et al. (\cite{clark18a}, \cite{clark19b}); for KMOS, details are presented by Clark et
al. (\cite{clark18c}).  For our purposes here, we note particularly that where multiple observations
had been made of a science target within the same epoch (in practice, within a few hours of each
other on the same night) their spectra were combined to give a higher S/N spectrum.  However,
distinct epochs were not then combined as described in Clark et al. (\cite{clark18a}), because our
goal here was the investigation of inter-epoch variability for the brightest objects.  
The final SINFONI spectra have a spectral resolving power ($\Delta\lambda/\lambda$) of $\sim9000$ and cover the wavelength range 2.02--2.45~$\mu$m.

\subsection{Stellar sample}

Multiple epochs of useful spectra were obtained for the majority of Arches sources brighter than F40
(i.e. up to number 40 in the list of Figer et al. (\cite{figer02}); objects fainter than this proved
to have radial velocity uncertainties comparable in size to their inter-epoch variability.
Additionally, stars B1 and B4 (numbers 1 and 4 in the list of Blum et al. \cite{blum01}) were
confirmed as massive cluster members in Clark et al. (\cite{clark18a}), and so were added to the
sample.  F39 was not included because it was only successfully observed on a single night in 2018
(it lies some distance south of the cluster, and was not surveyed in any previous runs; see Clark et
al. (\cite{clark19b}) for its spectrum and classification); F31 and 37 are located too close to
other, brighter stars for uncontaminated pixels to be extracted; F11 and F36 proved to be cool star
interlopers.  This leaves a total of 37 stars for which spectra were extracted on at least five
epochs.  Table~\ref{rvtable} summarises statistics on the number of epochs covered for the studied
targets, while Table~\ref{tab:appA1} gives the individual dates of observations of  each object.
Our RV sample consists of thirteen Wolf-Rayet stars (two WN7-8ha and eleven WN8-9ha subtypes), seven
O hypergiants (from O4-5 to O7-8Ia$^+$; the secondary of F2, classified from disentangled spectra in
Lohr et al. (\cite{lohr}), is not counted here), and seventeen O supergiants (O4-6.5Ia). Thus, while
we are complete for both WNLha stars and O hypergiants we only sample 17/30 of the supergiants
(Clark et al. \cite{clark19b}).

\subsection{Radial velocity measurements}

Radial velocities (RVs) were measured by cross-correlation using IRAF's \emph{fxcor} task.  To
obtain an initial and consistent measure of RV variability for all stars in the sample, in the
absence of suitable independent spectral templates, all possible pairs of epochs for a target were
cross-correlated with each other.  This yielded an $n \times n$ distance matrix, where $n$ is the
number of epochs observed, with zeroes along the diagonal where a spectrum was cross-correlated with
itself.  The sum of each row in this matrix then gave an estimate of the RV of the zero-element in
that row, relative to the middle of the range of RVs.  The sum of each column gave a second estimate
of the RV of the zero-element in that column, multiplied by -1.  A simple average was taken of these
two estimates for each velocity, and uncertainties were also determined as the average of the
uncertainties in the contributing cross-correlation measurements. We refrained from using weighted
averages as it was found to bias the results in favour of smaller RVs which could generally be
measured more precisely.  At this stage, SINFONI and KMOS spectra (where the latter were available)
were studied separately, to produce a consistent set of results.

Owing to the range of spectral types of the stars observed, from Wolf-Rayets through O hypergiants
to O supergiants, no single set of lines in the $K$ band could be used for all targets.  For
example, in the WNLh spectra, Br~$\gamma$ is the strongest emission line, while C\,{\sc iv} lines
(2.070--2.084~$\mu$m) are weak to non-existent; conversely, in the O supergiants, C\,{\sc iv} are
the strongest lines, with very weak Br$\gamma$ due to substantial wind filling of the line.
Moreover, many WNLh lines are significantly broadened or show P~Cygni profiles associated with wind
effects, while in O stars these features are not evident; there is also a significant blended line
(He\,{\sc i}, N\,{\sc iii}, C\,{\sc iii}, O\,{\sc iii} at 2.112--2.115~$\mu$m).  For these reasons,
fitting of Gaussian or Lorentzian profiles to individual lines was not judged suitable for obtaining
RVs across our sample. Cross-correlations were therefore carried out using three overlapping sets of
relatively narrow lines:
\begin{enumerate}
  \item He\,{\sc ii} at 2.037 and 2.189~$\mu$m, and N\,{\sc iii} at 2.247 and 2.251~$\mu$m for all
WNLh targets;
  \item The same lines as in (1) with the addition of C\,{\sc iv} at 2.07--2.084~$\mu$m for a subset
of WNLh stars in which these lines are stronger;
  \item He\,{\sc i} at 2.058~$\mu$m, C\,{\sc iv} at 2.07--2.084~$\mu$m, the blend at
2.112--2.115~$\mu$m, and He\,{\sc ii} at 2.189~$\mu$m for the O hypergiants and supergiants.
\end{enumerate}
Given our use of observed spectra of each target as templates, our RVs are necessarily relative
rather than absolute.  However, our approach to detecting variability in a target relies upon the
maximum significance of an RV difference between any two pairs of epochs, as used by various authors (Sana et al.
(\cite{sana13a},
Dunstall et al. (\cite{dunstall}),
Patrick et al. (\cite{patrick19}, \cite{patrick20}),
Ritchie et al. (\cite{ritchie21}):

\begin{equation}
\sigma_\mathrm{det} = \mathrm{max}\left(\frac{|v_i-v_j|}{\sqrt{\sigma_i^2+\sigma_j^2}}\right),
\end{equation}

\noindent where $v_{i,j}$ and $\sigma_{i,j}$ are the RVs and their associated uncertainties at epoch
$i$ or $j$.  Therefore, absolute RVs are not required.  

In order to identify variability associated with likely binarity, Sana et al. (\cite{sana13a})
required stars to exhibit both $\sigma_\mathrm{det}>4.0$ and at least one pair of measurements from
distinct epochs to satisfy $|v_i-v_j|>$20~km~s$^{-1}$. The former criterion ensures minimal false
positives; important given the  combination of large sample size but comparatively low observational
frequency  of  their study. The latter threshold was chosen to minimise contamination from pulsating
sources within the sample and was motivated by the behaviour of blue super-/hypergiants within
Westerlund 1 (Ritchie et al. \cite{ritchie}). A higher 25~km~s$^{-1}$ threshold is adopted by Ritchie
et al. (\cite{ritchie21}) to accommodate the larger pulsational amplitudes observed in the early-B
supergiant population of Westerlund 1, but these later-type objects are not present in the Arches.
Foreshadowing Sect. 3, we find no evidence for secular variability in the spectral morphologies of
our targets, 
as is observed in the lower-mass blue super-/hypergiant contingent within Westerlund 1 (Ritchie et al.
\cite{ritchie}, Clark et al. \cite{clark10}), and we regard the 20km~s$^{-1}$ threshold adopted by
Sana et al. (\cite{sana13a}) as appropriate for this study. 

The situation for the WNLh stars is different as the  emergent emission line spectrum originates in
an optically thick  wind; as a consequence we would not expect to observe pulsationally driven line
profile variability. However, a number of studies have revealed modification of line profiles in
(apparently) single WRs due to the presence of both small-scale wind clumping and large-scale
rotating structures (e.g. L\'{e}pine \& Moffat \cite{lepine}, St-Louis et al. \cite{stlouis}).
However the effect of the former is the superposition  of additional, small sub-peaks on the
stationary emission profile  (L\'{e}pine \& Moffat \cite{lepine}) - which would not be expected to
introduce global RV shifts. Fortunately, globally structured winds are only seen in a small subset
of WRs, none of which are of comparable spectral types to the Arches cohort (St-Louis et al.
\cite{stlouis}), while this phenomenon does not impart motion to the whole emission profile. For
completeness, Stevens \& Howarth (\cite{stevens99}) report significant line profile variability in
WR colliding wind binaries, but clearly we would not wish our detection threshold to exclude such
systems. Nevertheless, the potential for line profile variability to be ubiquitous amongst WRs still
requires us to chose an appropriate velocity cut for such stars. Fortuitously, Schnurr et al.
(\cite{schnurr08a}, \cite{schnurr09a}) searched for binaries amongst  the WNLh cohort of both
NGC3603 and R136 with an identical experimental set up (VLT/SINFONI) and wavelength range (K-band)
to that employed in this study, finding that apparently single  stars showed random variability of
$\Delta{\rm RV} <20$km~s$^{-1}$. 

Therefore, upon consideration of the above, we adopt thresholds of $\sigma_\mathrm{det}>4.0$ and
$|v_i-v_j|>$20~km~s$^{-1}$ for binarity for both the WNLh stars and O super-/hypergiants within the
Arches, allowing direct comparison to the studies highlighted above. After an initial identification
of candidate binaries via these criteria, RVs were measured again for each species individually, to
assess the most useful lines for determining precise velocities; moreover, the combined spectra of
non-variable Arches members with identical or very similar classifications were used as templates
for cross-correlation. Orbital periods were then sought, using both Lomb-Scargle periodograms (Lomb
\cite{lomb}, Scargle \cite{scargle}, Horne \& Baliunas \cite{horne}) and a form of string length
minimization (Dworetsky \cite{dworetsky}).

Given the limitations of our dataset in terms of sampling, false periods arising from noise and
aliasing may be problematic. In order to address this issue we employed $k-1$ cross validation to
assess the reality of any periodicities returned. In each resampled subset false periods in the
periodogram are expected to vary in strength and/or location, while a peak resulting from the true
orbital period will be robust against such a test, providing confirmation of its physical nature. 

\subsection{Photometry}

$K_s$ band time-series photometry of the Arches cluster (excluding B1 and F39) was obtained by
Pietrzynski et al.\footnote{ESO proposal 081.D-0480.} between 2008 and 2009 using NAOS-CONICA (NaCo)
on the VLT (Lenzen et al. \cite{lenzen}, Rousset et al. \cite{rousset}).  This study was briefly
reported in Markakis (\cite{markakis11}, \cite{markakis12}).  We reprocessed the raw archival data as
described in Lohr et al. (\cite{lohr}), and thus generated light curves for the brighter objects
which were present in all observations.  Non-variable sources were identified and combined to
provide a composite reference star, relative to which differential light curves could be constructed
for variable objects.

\section{Data presentation and results}

\begin{table*}
  \begin{center}
  \caption{RV variability results for Arches members, ordered by spectral classification.}     
  \label{rvtable}
  \begin{tabular}{l l c c c c c l l}
    \hline
    Star & Spectral Type & $\Delta$RV & $\sigma_\textrm{det}$ & Line & No. of  & X-ray & Radio properties & Alt. Star \\
    ref. & & (km~s$^{-1}$) & & set & epochs & detection? & & ref.\\
    \hline
        &         &     & & & & & \\
    B1  & WN8-9ha & 5.5 & 1.25 & 1 & 5                                        &  & thermal & WR~102bc\\
        &         & \textit{16.1} & \textit{1.72} & & \textit{7}                     &  & \\
    F1  & WN8-9ha & 20.9 & 2.18 & 2 & 14                                      & & thermal, variable & WR~102ad \\
        &         & \textit{5.9} & \textit{1.34} & & \textit{2}                      &  & \\
    F2  & WN8-9ha & \textbf{364.1} & \textbf{32.61} & 1 & 6                  & $\bullet$ & composite & WR~102aa\\
        &         & \textit{\textbf{280.7}} & \textit{\textbf{17.31}} & & \textit{7} & &  \\
    F3  & WN8-9ha & 9.8  & 1.97 & 1 & 7                               &  & thermal  & WR~102bb\\
    F4  & WN7-8ha & 19.4 & 5.36 & 1 & 10                                     & & thermal & WR~102al\\
    F5  & WN8-9ha & 15.4 & 5.63 & 1 & 7                                       &  & composite, variable & WR~102ai\\
    F6  & WN8-9ha & \textbf{65.4} & \textbf{7.99} & 2 & 10                     &$\bullet$ & composite, variable &   
WR~102ah \\
        &  & \textit{54.2} & \textit{2.99} & & \textit{7}                      & & \\
    F7  & WN8-9ha & \textbf{43.8} & \textbf{10.12} & 1 & 15                    &$\bullet$ & thermal, variable & WR~102aj\\
    F8  & WN8-9ha & \textbf{20.9} & \textbf{6.05} & 1 & 10                    &  & thermal & WR~102ag\\
        &         & \textit{14.9} & \textit{1.40} & & \textit{7}                  & & \\
    F9  & WN8-9ha & 27.9 & 3.44 & 2 & 15                                       &$\bullet$ & composite, variable & WR~102ae\\
        &         & \textit{13.6} & \textit{1.00} & & \textit{6}                  & & \\
    F12 & WN7-8ha & 19.0 & 2.33 & 2 & 10                       &  & non-thermal & WR~102af\\
    F14 & WN8-9ha & \textbf{32.8} & \textbf{8.39} & 2 & 11                     &  & unconstrained & WR~102ba\\
    F16 & WN8-9ha & \textbf{26.0} & \textbf{5.90} & 2 & 10                     &  & unconstrained & WR~102ak\\
\\
    F10 & O7-8 Ia$^+$ &  9.8 & 2.17 & 3 & 10                                   &  & - & [BSP2001] 30 \\
    F13 & O7-8 Ia$^+$ & 29.5 & 2.88 & 2 & 9                                    &  & - & [BSP2001] 31 \\
    F15 & O6-7 Ia$^+$ & \textbf{68.4} & \textbf{13.66} & 3 & 16                &  & - & [BSP2001] 8 \\
    F17 & O5-6 Ia$^+$ & 18.2 & 2.72 & 2 & 10                                   &  & - & [BSP2001] 29\\
    F18 & O4-5 Ia$^+$ & \textbf{32.2} & \textbf{4.56} & 3 & 10                 & & non-thermal, variable & [BSP2001] 20\\
    F27 & O4-5 Ia$^+$ & 30.9 & 3.16 & 3 & 14                   & & - & [BSP2001] 16\\
    F40 & O4-5 Ia$^+$ & 53.9 & 3.58 & 3 & 15                                   & & - & -\\
\\
    B4 & O5.5-6 Ia & 23.3 & 3.18 & 3 & 11                                      & & - & -\\
    F19 & O4-5 Ia &  -- & -- & 3 & 5                                           & & non-thermal &    [LGR2001] AR6 \\
    F20 & O4-5 Ia & 30.6 & 2.99 & 3 & 14                       & & - & -\\
    F21 & O6-6.5 Ia  & 31.7 & 3.15 & 3 & 15                                    & & - & [BSP2001] 7\\
    F22 & O5.5-6 Ia & 27.2 & 3.30 & 3 & 9                                      & & - & [BSP2001] 27\\
    F23 & O6-6.5 Ia & 20.8 & 2.58 & 3 & 6                                      & & - & [BSP2001] 2\\
    F24 & O4-5 Ia & 25.8 & 2.53 & 3 & 5                       & & - & -\\
    F25 & O4-5 Ia & \textbf{157.1} & \textbf{12.51} & 3 & 12 & & - & -\\
    F26 & O4-5 Ia & 53.3 & 2.78 & 3 & 8                                        & & unconstrained & [BSP2001] 18\\
    F28 & O4-5 Ia & \textbf{32.5} & \textbf{4.15} & 3 & 10                     & & - & - \\
    F29 & O5.5-6 Ia & \textbf{36.5} & \textbf{4.48} & 3 & 16                   & & - & [BSP2001] 9\\
    F30 & O4-5 Ia & 14.1 & 1.29 & 3 & 8                                        & & - & - \\
    F32 & O4-5 Ia & \textbf{43.9} & \textbf{5.24} & 3 & 10                     & & - & [BSP2001] 15\\
    F33 & O4-5 Ia & 35.9 & 3.14 & 3 & 10                                       & & - & [BSP2001] 13\\
    F34 & O4-5 Ia & 31.8 & 3.23 & 3 & 10                                       & & - & [BSP2001] 5\\
    F35 & O4-5 Ia & \textbf{207.3} & \textbf{12.74} & 3 & 16                   & & - & [BSP2001] 10\\
    F38 & O4-5 Ia & 28.8 & 3.39 & 3 & 9                        & & - & -\\
\\
  \hline
  \end{tabular}
    \end{center}

{Rows in italics correspond to KMOS observations.  Values in bold face meet our criteria for
significant variability associated with likely binarity. X-ray detections given by bullet points in
column 7 (Wang et al. \cite{wang}). Radio properties of individual stars derived from Lang et al.
(\cite{lang01}, \cite{lang05}) and Gallego-Calvente et al. (\cite{GC}) are summarised in column 8;
thermal sources are those with spectral indices consistent with expectations for a partially
optically thin wind ($\alpha{\sim}0.6$), non-thermal sources demonstrate $\alpha < 0.0$, while
composite sources lie between these extremes (Sect. 3.3). Three stars are only detected at one
frequency; hence their spectral indices are unconstrained.
Note that the targets can be accessed in SIMBAD via column one by replacing  \textquoteleft F\textquoteright with \textquoteleft [FNG2002]\textquoteright, or \textquoteleft B\textquoteright with \textquoteleft [BSP2001]\textquoteright.}
\end{table*}

We present the results of our initial RV variability measurements - determined independently for
SINFONI and KMOS observations - in Table~\ref{rvtable}. Before discussing these data in detail we
briefly note that none of the 37 targets showed evidence for long term, secular spectral
variability; we discuss this observational finding and the implications in Section~\ref{sub:sec_var} and Appendix B.

It proved possible to obtain $\Delta$RV values for all members of the sample except F19, which is a
close neighbour of the much brighter (and photometrically variable) F2 (Lohr et al. \cite{lohr}).
As a consequence we suspect the spectra of F19 were contaminated by F2 to varying extents in
different epochs: its cross-correlation function was not well-defined for any pairs of its five
epochs of spectra.  A few other objects had one or more epochs which proved to be strong RV
outliers. While these could derive from  highly eccentric binaries, such measurements generally
corresponded to poorer-quality observations with much larger uncertainties than the other epochs; as
such we proceed under the assumption that these values are spurious. The $\Delta$RV values obtained
after removal of these presumed outliers are also given in Table~\ref{rvtable}, and are used for
subsequent analysis.

\begin{figure}
\includegraphics[angle=0,width=\columnwidth]{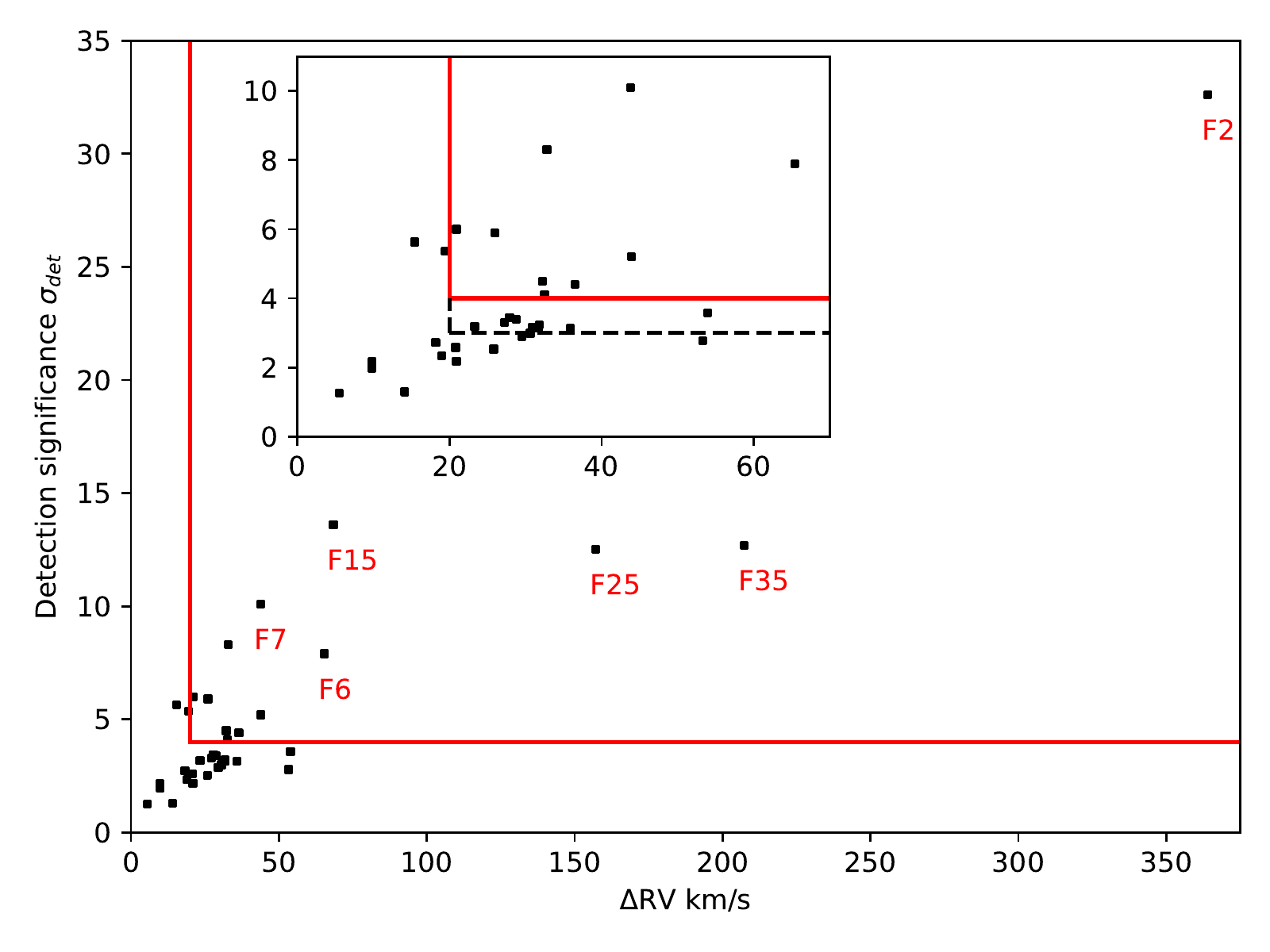}
\caption{Distribution of $\Delta$RV against $\sigma_\mathrm{det}$ for
  36 bright Arches members.  Our thresholds for significance and
  amplitude of variability suggesting binarity are marked with solid
  red lines. Stars for which searches for orbital solutions have been attempted 
 are labelled.
 The inset shows in more detail the majority of our targets.
 The black dashed line shows a less conservative,  $\sigma_\mathrm{det} >3.0$, threshold for binary detection.}
\label{deltarv}
\end{figure}

Fig.~\ref{deltarv} shows the distribution of the measured objects in
$\Delta$RV-$\sigma_\mathrm{det}$ space and their location with respect to the thresholds for
binarity we adopted: $\sigma_\mathrm{det}>4.0$ and $|v_i-v_j|>$20~km~s$^{-1}$. F2 stands out as
exhibiting the greatest amplitude of variability ($\Delta$RV$\sim$360~km~s$^{-1}$) and the highest
significance ($\sigma_\mathrm{det}\sim$33) of the SINFONI measurements, and the only significant
result determined from the KMOS dataset in isolation. No other object provides such clear-cut
evidence; however, thirteen targets meet our criteria for significant variability suggestive of
binarity: 6/13 WNLh stars, 2/7 O hypergiants and 5/16 O supergiants (Table~\ref{rvtable}).
This suggests an RV-derived binary fraction of 13/36 ($\sim36$\%) before any correction for observational biases.
As described in Sect.~2.3, the choice of $\sigma_\mathrm{det}>4.0$ reflects our desire to limit false positive detections in the dataset; that being said, given the sample size, statistically, one would also expect 0 false positives assuming $\sigma_\mathrm{det}>3.0$ as a threshold for binarity.
In this scenario, 22/36 ($\sim61$\%) meet the RV variability criteria for binarity.
This additional subset consists of 1 WNLh star, 2 O hypergiants and 6 O supergiants.

In addition to F2, the amplitude of RV variability of five cluster members - F6, F7, F15, F25, and
F35 - is sufficient to attempt a provisional determination of their orbital parameters. These stars
are discussed individually in Sect. 3.1. While the remaining seven sources - F8, F14, F16, F18,
F28, F29, and F32 - are also {\em bona fide} RV variables, a combination of fewer epochs of
observations and lower amplitude RV excursions precluded the identification of  
periodicities at this time. In the case of the supergiant cohort these limitations are compounded by
their weaker spectral features and lower S/N spectra. Despite their RV variability we highlight
that none of these latter sources appear to be SB2 binaries, nor do they show evidence for line
profile variability indicative of binarity.

In the case of the RV variables with $3.0 < \sigma_\mathrm{det} <4.0$, F40 stands out with $\Delta$RV$\sim$54~km~s$^{-1}$ and 15 epochs.
Analysis of this system strongly suggests that it is also a short-period binary, but, given the larger uncertainties for the RV measurements for this target (see Table~\ref{tab:appA1}), determination of orbital parameters would require further observations.

No further eclipsing or periodically varying systems were identified in the photometric dataset
analysed; where appropriate these data are discussed below for individual objects. Finally we may
make use of additional, extant  X-ray and radio observations to identify further binaries; these
datasets - and the conclusions deriving from them - are discussed in Sect. 3.3.

\subsection{Individual sources} 

\subsubsection{F6}

\begin{figure}
\includegraphics[width=\columnwidth]{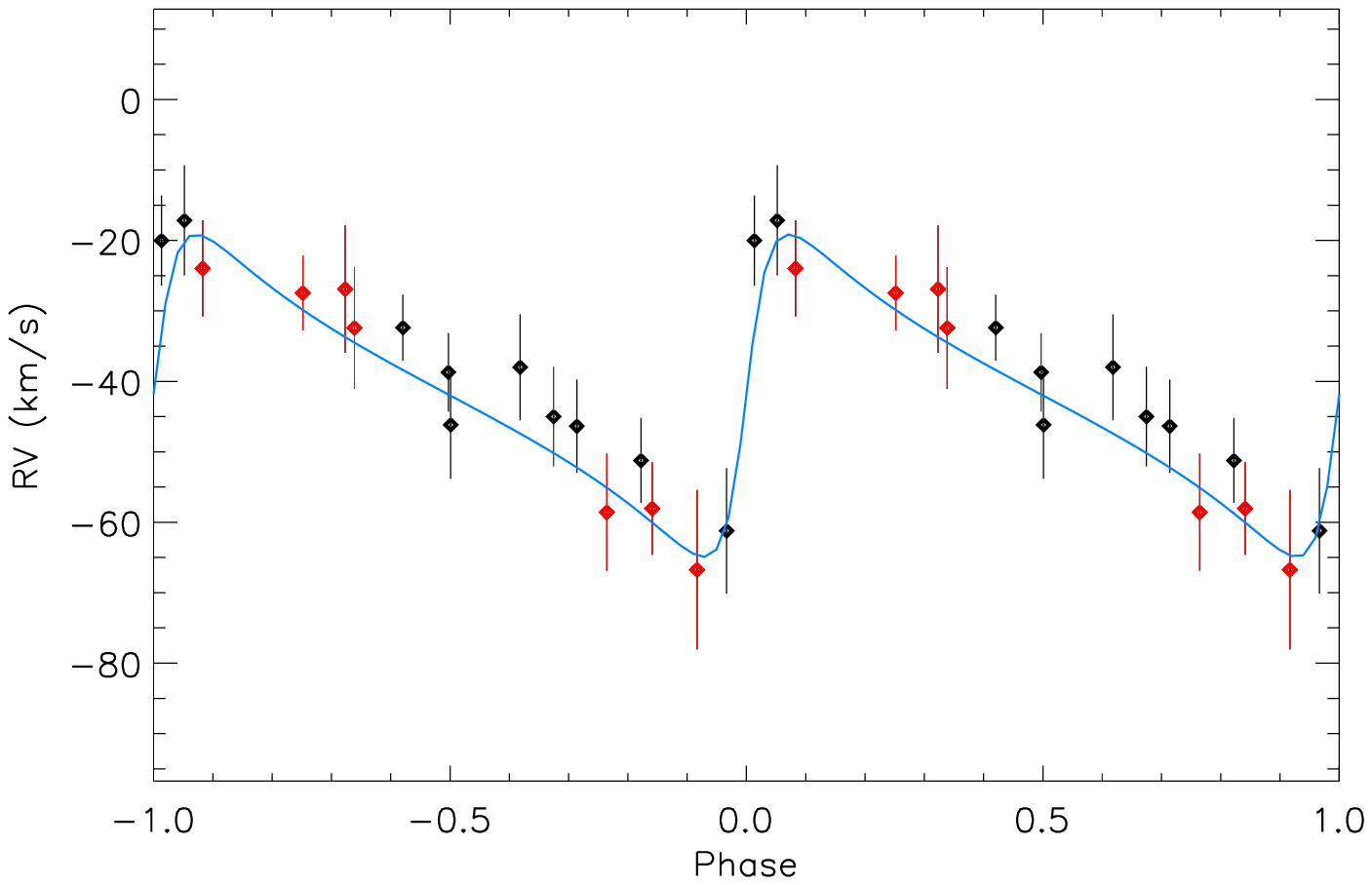}
  \caption{RV curve for F6, cross-correlated relative to F1, measured from three spectral regions containing He\,{\sc ii} lines, folded on
    $P$=13.378~d.  RVs and uncertainties from SINFONI spectra are shown
    in black; those from KMOS are shown in red.  The blue line
    corresponds to a possible model system with $e$=0.6,
    $\omega$=270\degr.}
  \label{f6rvplot}
\end{figure}

F6 (WN8--9ha) is perhaps the brightest source in the Arches, and is projected against the cluster
core. He\,{\sc ii} lines at 2.037, 2.189 and 2.346~$\mu$m and C\,{\sc iv} lines at 2.070 and
2.078~$\mu$m were used to determine RVs, and F1 was selected as a cross-correlation template: this
object is also classified as WN8-9ha and exhibits a very similar appearance to F6 in terms of line
strengths and profiles, but does not exhibit significant RV variability. 

We find $\Delta$RV$\sim$65~km~s$^{-1}$ and $\sigma_\mathrm{det}\sim$8 from ten SINFONI
spectra, while seven KMOS spectra from 2014 independently yield $\Delta$RV$\sim$55~km~s$^{-1}$.
Fig.~\ref{f6rvplot} shows the RV curve derived from the He\,{\sc ii} line set, folded to a period of
13.378$\pm$0.004~days derived from a string-length search\footnote{No periods with false-alarm
probabilities below 10$\%$ were found using the Lomb-Scargle method; this is perhaps unsurprising
given that the RV curve appears strongly non-sinusoidal in shape}. We find a full amplitude of $\sim50\pm5$~km~s$^{-1}$ and orbital eccentricity $e\sim0.6$. Similar results are obtained from
the C\,{\sc iv} line set, the strong blended feature at 2.112--2.115~$\mu$m, and a combination of
C\,{\sc iv} lines, He\,{\sc ii} at 2.189~$\mu$m and N\,{\sc iii} at 2.247 and 2.251~$\mu$m.  Checks
were also made using F5 (another relatively RV-invariant WN8-9ha star) as a template, yielding
consistent, if somewhat noisier, results.

This solution suggests a relatively low-mass companion, and/or a moderate angle of inclination,
which may in turn explain why F6 exhibits relatively little photometric variability and no
significant periodicity in its light curve. There is, however, possible evidence for line profile
variability, with weaker Br~$\gamma$ and a less marked P~Cygni profile for the He\,{\sc ii}
2.189~$\mu$m line in a maximum-RV spectrum, as compared with a phase 0 spectrum, in which both
components' features would be expected to be aligned. Nevertheless, the RV data presented here
strongly support the identification of F6 as a massive binary, and its high eccentricity would imply
a pre-interaction system. 

\subsubsection{F7}

\begin{figure}
\includegraphics[width=\columnwidth]{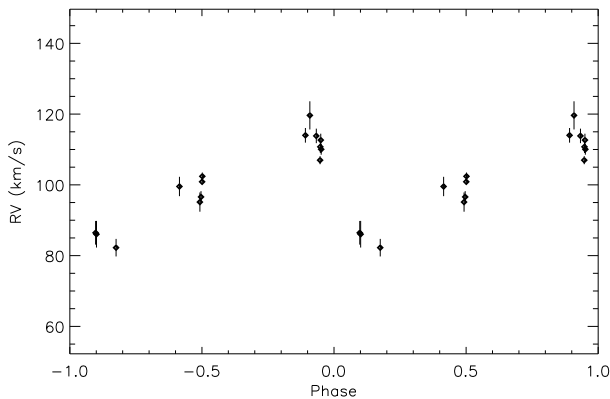}
\caption{SINFONI RV curve for F7, cross-correlated relative to F5, measured from the
  Br~$\gamma$ line, and folded on $P$=1184.46~d.}
\label{f7rvplot}
\end{figure}

F7 is another very bright WN8-9h star  in the cluster core. RVs were measured using F5 as a
template, and cross-correlating over a range of discrete wavelength regions. The best results were
produced by the very strong Br~$\gamma$ line, revealing variability with a significance of
${\sigma}_{\rm det}\sim10$ and an amplitude of $\sim$40~km~s$^{-1}$. It is notable that the RVs within each season of observation are very similar\footnote{F7 was observed seven times in 2013, three times in 2017, and five times in 2018.}, implying a long period system and effectively reducing to three the number of separate RV measurements; systematic errors can be excluded, as other nearby bright targets extracted from the same data cubes do not exhibit the same RV shifts. 
The orbital period is not well constrained by the current data; the shortest
plausible period is around 3.24~years (see Fig.~\ref{f7rvplot}), but the true period may be
significantly longer given the degeneracies in the interpretation of the RV curve with three effective RV measurements, which means that longer orbital periods are also possible. Photometric data shows very little variability (0.05~mag), and there are also
no obvious line profile changes between seasons.  

\subsubsection{F15}
\begin{figure}
  \includegraphics[width=\columnwidth]{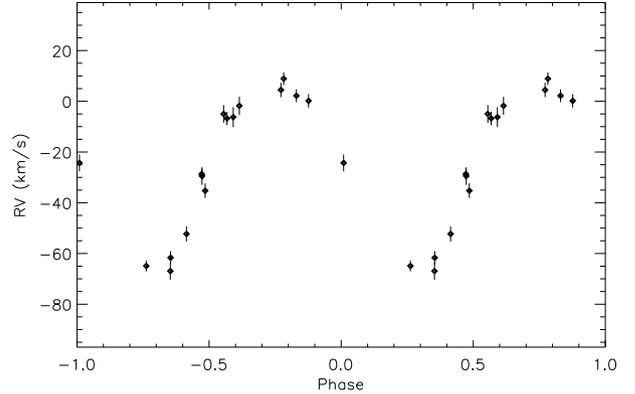}
  \caption{SINFONI RV curve for F15, cross-correlated relative to F10, measured in the region of
    the blended spectral feature at 2.112--2.115~$\mu$m, and folded on
    $P$=83.95~d.}
  \label{f15rvplot}
\end{figure}

F15 (O6-7 Ia$^+$) lies near the centre of the cluster, and has the second highest detection
significance evidence in our sample. Like the other O hypergiants in the Arches, F15's spectrum
exhibits a number of relatively strong and narrow features, especially the C\,{\sc iv} lines at
2.070 and 2.078~$\mu$m, He\,{\sc i} at 2.059~$\mu$m, and the blend at 2.112--2.115~$\mu$m (Sect. 4).
Improved RVs were determined in these wavelength regions using F10 (O7-8~Ia$^+$) as a template,
since it showed very little evidence for variability itself; F17 (O5-6~Ia$^+$) was also used as a
check comparator.

Period searches using both Lomb-Scargle and string length minimization approaches give a period of
83.9~days with false-alarm probability $<1\%$. The cleanest folded RV curve (shown in
Fig.~\ref{f15rvplot}) was produced using the blend region at 2.112--2.115~$\mu$m, with the C\,{\sc
iv} lines also producing a very similar result.  Using these two sets of RVs, we converged on a
best consensus period of 83.95$\pm$0.15~days. The RV curve is approximately sinusoidal, and has a
full amplitude around $75\pm5$~km~s$^{-1}$. No notable line profile changes are observed across the
orbital cycle. F15 is a low amplitude photometric variable ($\sim0.06$~mag), with no evidence of
periodic modulation, which is unsurprising given its period and the limited duration and temporal
sampling of the dataset. The orbital period of nearly three months suggests a detached
pre-interaction system, with the relatively low amplitude and lack of photometric or line profile
variability suggesting that the secondary is of lower mass and luminosity, and/or low inclination.

\subsubsection{F25}

\begin{figure}
  \includegraphics[width=\columnwidth]{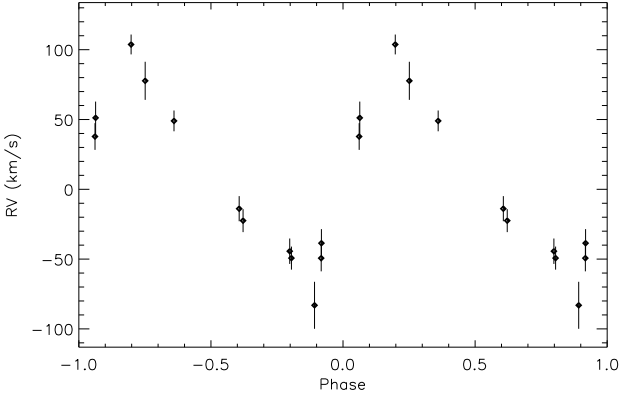}
  \caption{SINFONI RV curve for F25, cross-correlated relative to F23, measured in the
    regions $\sim$2.05--2.13~$\mu$m and 2.18--2.20~$\mu$m, and folded on
    $P$=6.643~d.}
  \label{f25rvplot}
\end{figure}

F25 (O4-5~Ia) is located southwest of the cluster core in the vicinity of F2. Like the other O
supergiants, it is less luminous than the WNLs and O hypergiants in the cluster, and its spectral
features are relatively weak; consequently the individual epochs are noisier, making it more
difficult to obtain reliable RVs. F23 was selected as a template and the regions
$\sim$2.05--2.13~$\mu$m and 2.18--2.20~$\mu$m were included in the cross-correlation in order to
improve RV measurement. 

The most compelling orbital solution, shown in Fig.~\ref{f25rvplot}, suggests a period of
$6.643\pm0.002$~days, with pronounced eccentricity ($e=0.35\pm0.09$) and a full amplitude of
$\sim165\pm15$~km~s$^{-1}$, although an alternative candidate period of $\sim$5.45~days with similar
eccentricity and amplitude cannot be excluded\footnote{A number of other candidate periods below ten
days have false-alarm probabilities 5--10$\%$.}. Both solutions imply a pre-interaction system with
a massive companion. F25 exhibits a moderate degree of photometric variability ($\sim$0.2~mag), but
no obvious periodicity was detected in its light curve. Further high S/N observations will be
required to provide a unique solution. 

\subsubsection{F35}
\begin{figure}
  \includegraphics[width=\columnwidth]{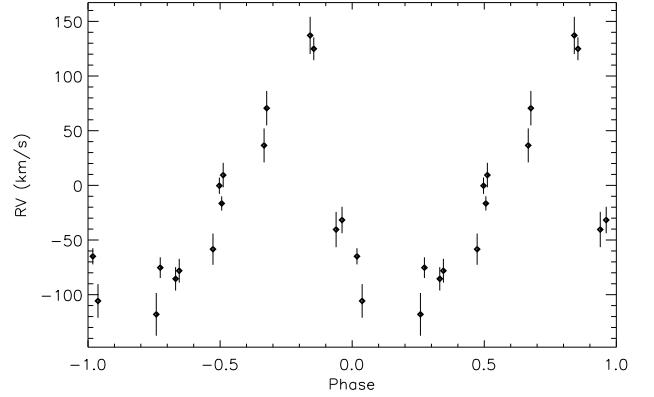}
  \caption{SINFONI RV curve for F35, cross-correlated relative to F23, measured in the
    regions $\sim$2.05--2.13~$\mu$m and 2.18--2.20~$\mu$m, and folded on
    $P$=6.043~d.}
  \label{f35rvplot}
  \end{figure}

Classified as O4-5~Ia and lying near the centre of the cluster, F35 is only slightly brighter than
F25 and hence exhibits similar limitations in determining reliable RVs. F23 was again used as a template, with the same wavelength regions for cross-correlation as were adopted for F25. 

A best-fit period of $6.043\pm0.003$~days was found, with eccentricity $0.44\pm0.06$ and a full
amplitude $260\pm10$~km~s$^{-1}$ that is the second largest in our sample, after F2. This solution is
plotted in Fig.~\ref{f35rvplot}. Unlike F25, no strong alternative candidate periods are found.
F35's photometric variability is limited (0.08~mag) and shows no obvious periodicity near six days.
However, comparison of spectral line profiles from observations made during the assumed secondary
eclipse and the two quadratures of the proposed $\sim$6.043~d orbital period does suggest a
consistent shape difference: when the two assumed binary components have a common RV (around phase
0.5), the profile is narrower than at either quadrature (when the components are maximally separated
in wavelength). 

Although further epochs would be desirable to confirm the period of F35, its very large amplitude of
RV variability and phase-dependent line profile changes support its identification as another
eccentric, short-period massive binary in the Arches. The broadening of spectral lines at
quadratures suggests that the companion is of similar spectral type to the O supergiant primary; the
high RV amplitude also supports a companion of comparable mass to the primary.

\subsection{Ancillary radio and X-ray observations} 

While spectroscopic and photometric methodologies provide robust binary detections, other indirect
diagnostics also exist. Specifically, shocks in the wind collision zones of massive binaries produce
high temperature plasma that leads to excess hard, thermal X-ray emission over that expected from
single stars (Stevens et al. \cite{stevens}, Pittard \& Dawson \cite{pittard}) as well as a
population of relativistic electrons that yield non-thermal synchrotron emission at radio
wavelengths (Williams et al. \cite{williams}, Chapman et al. \cite{chapman}, Dougherty \& Williams
\cite{dougherty}, Van Loo et al. \cite{vanloo}).

\subsubsection{X-ray observations}

Wang et al. (\cite{wang}) report four X-ray detections of WNLha stars within the Arches: F6 (=A1S;
$L_X=1.1\times 10^{34}$~ergs$^{-1}$,  $kT\sim 2.2^{+0.4}_{-0.3}$keV),  
F7 (=A1N; $L_X=7.2\times 10^{33}$~ergs$^{-1}$,  $kT\sim 1.8^{+0.2}_{-0.2}$keV), F9 (=A2;
$L_X=4.6\times 10^{33}$~ergs$^{-1}$,  $kT\sim 2.5^{+0.4}_{-0.3}$keV), and F2 (= A6; no analysis
undertaken)\footnote{While finding an acceptable single temperature fit to the spectra, Wang et al.
(\cite{wang}) report that a two temperature fit comparable to that employed for the  comparable
Galactic Centre binary CXO J1745-28 (Table~\ref{pubin}; Mikles et al. \cite{mikles06}, Clark et al.
\cite{clark09b}) would lead to an increase in flux by over an order of magnitude.  The extreme
sensitivity of X-ray  luminosity to the model assumed is a result of the large column density
towards Galactic Centre sources; as such, we consider the luminosities adopted for these systems to
be lower limits. Moreover this finding  also acts as a caution when comparing the properties of
sources modelled with different assumptions and also stars observed at very different absorption
columns (cf. Table~\ref{tab:appC1}).}. Unfortunately, presumably due to a combination of the compact
nature of the cluster and the spatially variable diffuse background emission, no upper limits were
provided for the remaining cluster members. 

As described previously, F2, F6 and F7 all appear {\em bona fide} binaries by virtue of their RV
variability.  Our extensive RV observations of F9 (15 SINFONI epochs from 2005, 2011, 2013, 2017 and
2018, and 6 KMOS epochs from 2014) revealed no  evidence for variability, while  it is a low
amplitude ($<0.04$~mag) aperiodic photometric variable. Nevertheless, foreshadowing discussion of
its radio data (Sect. 3.2.2), we  conclude that F9  is a massive colliding wind binary due to the
similarity  of its X-ray properties to those of F2, F6, and F7 (and indeed other confirmed WNLha
binaries; Appendix C), with the lack of reflex motion potentially due to an unfavourable orbital
inclination. 

A number of other WNLha stars that appear strong binary candidates - for example the RV variables
F8, F14, and F16 - are not detected as X-ray sources. One might suppose these non detections derive
from either environmental issues (i.e. source blending/confusion in the confines of the cluster) or
the intrinsic properties of the binaries themselves. Specifically, while hard, overluminous X-ray
emission has traditionally been adopted as a key colliding wind diagnostic, Oskinova
(\cite{oskinova05}) and Rauw et al. (\cite{rauw15}) suggest that not all such systems conform to
this expectation. We discuss this further  in Appendix C, but note that this assertion is
exemplified by observations of Westerlund 1 (Clark et al. \cite{clark19c}), where binaries may (i)
exhibit the properties of single stars, (ii) be soft X-ray sources that are overluminous with
respect to expectations for the primary or (iii) appear as `canonical' hard, overluminous sources;
we note, however, that the hardest and brightest X-ray detections remain predominantly binaries. One
might suppose that the X-ray properties of colliding wind systems are a sensitive function of both
the nature of the stellar components (specifically wind momenta) and orbital configuration
(separation and eccentricity); this is particularly important  in the context of the Arches, where
the extreme column density to the cluster would preclude detection of binaries exhibiting soft X-ray
spectra, even if significantly overluminous. 

\subsubsection{Radio observations}
The Arches cluster has been the subject of three radio surveys of increasing sensitivity (Lang et al
\cite{lang01}, \cite{lang05} and  Gallego-Calvente et al. \cite{GC}). In total 16 cluster members
have been detected across these studies, with their properties summarised in Table~\ref{rvtable}.
Following Wright \& Barlow (\cite{wright}) thermal emission from a partially optically-thick,
isothermal stellar wind should have a characteristic spectral index, $\alpha\sim0.6$ (where $S_{\nu}
\propto \nu^{\alpha}$), with increasing optical depth leading to a steepening spectrum. Conversely,
non-thermal emission will have a negative spectral index, $\alpha <0.0$, with composite sources
comprising both thermal and non-thermal components being intermediate between these extremes.
Following this taxonomy, the radio emission associated with six stars appears thermal in origin (B1,
F1, F3, F4, F7, and F8), composite in four (F2, F5, F6, and F9), non-thermal  in three (F12, F18,
and F19) and unconstrained for the remaining three sources (F14, F16, and F26),  
which were only  detected at a single frequency.

We first consider the apparently thermal sources, and of these none show evidence for binarity with
the sole exception of F7, which is a confirmed binary by virtue of RV variability and X-ray
emission.  F7 is an important exemplar since it is known that massive binaries can be radio-variable, with occultation of the wind collision zone at certain orbital phases leading to varying
fluxes and spectral indices; indeed such sources sometimes present as thermal in origin (cf. WR140;
Williams et al. \cite{williams}). As such, we may not immediately conclude that the remaining
thermal sources within the Arches are single stars; in this regard it is notable that Lang et al.
(\cite{lang05}) report  F1 as radio-variable (although this behaviour could also arise from changes
in wind properties). 

Turning next to the composite systems, F2 and F6 are  unambiguously binary,  while F9 is a
bright, hard X-ray emitter - we consider the combination of X-ray and radio properties for this
source  as strong evidence for binarity, further noting that Gallego-Calvente et al. (\cite{GC})
report it to be radio-variable. Finally, no evidence for RV variability is visible in the seven
epochs of observations of F5; therefore despite the variable, composite nature of its radio emission
we refrain from concluding a binary nature for it at this juncture. 

F12, F18, and F19 all appear strong binary candidates on the basis of their non-thermal emission,
with Gallego-Calvente et al. (\cite{GC}) further noting that F18 appears radio-variable. Our
spectroscopic  survey reveals that F18 is also RV-variable,  while insufficient data exists to draw
any conclusions for F19 and no evidence for variability exists in the 12 epochs of observations  of
F12. Finally, we appraise the three stars displaying unconstrained spectral indices. Both F14 and
F16 appear to be binaries on the basis of RV variability, although  their radio fluxes are
consistent with expectations for emission originating solely in their stellar winds (Martins et al.
\cite{martins08}, Gallego-Calvente et al. \cite{GC}). Conversely no evidence for binarity exists for
F26; however the mass loss rate inferred from its radio flux is significantly in excess of that
expected for a star of its spectral type. Hence we suspect that a non-thermal component may be
present and that it is also a colliding wind binary (with a similar discrepancy observed for the
confirmed binaries F6 and F18; Gallego-Calvente et al. \cite{GC}).

\subsection{Synopsis}

From a sample of 37 objects  our spectroscopic  survey has identified a total of 13 RV variable
cluster members  that, as a consequence, are strong binary candidates (Table~\ref{rvtable}). To
these we may add F9 on the basis of its combined X-ray and radio properties, and the non-thermal
sources F12 and F19,  while  three  further stars - F1, F5 and F26 -  possess radio properties that
are, at a minimum, consistent with binarity. Excluding the latter three stars at this point implies
a conservative minimum binary fraction for the sample of  $\gtrsim43$\%, noting that no correction
has  been applied to the spectroscopic survey to account for differences in observational cadences
for individual stars, nor the lack of sensitivity to high-inclination systems. 

While the spectroscopic survey is complete for the 13  WNLha stars and seven O hypergiants, 14 of
the fainter cluster supergiants were excluded due to their lower S/N spectra; given that binaries
are likely to be overluminous this potentially introduces an observational bias. Likewise while all
stars are expected to be massive ($\gtrsim40M_{\odot}$; Sect. 4 and Clark et al. \cite{clark18a}),
we expect the most luminous WNLha stars to be considerably more massive than the faintest
supergiants (cf. Lohr et al. \cite{lohr}). Consequently, one would expect them to support stronger
winds which might be expected to aid in the detection of emission deriving from wind collision
shocks, introducing another bias. As such, it is instructive to consider the binary fractions for the
homogeneous and complete subsets of WNLha stars and  O hypergiants, finding them to be $\gtrsim62$\%
and $\gtrsim29$\% respectively, which yields a combined  binary fraction of $\gtrsim50$\% for this
cohort (noting that these numbers exclude the WNLha stars F1 and F5). We discuss these values
further in Sect. 5, where we place them into a wider context. 

\section{Model atmosphere analysis}

\begin{table*}
\caption{Best-fit stellar parameters for the arches RV systems from
CMFGEN modelling and SED fitting.}
\label{tab:modelrv}
\begin{center}
\tabcolsep1.5mm
\small
\begin{tabular}{l l c c c c c c c c c c c c} 
\hline\hline
Star  &Spec.        & Ak$_{\rm s}$ &  \Teff       & log(\Lstar)    & \Rstar     & log(\Mdot)     & \vinf    & $\beta$ &f$_{\rm cl}$&       Y                & X$_{\rm N}$ & X$_{\rm C}$ & X$_{\rm O}$              \\
      & Type        &              &  kK          & \Lsun$^1$      &  \Rsun     & \Msunyr        & km~s$^{-1}$     &         &            &                        & \multicolumn{3}{c}{mass fraction ($\times 10^{-3}$}) \\
\hline				             
F6    & WN8-9ha     &     3.01     & 33.3$\pm1.3$ &  6.32$\pm0.05$ & 43.4       & -4.48$\pm0.06$ &  1400    &  1.10   &   0.10     & 0.33$^{+0.08}_{-0.08}$ &  20         &  0.75       & $\leq$ 1.0               \\
      &             &     3.63     &              &  6.57$\pm0.05$ & 50.0       & -4.30$\pm0.06$ &          &         &            &                        &             &             &                          \\
      &             &     1.97     &              &  5.89$\pm0.05$ & 26.5       & -4.81$\pm0.06$ &          &         &            &                        &             &             &                          \\
F7    & WN8-9ha     &     3.01     & 33.9$\pm1.3$ &  6.27$\pm0.05$ & 39.7       & -4.70$\pm0.06$ &  1250    &  1.05   &   0.08     & 0.45$^{+0.08}_{-0.08}$ &  22         & 0.24        & $\leq$ 3.0               \\
      &             &     3.59     &              &  6.52$\pm0.05$ & 52.7       & -4.52$\pm0.06$ &          &         &            &                        &             &             &                          \\
      &             &     1.97     &              &  5.57$\pm0.05$ & 24.5       & -5.01$\pm0.06$ &          &         &            &                        &             &             &                          \\
F15   & O6-7 Ia$^+$ &     2.78     & 32.0$\pm1.5$ &  5.90$\pm0.06$ & 29.0       & -5.33$\pm0.08$ &  2100    &  1.35   &   0.10     & 0.17$^{+0.05}_{-0.03}$ &  6.6        &  2.0        &  11                      \\
      &             &     3.34     &              &  6.14$\pm0.06$ & 38.0       & -5.16$\pm0.08$ &          &         &            &                        &             &             &                          \\
      &             &     1.82     &              &  5.51$\pm0.06$ & 18.4       & -5.63$\pm0.08$ &          &         &            &                        &             &             &                          \\
F25   & O4-5 Ia     &     3.81     & 40.0$\pm2.0$ &  5.93$\pm0.06$ & 19.1       & -6.06$\pm0.13$ & 2200$^2$ &  1.15   &   0.10     & 0.10$^{+0.03}_{-0.02}$ &  5.5        &  5.3        &  14                      \\
      &             &     4.56     &              &  6.23$\pm0.06$ & 27.1       & -5.83$\pm0.13$ &          &         &            &                        &             &             &                          \\
      &             &     2.52     &              &  5.51$\pm0.06$ & 18.4       & -6.44$\pm0.13$ &          &         &            &                        &             &             &                          \\
F35   & O4-5 Ia     &     2.80     & 40.0$\pm2.0$ &  5.77$\pm0.07$ & 16.0       & -6.17$\pm0.13$ & 2200$^2$ &  1.15   &   0.10     & 0.10$^{+0.03}_{-0.02}$ &  4.8        &  7.8        &  14                      \\
      &             &     3.36     &              &  6.00$\pm0.07$ & 20.9       & -6.00$\pm0.13$ &          &         &            &                        &             &             &                          \\
      &             &     1.85     &              &  5.38$\pm0.07$ & 10.2       & -6.46$\pm0.13$ &          &         &            &                        &             &             &                          \\
\hline
\end{tabular}
\normalsize
\end{center}
{Three different parameter sets are displayed as a function of the extinction law
adopted: the uppermost is our favoured two $\alpha$ prescription, the middle a Moneti law and the
latter a single $\alpha$ law (Sect. 4). \Teff\ and \Rstar\ correspond to $(\tau_{\rm Ross}=2/3)$.
f$_{\rm cl}$ and $\beta$ are the clumping factor and velocity-law parameters as given in Najarro et
al. (\cite{najarro09}). Y is the He/H ratio by number and CNO abundances are provided as mass
fractions. Uncertainties in the extinction law and distance (8~kpc) to the cluster are not
considered. Therefore, \Lstar\ estimates only propagate errors in \Teff\ assuming a fixed
observed flux in the Rayleigh-Jeans-like NIR bands (i.e. $\propto$\Teff~\Rstar$^2$). Errors in the
terminal velocities of F6 \&\ F7 are $\approx$150~km~s$^{-1}$ and $\approx$200~km~s$^{-1}$ for F15 while a value of
\vinf=2200~km~s$^{-1}$ has been assumed for the O-supergiants F25 and F35.   
Nitrogen abundance uncertainties are $\sim 0.15$dex for the WNh stars
and $\sim 0.20$dex for the O stars, while we estimate a $\sim 0.30$dex
uncertainty in the Carbon abundance. Only upper limits could be set on the Oxygen abundance
of F6 and F7 WNh stars, while a $\sim 0.30$dex uncertainty is obtained for the O stars F15, F25 and F35.}
\end{table*}

\begin{figure*}
\includegraphics[width=\textwidth]{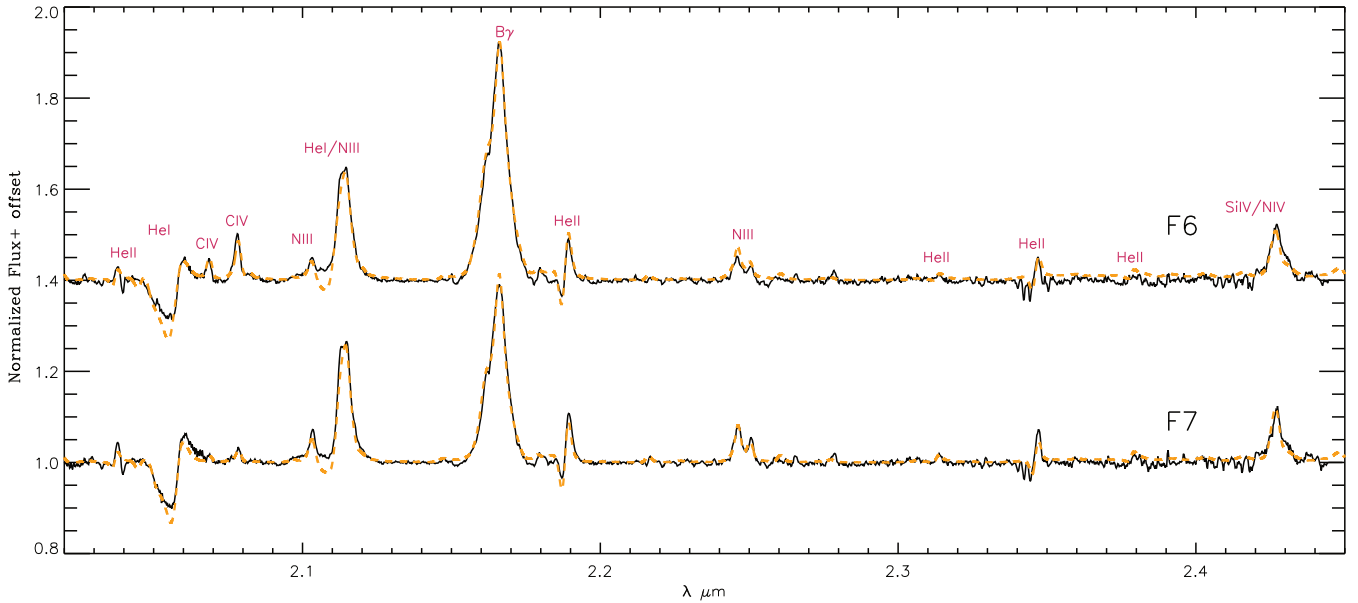}
\caption{Comparison of model fits (orange dotted lines) to observational spectra of WNLh primaries
(black solid lines). Major spectral transitions are indicated.}
\label{mod_wnh}
\end{figure*}

We carried out a quantitative spectroscopic analysis of these five systems to address three issues
relevant to their binary nature. Firstly we searched for signatures of previous or ongoing binary
interaction, such as the simultaneous presence of enhanced carbon and nitrogen (cf. HD 153919 and
Wd1-13; Clark et al. \cite{clark02}, \cite{clark14}). Secondly, we checked for subtle observational
signatures of the secondary. Finally we sought to obtain evolutionary masses and other fundamental
properties for the dominant component(s) of each binary. To model the stars we followed the
procedure presented in Lohr et al. (\cite{lohr}) making use of two grids (one for WNL and one
O-stars) created with CMFGEN (Hillier \&\ Miller \cite{hil98}, \cite{hil99}).  In their analysis of
the F2 system Lohr et al. (\cite{lohr}) were able to undertake spectroscopic disentangling, allowing
them to analyse each component separately. However, this was not possible for the five systems in
question, and so we modelled the mean  of their stacked spectra (once corrected from radial velocity
shifts). 

We made use of the line profiles of H, He, C, N, O and Si present in the K-band spectra of the objects together with the observed photometry to constrain their stellar properties (e.g. Najarro et al. \cite{najarro04}, Martins et al. \cite{martins08} and Lohr et al. \cite{lohr}).
The effective temperature was obtained from the He\,{\sc ii}/He\,{\sc i} ionization balance by means of the He\,{\sc ii} 2.037, 2.189 and 2.346~$\mu$m and the He\,{\sc i} 2.059~$\mu$m, and 2.112-3~$\mu$m line profiles. 
The relative strength between the He and Br~$\gamma$ lines was utilized to constrain the He/H fraction while the mass-loss rate was inferred from their absolute strengths.
For the WNL stars, the terminal velocity is obtained from  He\,{\sc i} 2.059~$\mu$m, while the full Br~$\gamma$ profile constrains $\beta$ and its red emission determines the clumping factor, f.
For the OIf$^+$ star we find a degeneracy between \vinf\ and f and adopt a value of f=0.1.
In the case of the two OIf objects we fixed f and vinf (see Table~\ref{tab:modelrv}).
The N\,{\sc iii} diagnostic lines at 2.103, 2.115, 2.247 and 2.251~$\mu$m together with the N\,{\sc iv}/Si\,{\sc iv} feature at $\sim$2.43 (Clark et al.~\cite{clark18a}) constrain quite well the nitrogen abundance (Najarro et al. \cite{najarro04}) specially for the WNLs (see Table~\ref{tab:modelrv}). 
The carbon content is obtained from the  C\,{\sc iv} lines at 2.070 and 2.078~$\mu$m. 
The emission component of the strong He\,{\sc i}/C\,{\sc iii}/N\,{\sc iii}/O\,{\sc iii} blend at 2.112--2.115~$\mu$m is expected to be dominated by nitrogen in the WNL stars and by oxygen in the O stars (Geballe et al.~\cite{2006ApJ...652..370G}).
Therefore, when combined with the rest of diagnostic K-band lines it constrains quite well the oxygen abundance of the O stars while providing just an upper limit for the WNLs.

To estimate the system luminosities, we utilised the HST photometry compiled in Clark et al.
\cite{clark18a} and performed Levenberg-Marquardt fits for different extinction prescriptions
scaling the SEDs from the spectroscopic modelling.
The availability of six different photometric measurements between 1.0 and 2.5~$\mu$m enable a study of a wavelength-dependence of the extinction curve.
This has been explored by Lohr et al \cite{lohr} in their analysis of F2 and investigated by Nogueras-Lara et al. \cite{2020A&A...641A.141N} for their extinction map of the Galactic Centre. 
Based on the analysis of our full Arches data sample and the resulting location in the HR diagram of all cluster members (Najarro et al, in prep.)
we opted finally for a two-$\alpha$ extinction law of the form $A_{\lambda}=A_{k0}
(\frac{\lambda_{k0}}{\lambda})^{\alpha}$ where $\alpha=\alpha_1 + \frac{(\alpha_2-\alpha_1)}{(log
\lambda_2- log \lambda_1)} (log \lambda -log \lambda_1)$ with $\alpha_1$ and $\alpha_2$ defined at
$\lambda_1$ and  $\lambda_2$ in the $J$ and $K$ bands respectively.
In our case we chose $\lambda_1=1.282$ and  $\lambda_2=2.166$ and obtained $\alpha_1=1.79$ and $\alpha_2=1.600$. 
This extinction law results in luminosities roughly $\sim$0.25~dex below those which are obtained with the
Moneti et al. (\cite{moneti01}) law and $\sim$0.4~dex above those derived with a $\alpha=2.3$ law
approach (Nogueras-Lara et al. \cite{nogueras18}).

\begin{figure*}
\includegraphics[width=\textwidth]{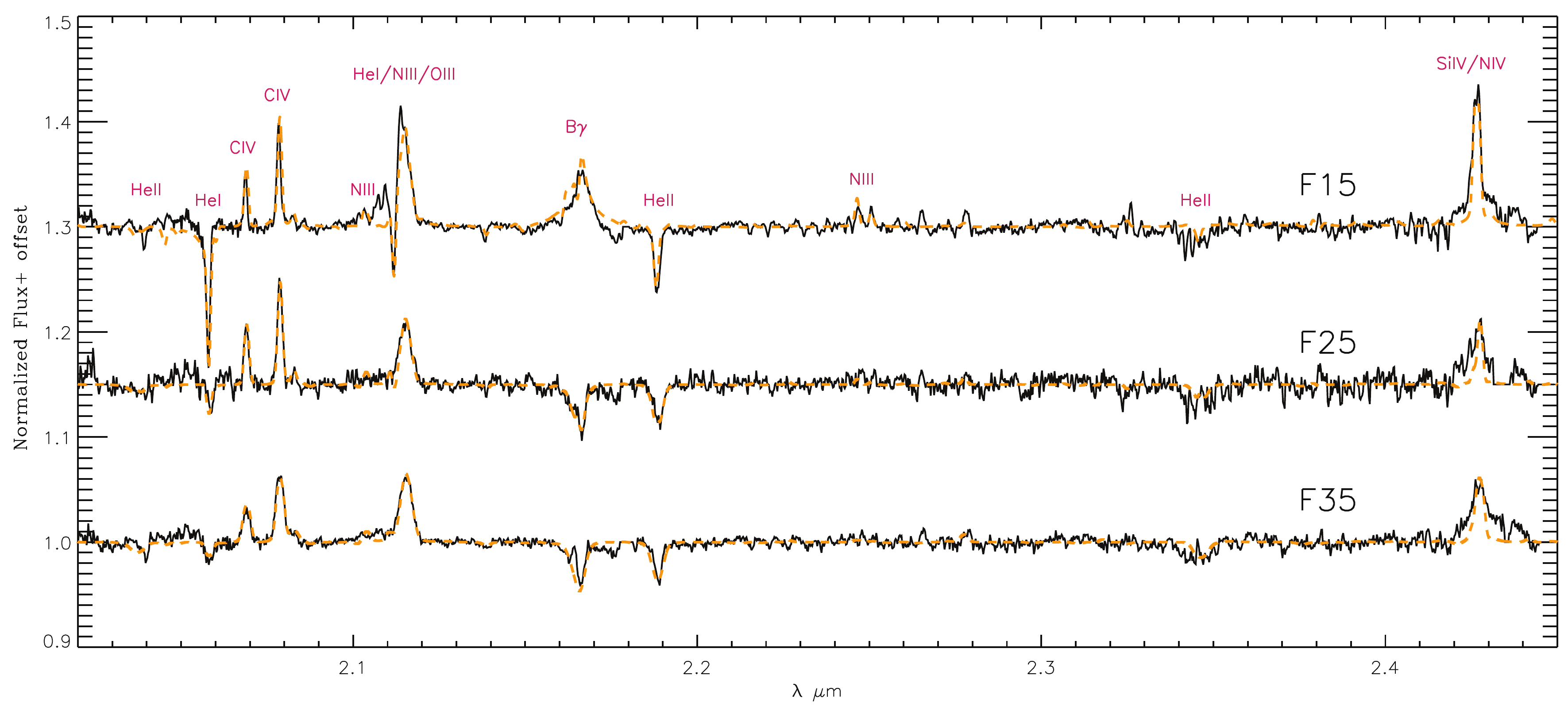}
\caption{Comparison of model fits (orange dotted lines) to observational spectra of O super-/hypergiant primaries (black solid lines). Major spectral transitions 
are indicated.}
\label{mod_oif}
\end{figure*}

\subsection{System properties and comparison to previous analyses}

Stellar properties of the two WN9h (F6 \& F7), the O-hypergiant (F15) and the two O-supergiants (F25
\& F35) are presented in Table~\ref{tab:modelrv}. For each of the objects the first line provides
the stellar parameters assuming the two-$\alpha$ extinction law from above, while the second and
third rows display the $A_k$, luminosity, radius and mass-loss rates obtained employing Moneti's and
$\alpha=2.3$ extinction laws respectively. A deeper, more detailed analysis of the stellar
properties of these systems will be presented in a future work, where a study of the whole Arches
sample will be discussed. 

The resultant spectral fits for the WNLh stars (F6 \& F7) and the O-stars (F15, F25, \& F35) are
displayed in Fig.~\ref{mod_wnh} and Fig.~\ref{mod_oif} respectively. The excellent fits with a
single atmospheric model of the full systems support our previous claims of either less massive
companions (F6, F7 and F15) or systems of comparable mass and similar spectral type (F25 and F35).
No spectral features from the secondaries are apparent from comparison of our synthetic spectra to
the observations, with the sole exception of the blue flank of the He\,{\sc I}/N\,{\sc
iii}~2.112-2.115~$\mu$m feature in F15 where unexpected emission is potentially present.

Stellar properties for these sources are presented in  Table~\ref{tab:modelrv}. The WNLh stars are
found to be exceptionally luminous and support powerful stellar winds while, as expected, the O
super-/hypergiants appear less luminous and drive lower density, higher velocity outflows. The
carbon, nitrogen and oxygen abundances derived for all objects are consistent with equilibrium
values for their respective evolutionary stages and  no simultaneous enhancement of carbon and
nitrogen - suggestive of binary interaction (cf. Clark et al. \cite{clark14}) - are observed. The
only possible exception is F6, where carbon is less depleted; however this finding is consistent
with the moderate  He enrichment (Y=0.33). We highlight that our CNO estimates are consistent with
an $\alpha$-enhanced metallicity pattern for the Arches as also found by Najarro et al.
(\cite{najarro09}) in the Quintuplet cluster. 

With the exception of F25, these stars were also analysed by Martins et al. (\cite{martins08}),
while  F15 had been previously modelled by Najarro et al. (\cite{najarro04}). Both groups adopted a
similar approach, with the main difference posed with the handling of extinction:  Najarro et al.
used a variable extinction description  according to Moneti's law, while Martins et al. chose a
fixed $A_{\rm K}=2.8$ value. We find very good agreement with the \Teff, \Lstar and \vinf\ values
from Martins et al. (their Table~2) for the WNLh stars F6 and F7; however, helium and carbon
abundances for both stars are enhanced by, respectively, $\sim$50\% and $\sim$100\%, while nitrogen
in F6 demonstrates a $\sim$75\% enhancement with respect to this work. As for the O hypergiant,
F15, our derived \Teff, \Lstar\ and helium abundances land in between the values provided by both
studies and, while we find a consistent nitrogen abundance, our value for carbon is significantly
higher than that of Martins et al., being driven by the difference in \Teff, and hence   closer to
the estimate of Najarro et al. 

F35 demonstrates the largest discrepancy  between both studies;  where Martins et al. find
\Teff=33.5~kK, we obtain \Teff=40.0~kK. We attribute this divergence to the weight given to
the diagnostic He\,{\sc i}/N\,{\sc iii}~2.112-2.115~$\mu$m feature. The absence of an observed
absorption component points towards a higher \Teff\ (earlier type) than the one obtained by Martins
et al.

\subsection{Location in the HR diagram}

\begin{figure}
 \includegraphics[width=\columnwidth]{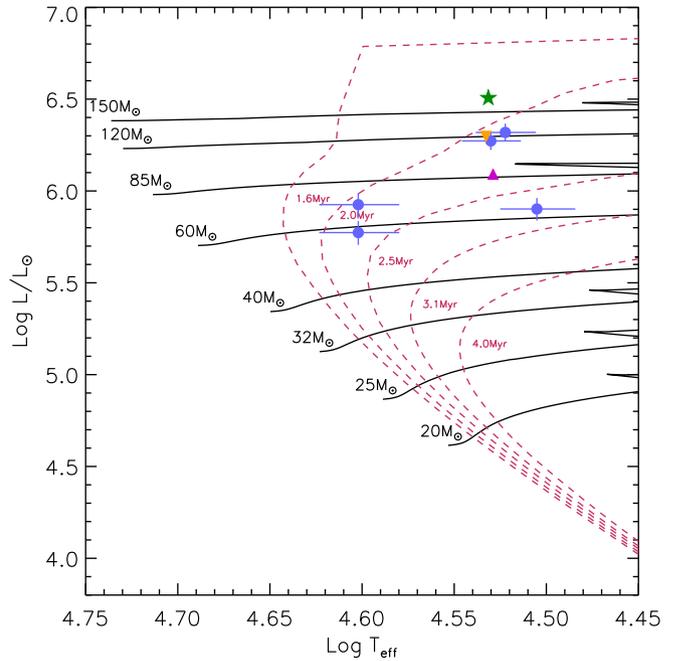}
 \caption{HR diagram displaying the RV arches objects together with non-rotating Geneva model and isochrones. Stellar luminosities are those from the two-$\alpha$ extinction fits in Table~\ref{tab:modelrv}. We also plot the position of the SB2 binary F2 (green-star) as well as the locations of the  WN9h primary and O5-6 Ia$^+$ secondary components (yellow and purple triangles respectively) from the unconstrained binary modelling by Lohr et al. (\protect\cite{lohr}).}
 \label{hr_rv}
\end{figure}

Figure~\ref{hr_rv} shows the location of these systems in the HR diagram, together with non-rotating
Geneva model and isochrones (Ekstr\"om et al.~\cite{ekstrom12}). We also show the position of both
the individual components of the SB2 binary F2 along with its  integrated properties (Lohr et al.
\cite{lohr}). Bearing in mind that both F6 and F7 appear dominated by their primaries, their near
coincidence in the HR diagram with the WNLh primary of F2 endorses the independent results from the
latter work. 

From Fig.~\ref{hr_rv} we see that non-rotating models indicate initial masses of $\sim$100-120\Msun\
for the WNh primaries of F2, F6, and F7. Values around 60\Msun\ are suggested for the O stars,
although the resultant mass range might extend downwards to   $\sim40-50$\Msun\ if the secondaries
in the supergiant binaries F25 and F35 contribute significantly to their integrated fluxes. Taken in
isolation these results should be treated with caution,  as they display a critical dependence on
the final extinction law assumed. However, in view of our analysis of the whole sample (Najarro et
al. in prep.), we anticipate that they should be close to the final values. Moreover we highlight
that initial masses ranging from $\sim$60-120\Msun\ are fully consistent with the dynamical masses
determined for the components of F2 (Table~\ref{pubin}; Lohr et al. \cite{lohr}), buttressing the
reliability of our evolutionary values.

Finally, the properties of F2, F6, F7, F25, and F35  are consistent with a cluster age of
2.0-2.5~Myr, as suggested by a number of studies (Figer et al.~\cite{figer02}, Najarro et
al.~\cite{najarro04}, Clark et al.~\cite{clark18a} and is consistent with the results of Najarro et al. (in prep.)).
The location of F15 in the HR diagram
suggests a slightly larger age of $\sim$2.8~Myr, although it is still consistent with the range
defined by the former stars upon consideration of the relevant observational and modelling
uncertainties.

\section{Discussion}

The synthesis of multi-wavelength observational data and quantitative model atmosphere analysis
reveals that the Arches cluster is characterised by a rich population of  very massive binaries.
Previous spectroscopic binary surveys have targeted field stars (e.g. Mason et al. \cite{mason},
Chini et al. \cite{chini}, Sota et al. \cite{sota}), a number of comparatively low mass clusters
(Sana et al. \cite{sana12}) and large stellar aggregates such as the Cyg OB2 association (e.g.
Kobulnicky et al. \cite{kobulnicky}) and 30 Doradus (Sana et al. \cite{sana13a}, Dunstall et al.
\cite{dunstall}, Almeida et al. \cite{almeida}). While these studies provide a significantly larger
sample size than that furnished by the Arches, the stars considered are heterogeneous in terms of
age and hence evolutionary phase. While the cohort derived from Westerlund~1 is thought to be
co-eval (Ritchie et al. \cite{ritchie}, Clark et al. \cite{clark20a}) it is older than the Arches,
and hence samples lower mass OB and WR stars. Indeed, with the exception of 30 Dor, none of these
surveys reach a comparably rich contingent  of very massive stars. As a consequence  the Arches
currently provides a unique (and complementary) perspective on the  properties of such objects.

Critically, we find  that Arches members exhibit a remarkably high binary fraction: $\geq50$\% for
$M\gtrsim60M_{\odot}$, rising to $\gtrsim62$\% if only the more massive and evolved WNLh subset are
considered.  Although we consider the  parameters for the binaries considered in Sect. 3 to be
provisional, and contingent on future observations, it appears likely that a number have (highly)
eccentric orbits - F2, F6, F25, and F35 (Table~\ref{rvtable}) - indicative of a pre-interaction
evolutionary phase. Likewise, an orbital period apparently in excess of 3 years would appear to
exclude prior interplay between  the components of F7, although we cannot reject the possibility
that the primary is itself a merger remnant. Crucially however, with periods $<14$d, the compact
nature of both F2 and F6 likely precludes such a scenario. As such, we are left to conclude that
their components are not the product of mass transfer or merger,
although the short orbital periods of many of the Arches binaries imply that
future binary interaction is unavoidable (Sect. 3 and Table~\ref{pubin}).
Such a conclusion is at
least consistent with the homogeneity of the spectra of the most massive stars within the Arches
(Clark et al.~\cite{clark18a}), although we caution that is possible that putative blue stragglers
might not differ significantly in terms of spectral morphology from less massive stars in the upper
reaches of the HR diagram.

Schneider et al. \cite{schneider} argued that the most massive 9\,$\pm$\,3 stars in the Arches cluster are the rejuvenated products of binary interactions. The conclusions outlined above appear at odds with this result.
As discussed in Paper~II, one explanation for this discrepancy is that Schneider et al. \cite{schneider} assumed an age of 3.5\,Myr for the Arches cluster, which now appears to be at the upper end of the age range based on the age inferred from spectroscopic analyses (e.g. Papers~I,~II) and the results presented here (which both favour ages in the range 2-2.5\,Myr).
Assuming an age of 2.5\,Myr, a cluster mass of 1.5$\times 10^4$\,M$_\odot$ (Clarkson et al.~\cite{clarkson}) and an initial binary fraction of 100\%, interpolating Fig. 5 of Schneider et al. \cite{schneider} would result in roughly an average of roughly 2\,$\pm$3 of the most massive stars being the result of rejuvenated binary interaction in the Arches cluster.

Based on the above, we suggest that the current most massive stars in the Arches cluster may not be the products of binary mass-transfer or mergers of lower mass stars.
That makes the Arches cluster, and these stars, crucial indicators of single stellar evolutionary models in the $>50$\,M$_\odot$ regime and highlights the urgency of follow-up observations to better characterise their orbital properties and better establish their evolutionary histories.

\subsection{Spectral variability}     \label{sub:sec_var}

With the first high S/N and resolution spectra of members of the Arches obtained in 1999 (Figer et al. \cite{figer02}), a second epoch in 2005 (Martins et al. \cite{martins08}) and extensive monitoring between 2011-2018 (Table~\ref{tab:appA1}), an observational baseline of $\sim2$ decades permits a search for the characteristic long term spectral variability. 
To our knowledge, this is the first study that permits the assessment of long term variability in WNh/O Ia$^+$ stars.
Despite such a baseline, none of the 37 targets display any evidence of spectral variability given our S/N and resolution limitations, as might be expected if the WNh/O Ia$^+$ cohort were close to the S~Doradus instability strip.
In Appendix~\ref{AppB}, we discuss the implications of the lack of spectral variability observed in the current sample, with respect to the secular variability observed in LBVs.
This seems unlikely to the be the result of observational limitations given the variability exhibited in the LBV spectra of the Quintuplet cluster (Clark et al. \cite{clark18b} and references therein).

\subsection{The Arches population in context: fundamental properties of WNLha stars and O hypergiants}

\begin{figure}
\includegraphics[angle=-90, width=\columnwidth, trim=1cm 0 0 3cm, clip]{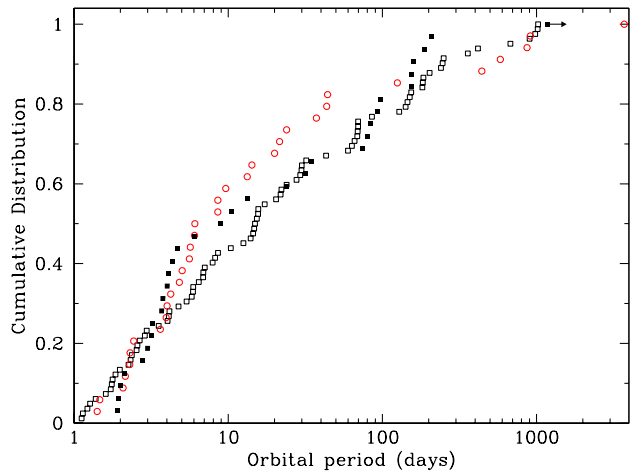}
\caption{Plot of the cumulative distributions of orbital periods from the sample of WNLha and O super-/hypergiants given in Table~\ref{pubin} (filled squares), O stars in Galactic clusters (red circles; Sana et al. \protect\cite{sana12}) and in 30 Dor (open squares; Almeida et al. \protect\cite{almeida}). Given their uncertain periods, RMC136c and F25 are excluded from this plot, while the value for F7 is a lower limit.}
\label{fig:Pcumdist}
\end{figure}

It is instructive to place the properties of the binary population of the Arches into a wider
context. In Table~\ref{pubin} we summarise the orbital properties of other WNLha (and the related
Of/WN stars) and O hypergiants within the Galaxy and Large Magellanic Cloud (noting that a lack of
comparator systems prevents extending this exercise to the cluster mid-O supergiants), while in
Appendix C we compare the properties of the subset of WNLha stars found within stellar aggregates.
In doing so we note that slightly earlier subtypes are present within R136/30 Doradus (WN5-6ha and
O2-2.5 If/WN5-6) and some Milky Way clusters (e.g the WN6ha stars within NGC3603); it is not clear
whether this disparity results from different initial masses, ages or metallicities, or a
combination of these properties.

\subsubsection{Masses and evolutionary state}

Consideration of the dynamical masses listed in Table~\ref{pubin} supplemented, where available,  by
evolutionary masses determined from quantitative modelling  (cf. Table~\ref{tab:appC1}) confirm the
conclusions derived from the Arches population: that the WNLh stars and   mid-O hypergiants appear
to form a homogenous class of  uniquely massive objects.  Masses range from $\sim50M_{\odot}$ (WR29
and RMC145; Table~\ref{pubin}) to $>>100M_{\odot}$ (the WN5ha components of R136;  Crowther et al.
\cite{crowther16}) for the former, with the latter potentially being somewhat lower
(Table~\ref{pubin})\footnote{The high velocity runaway WR148 may be an exception to this  conclusion
(Table~\ref{pubin}), although its dynamical mass determination is based on an assumed mass for the O
secondary and, moreover, is dependent on the uncertain inclination of the system (Munoz et al.
\cite{munoz}). Conversely, WR79a, the star with the lowest bolometric luminosity
(Table~\ref{tab:appC1} has an evolutionary mass of $\sim60M_{\odot}$.}.  Their masses therefore
exceed those of O dwarfs (Clark et al. \cite{clark18a}), consistent  with the suggestion
that they are relatively unevolved core-H burning stars populating the upper reaches of the stellar
mass function, whose proximity to the Eddington limit leads to inflated envelopes and the
development of powerful stellar winds (Hamann et al. \cite{hamann06}, Schnurr et al.
\cite{schnurr09a}, Crowther et al. \cite{crowther10}, Gr\"{a}fener et al. \cite{grafener21}). This
interpretation is buttressed by the analysis of Hamann et al. (\cite{hamann19}) who show that the
most luminous WN stars are exclusively WNL stars with pronounced signatures of hydrogen in their
spectra.

Such conclusions are consistent with the observational finding that WNLh stars appear confined to
the youngest coeval clusters within both galaxies ($<4$Myr; Appendix C and refs. therein), being
present within Mercer 30 ($4.0^{+0.8}_{-0.8}$Myr; de la Fuente et al. \cite{dlF})  and Mercer 81
($3.7^{+0.4}_{-0.5}$Myr; Davies et al. \cite{davies12a}) and seen in relative profusion in still
younger clusters such as R136 ($1.5^{+0.3}_{-0.7}$Myr; de Koter et al. \cite{1998ApJ...509..879D}; Crowther et al. \cite{crowther16}), but
absent from the slightly older Westerlund 1 ($\sim5-6$Myr; Negueruela et al. \cite{iggy}, Clark et
al. \cite{clark20a}). 

In this regard we highlight  that WR121a (WN7ha) is found in the core of the deeply embedded cluster
within the Giant H\,{\sc ii} region W43 (Blum et al. \cite{blum}), while MMT58 (O2-3.5
If$^*$/WN5-6ha) is located within a compact H\,{\sc ii} region on the periphery of the star-forming
region surrounding NGC3603 (Roman-Lopes \cite{RL13}). Their environments suggest extreme youth for
both stars - with Roman-Lopes (\cite{RL13}) estimating a dynamical age of $<600,000$yr for the
latter system -   allowing limited opportunity for the prior interaction of the binary components of
both systems. Furthermore both stars are found to be in very compact  configurations ($P_{\rm
orb}\lesssim4$d; Table~\ref{pubin}) which would appear to disfavour their formation  via prior
mergers (cf. Arches F2 and F6). As a consequence we conclude that such very massive stars are
able to form monolithically, without requiring  binary mass transfer or merger on or after the main
sequence to attain their final masses, a hypothesis also suggested by the properties of the Arches
binary cohort. Nevertheless, it is clear that a subset may also form via a binary channel, as
evidenced by the presence of the highly (over)luminous mid-O hypergiant blue stragglers Wd1-27 (O7-8
Ia$^+$) and Wd1-30a (O4-5 Ia$^+$) within Westerlund 1 (Clark et al. \cite{clark19a}). 

\subsubsection{Binary properties and frequency of occurrence}

We now turn to the physical properties of the  subset of the WNLh and O hypergiant population for
which binary solutions have been obtained (Table~\ref{pubin}). In doing so we are conscious of the
observational biases introduced by the differing sensitivities, durations and cadences of the
observations used in both the construction of this sample, and in comparison to the wider surveys of
lower mass stars mentioned above. Accounting for this would require Monte Carlo simulations for
every individual  study in order to quantify their completeness as a function of orbital parameters,
beyond the scope of the work.

Mindful of this, we first turn to orbital separation and in Fig.~\ref{fig:Pcumdist} we plot the cumulative
distribution of orbital periods for the stars listed in Table~\ref{pubin}, alongside the population
of Galactic O stars of Sana et al. (\cite{sana12}) and the corresponding cohort in 30 Dor (Almeida
et al. \cite{almeida}). We note that the latter two distributions are constructed for stars that, on
average, are expected to be of lower mass than those considered here; dynamical and evolutionary
mass estimates suggest  $M\lesssim47M_{\odot}$ for the 30 Dor sample (Mahy et al. \cite{mahy20a},
\cite{mahy20b}), while  that  of Sana et al. (\cite{sana12}) is predominantly populated by late O
dwarfs and giants of moderate luminosity. Furthermore, in undertaking this comparison we explicitly
highlight that, as with the latter authors, we have not applied any bias corrections to these data.
As a consequence detailed quantitative analysis is premature but, following the argument of Almeida
et al. (\cite{almeida}) that  one might expect differing studies to suffer similar biases, a
qualitative comparison is potentially instructive.

The three cumulative distributions look to be broadly similar. To the naked eye, the one constructed
for WNLh stars and O hypergiants  and that derived for Galactic O stars by  Sana et al.
(\cite{sana12}) show a close correspondence, appearing to be overabundant in binaries with $P_{\rm
orb}\sim4-40$d with respect to that of Almeida et al. (\cite{almeida}) for O stars within 30 Dor;
however these authors show that there  are no statistically significant differences between these
distributions. As might be expected we identify a predominance of comparatively short orbital
period systems, with $\sim50$\% of binaries listed in Table~\ref{pubin} found with periods of less
than a week, $\sim70$\% less than a month and $\sim97$\% less than a year (the ratios being
identical if just the subset of WNLha stars are considered). These numbers appear  comparable to the
findings of Almeida  et al. (\cite{almeida}) and Sana et al. (\cite{sana12}), who report percentages
of 40\%, 70\%  and 90\%; and 50\%, 74\% and 85\% for these period intervals for, respectively,
O-star binaries within 30 Dor and Galactic clusters. 

Evidently, the distributions are in part representative of the  observing strategies employed, which
disfavour the  characterisation of binaries with multi-year periods (cf. F7; Sect. 3). Conversely,
while the identification of short period systems appears favoured, such  very massive and compact
binaries still have to exist in order to be detected. In this regard we highlight that the shortest
period binaries within the WNLha sample are the $P_{\rm orb}\sim1.9$d systems BAT99-32 and MMT58; in
contrast the 30 Dor O star sample hosts 10/82 binaries with shorter orbital periods than this, with
a minimum  $P_{\rm orb}\sim1.1$d (VFTS066 and 352). One could attribute this apparent discrepancy to
binary interaction/merger amongst the WNLh cohort, but we note that the short period 30 Dor O star
binaries  are exclusively of luminosity class III-V; consequently an alternative (or complementary)
explanation would be that the larger radii of the WNLh stars prohibits the formation of such  tight
binaries. Indeed, by way of an illustration, quantitative estimates of the stellar radii and masses
of the components of Arches F2 (Lohr et al. \cite{lohr}) and WR20a (Rauw et al. \cite{rauw05})
preclude the component stars being found in similarly  compact configurations. 

Sana et al. (\cite{sana12}) conclude that the rich population  of comparatively short period
binaries implies that fully 70\% of the O stars they consider will interact at some point in their
life cycle. Despite the similarity in orbital period distributions, uncertainties in the evolution of
massive ($\gtrsim60M_{\odot}$) stars - due to significant mass loss driven by stellar winds and
instabilities  - mean that this quantitative prediction is not immediately applicable to the sample
considered here. However, if such stars pass through a comparatively `cool', physically extended
LBV-like phase (cf. the evolutionary predictions for a $M_{\rm initial}\sim60M_{\odot}$ star by Groh
et al.  \cite{groh}) then one would also expect a large proportion will undergo binary interaction. 

Moving on to orbital eccentricity, only a subset of WNLha and O super-/hypergiants have
observational determinations, but the proportion with essentially circular orbits ($e<0.1$) appears
broadly comparable to that of the 30 Dor O star sample (48\% versus 40\% respectively; Almeida  et
al. \cite{almeida}). Once again this is potentially explicable by a combination of tidal interaction
and/or mass transfer, plus the inability for short period highly eccentric configurations to
accommodate such physically massive stars. 

Explicitly excluding the post-supernova X-ray binary HD153919, of those systems for which (lower
limits to) dynamical mass estimates are available, it is apparent that the secondaries are also
highly luminous stars; such a preponderance of massive companions is also seen in both 30 Dor
(Almeida  et al. \cite{almeida}) and  Westerlund 1 (Clark et al. \cite{clark20a}; Ritchie et al.
\cite{ritchie21}). This is likely to be in part an observational bias; in order for mass
determinations to be made one requires SB2  systems. As a consequence   the companion has to be of
comparable luminosity - and hence mass - to the primary in order to fulfil this requirement.
Nevertheless, a number count  of  14/33 SB2 systems appears difficult to explain with a mass
distribution for the companion stars deriving from  random sampling of  a canonical IMF.

Finally we turn to the binary frequency for WNLh stars. In order to assess this we adopt a different
approach - mirroring that of Sana et al. (\cite{sana12}) - and limit ourselves to a sample derived
from examples within stellar clusters. This has four  advantages: (i) mindful of the discussion in
Sana et al. (\cite{sana13a}), it limits the potential bias towards the detection of (overluminous)
binaries  in a magnitude limited sample; (ii) given the chosen clusters are well studied the
resultant populations are better defined inasmuch as we are complete for such stars; (iii) in many
cases the cluster properties (distance and age) have been determined, allowing quantitative
determination of stellar properties; (iv) a number of clusters have been subject to radio and X-ray
observations, which provide additional binary diagnostics. Unfortunately, however, a lack of data
precludes analysis of analogues of the Arches mid-O super-/hypergiants via such a methodology.

A detailed breakdown of properties of this population is provided in Appendix C, where we consider
11 Galactic clusters (excluding the Arches) as well as R136 and the 30 Dor star-forming region
within the Large Magellanic Cloud. This yields a total of 21 WNLh (and related objects) within R136
and 30 Dor and 24 within the Galactic clusters, for which one or more observational diagnostics for
binarity  are available. As with the Arches we give primacy to RV survey data, though
also employ extant radio and X-ray data to identify  colliding wind systems; in order to achieve
this goal we also  provide a thorough reappraisal of the X-ray properties of single and binary WNLh
stars (Appendix C). Despite our best efforts we  consider the resultant  counts to be conservative
lower limits; this is due to  the aforementioned  limitations of RV surveys and the fact that some
stars have not been subject to such efforts, with an assessment of binarity instead relying solely
on radio and/or X-ray data. Nevertheless, we find binary fractions of 10/21 ($\sim48$\%) for
examples within R136 and 30 Dor and 15/24 ($\sim63$\%) for Galactic clusters, both of which compare
favourably to our estimate for the  Arches (8/13 ($\sim62$\%); Sect. 3). In total this suggests a
minimum binary fraction of 33/58 ($\sim57$\%) for WNLh stars which, following the preceding
discussion,  appear to comprise objects with $M_{\rm init}\gtrsim50M_{\odot}$.

Because of the different observational approaches adopted across various studies - and in particular
the heterogenous nature of the dataset constructed here -  direct, quantitative  comparison of the
binary frequencies returned is potentially misleading. Nevertheless, previously published surveys
suggest a  landscape in which binarity is pervasive amongst OB  stars. Allowing for biases in their
RV dataset,  Kobulnicky et al. (\cite{kobulnicky}) infer a fraction of $\sim55$\% for OB binaries
with $P_{\rm orb}<5000$d within the Cygnus OB2 association,  Sana et al. (\cite{sana12}) return
$\sim69\pm9$\% for O stars within a subset of Galactic clusters, and Sana et al. (\cite{sana13a})
quote  $\sim54\pm4$\% for O stars within 30 Doradus. Our results  appear fully consistent with such
a picture and extend the predominance of binarity to the most massive stars in the local Universe.

\begin{table*}
\caption{Published parameters for binaries with WNLha  and mid-O super-/hypergiant primaries,
 ordered by increasing orbital period}
\label{pubin}
\begin{center}
\begin{tabular}{l l c c c l l}
\hline\hline
\noalign{\smallskip} 
Name & Spectral types & $P_{orb}$ & $e$ & $M_1+M_2$ & Aggregate & 
References\\
 & & (d) & & (M$_{\sun}$) &  & \\
\hline
& &  & &  &  & \\

BAT99-32 & WN6(h) + ?       & 1.91 & 0.0 & - &-                                          & (1) \\
MMT58  & O2-3.5 If$^*$/WN5-6ha + ? & 1.936 & 0.0 (assumed) & 84.5 + 31 & NGC3603           & (2)  \\
BAT99-6 & O3If$^*$/WN7 +OB +SB1 & 2.0 & 0.0 & - & - & (3) \\
RMC135 & WN7h + ?  & 2.112 & 0.0 (assumed)   & - & 30 Dor                               & (1) \\
RMC140b & WN5h + O & 2.76 & 0.23 & - & 30 Dor                                           & (1)  \\
BAT99-77 & WN7ha  + ?       & 3.00 & 0.32 & - &-                                         & (1) \\
WR29 & O + WN7h & 3.164 & 0.0 (assumed) &  53.3$\pm3.5$ + 42.1$\pm3.5$ &     -           & (4,5) \\
BAT99-12 & O3If$^*$/WN6 + ? & 3.23 & 0.34 & - & -                                        & (1)  \\
WR20a & WN6ha + WN6ha & 3.7 & 0.0 & 83$\pm5$ + 82$\pm5$ & Wd2                           & (6,7,8)  \\
NGC3603-A1 & WN6ha + WN6ha & 3.77 & 0.0 &  116$\pm31$ + 89$\pm16$ & NGC3603             & (9,10) \\
WR121a     & WN7ha + ?     & 4.1 & 0.0 (assumed) & - & W43                             & (11) \\

WR148   & WN7ha + O5V         & 4.32 & 0.0 & (33+37) & -                                        & (12) \\

Mk30     & O3If$^*$/WN6 + ? & 4.70 & 0.20 & - & 30 Dor                                       & (1) \\
RMC136c & WN5h + ? & $\sim8.4$ & 0.0 (assumed) & - & 30 Dor & (13) \\   
NGC3603-C  & WN6ha  + ?    & 8.89 & 0.30 & - & NGC3603                                  & (9,10)  \\ 
F2 & WN8--9h + O5--6~Ia$^+$ & 10.5 & 0.075 & 82$\pm$12 + 60$\pm$8 & Arches               &  (14)  \\
F6 & WN8--9h + ? & 13.378 & 0.6  & - &  Arches                                           & (this work)  \\
WR12    & WN8h + ?         & 23.9 & 0.0   &- &-                                & (15)  \\
WR21a & O3/WN5ha + O3Vz((f*)) & 31.672 &0.695 & 64.4$\pm4.8$ + 36.3$\pm1.7$ & -         & (16,17,18)\\
      &                       &      & & (103.6$\pm10.2$ + 58.3$\pm3.7$) &  & \\

RMC144 & WN6h + WN6h & 70.4 & 0.51  & $>48.6\pm1.8$ + $>45.7\pm1.9$ & 30 Dor    & (19) \\
     &           &       &       &  ($74\pm4 + 69\pm4$)  &      &    \\

WR22 & WN7+abs + O9 & 80.3 & 0.598 & $>55.3\pm7.3$ + $>20.6\pm1.7$ & Carina                  & (20,21,22) \\
Mk37 & O2.5If*/WN6 + ? & 92.6 & 0.0 (assumed) & - & 30 Dor                               &  (1) \\
Mk34 & WN5h + WN5h & 155.1 & 0.68  & $>65\pm7$ + $>60\pm7$ & 30 Dor                        & (23)  \\
     &           &       &       &  ($139^{+21}_{-18} + 127^{+17}_{-17}$)  &      &    \\
RMC145 & WN6h + O3.5~If$^*$/WN7 & 159 & 0.78  & $53^{+40}_{-20} + 54^{+40}_{-20}$  & 30 Dor& (24,25)   \\
J1745-28 &  WN9h + ? &  189 & -   & -            &           -                           &  (26,27)  \\
WR25 & O2.5If$^*$/WN6 + O & $207.7$ & 0.56  & 75$\pm7$ + 27$\pm3$ & Carina          & (28,29)  \\
F7 & WN8--9h + ? & $\gtrsim$1184.46 & -  & - &  Arches                                   & (this work) \\
& &  & &  &  & \\
& &  & &  &  & \\
HD153919 & O6.5Iaf$^+$ + cc &3.412  & 0.0 (assumed) & $58^{+11}_{-11} + 2.44^{+0.27}_{-0.27}$ & -  & (30)\\
Cyg OB2 B17 & O7Iaf + O9Iaf & 4.022 & 0.0 (assumed) & $60^{+5}_{-5} + 45^{+4}_{-4}$ & Cyg OB2 & (31) \\
F25        & O4-5 Ia + ? & $\sim5-9$? & $>0.0$    &  - & Arches & (this work) \\      
F35       & O4-5 Ia + ? & 6.043    & 0.44          & - & Arches & (this work) \\
HD166734 & O7If + O9I(f) & 34.5 & 0.618 & $39.5^{+5.4}_{-4.4} + 33.5^{+4.6}_{-3.7}$ & - & (32) \\
F15     & O6-7 Ia$^+$ + ? & 83.95 & 0.0 & - & Arches & (this work) \\  
LS III+46 11 & O3.5If$^*$ + O3.5 If* & 97.2 & 0.56 &  $>38.8\pm0.8$ + $>35.6\pm0.8$ & Berkeley 90 & (33)  \\
RMC139 &O6.5Iafc + O6Iaf & 153.9 & 0.382 & $>69.4\pm4.1$ + $>53.9\pm3.1$ & 30 Dor & (34,35)\\
& &  & &  &  & \\
\hline
\end{tabular}
\end{center}
{Component masses given in parenthesis for WR21a and WR148 are determined under the expectation  of  a canonical mass for the 
O3Vz and O5V secondaries (Tramper et al.  \cite{tramper16}, Munoz et al. \cite{munoz}), while those
for Mk34 are inferred from  spectroscopic analysis (Tehrani et al. \cite{tehrani}). References used in the construction of this table are: 
(1) Schnurr et al. (\cite{schnurr08b}); 
(2) Jaque Arancibia et al. (\cite{JA});
(3) Niemela et al. (\cite{niemela01});
(4) Niemela \& Gamen (\cite{niemela00});
(5) Gamen et al. (\cite{gamen09});
(6) Rauw et al. (\cite{rauw04}); 
(7) Bonanos et al. (\cite{bonanos04}); 
(8) Naze et al. \cite{naze08}); 
(9)  Moffat et al. (\cite{moffat04}); 
(10) Schnurr et al. (\cite{schnurr08b}); 
(11) Arora \& Pandy (\cite{arora});
(12) Munoz et al. (\cite{munoz});
(13) Schnurr et al. (\cite{schnurr09b});
(14) Lohr et al. (\cite{lohr}); 
(15) Fahed \& Moffat (\cite{fahed});
(16) Benaglia et al. (\cite{benaglia05}); 
(17) Niemela et al. (\cite{niemela08}); 
(18) Tramper et al.  (\cite{tramper16});
(19) Shenar et al. (\cite{shenar21}); 
(20) Rauw et al. (\cite{rauw96});
(21) Schweickhardt et al. (\cite{schweickhardt99}); 
(22) Gr\"{a}fener \& Hamann (\cite{grafener08}); 
(23) Tehrani et al. (\cite{tehrani});
(24) Schnurr et al. (\cite{schnurr09a}); 
(25) Shenar et al. (\cite{shenar17}); 
(26) Mikles et al. (\cite{mikles}); 
(27) Clark et al. (\cite{clark09b});
(29) Gamen et al. (\cite{gamen08}); 
(29) Hur et al. (\cite{hur});
(30) Clark et al. (\cite{clark02});
(31) Stroud et al. (\cite{stroud});
(32) Mahy et al. (\cite{mahy17});
(33) Ma\'{i}z Apell\'{a}niz et al. (\cite{MAIZ});
(34) Taylor et al. (\cite{taylor});
(35) Mahy et al. (\cite{mahy20a})
} 
\end{table*}

\section{Concluding remarks and future prospects}

In this paper we have presented the results of a long term spectroscopic monitoring campaign of the
massive members of the Arches cluster. With the inclusion of previous published data the baseline of
observations extends to $\sim19$yrs yet no stars show evidence for long term secular variability,
such as that seen in LBVs. This is surprising since both observational data and theoretical
simulations suggest that comparably  massive stars should encounter this evolutionary phase
(Appendix B). Nevertheless, this lack of variability permits the primary science goal of the
programme - an assessment of the binary frequency and properties of very massive stars. We find 13 RV variables from a total of 37 stars ($\sim35$\%), which rises to $\sim60$\% if the RV threshold is relaxed slightly; of these, provisional orbital solutions are
reported for five stars and constraints for a further two. Orbital periods range from $\sim6$d to
$\gtrsim1184$d, with a number showing pronounced eccentricity (Sect. 3.1).
 
Examination of X-ray and radio data in order to identify the radiative signature of colliding wind
systems identifies  three additional candidate binaries, while the  
composite thermal+non-thermal radio spectra of a further three are at least  consistent with binarity (Sect. 3.2). Excluding the latter stars at this time - and prior to correction for 
observational biases - we report a conservative lower limit to the cluster binary fraction of 16/37 ($\sim43$\%). However we are both incomplete for the O supergiant cohort and 
less sensitive to variability  in these stars by virtue of their lower S/N spectra, so a fraction of 10/20 for the combined - and completely sampled - WNLh and O hypergiant populations is more instructive (rising to 8/13 for the former stars alone; Sect. 3.3). 

Dynamical mass estimates for the sole eclipsing SB2 binary F2 suggest extreme current masses for
both WN8-9ha ($82\pm12M_{\odot}$) and O5-6 Ia$^+$ ($60\pm8M_{\odot}$) components. Quantitative model
atmosphere analysis for the remaining binary candidates subject to orbital analysis is consonant
with this finding, allowing us to infer evolutionary initial masses of $\sim100-120M_{\odot}$ for
the WNLh stars and $\gtrsim60M_{\odot}$ for the O stars. Both findings are consistent with an age of
$\sim2.0-2.5$Myr for the Arches. 

Several conclusions follow directly  from  consideration of these data. The combination of short
orbital periods and significant eccentricities of  F6 and F35 strongly suggests that they are
pre-interaction systems (Table~\ref{pubin}); as a consequence it would appear that the very massive
stars within the Arches do not form exclusively via binary mass transfer and/or merger of lower mass
stars (cf. Schneider et al. \cite{schneider}). Indeed we find no evidence for ongoing or recent
binary interaction amongst any cluster members (Clark et al. \cite{clark18a}), although we cannot
exclude the possibility that a subset of stars are the product of such an evolutionary channel.
Secondly, star formation in the extreme conditions prevalent in the CMZ yields very massive (binary)
stars in a similar manner to more quiescent regions of the Galaxy. This is of particular interest
since  the environment of the CMZ is thought to closely  resemble that of  high-redshift star-forming galaxies (Kruijssen \& Longmore \cite{dkl}), suggesting that star formation may proceed in a
comparable manner in such objects.  

The properties of the Arches members are better understood by being placed in a wider context, and
we provide a summary of the properties of all known WNLha and O-type hypergiant binaries in
Table~\ref{pubin}. This enlarged dataset shows that such stars are exceptionally massive, with
dynamical and evolutionary mass determinations suggesting a lower limit of $\sim50M_{\odot}$, with
extreme examples greatly exceeding $100M_{\odot}$. We highlight that  such a mass range  is in
excess of that sampled in previous studies of OB  binaries  (Sana et al. \cite{sana12}, Kobulnicky
et al. \cite{kobulnicky}, Almeida et al. \cite{almeida}). Nevertheless, within observational
uncertainties and before correction for survey biases,  we find no evidence that the physical
properties of the WNLh and O hypergiant binaries - orbital period, eccentricity and mass ratio
distributions - significantly differ from those of lower mass OB stars. The sole exception to this
is an apparent absence of  WNLh and O hypergiant binaries with very short orbital periods
($\lesssim1.9$d), potentially due to a combination of binary  evolution depleting the tightest
systems, and their physical extent, which precludes their accommodation within such compact
configurations. As with the Arches, the  pronounced eccentricities of a number of short period
systems suggests they are in a pre-interaction state. In this regard the compact binaries MMT58 and
WR121a are of  particular interest due to  the extreme youth implied for them  as a result of  their
location within active star forming regions (Sect. 5.1). Taken as a whole these observations bolster
the conclusion from the Arches cohort that very massive WNLh stars do not solely form via a binary
channel.  

In order to assess the frequency of occurrence of binarity amongst WNLh stars we followed the
approach of Sana et al. (\cite{sana12}) and reviewed the properties of those found in stellar
clusters utilising a combination of radial velocity, radio and X-ray data. This required a
re-analysis of extant X-ray data. This revealed that while single WNLh stars appear to be
intrinsically faint, binaries hosting them appear to be amongst the most luminous of all
non-accreting stellar systems, with $\sim76$\% confirmed examples  lying above a  $(L_{\rm x}/L_{\rm
bol})\sim10^{-7}$ threshold; we consequently adopt this as an identifier for CWBs.

Utilising all diagnostics at our disposal  we identify a total of 33/58 binaries amongst the
complete sample, noting that is almost certainly a lower limit to the true frequency given the
limitations of the datasets utilised (Sect. 4 and Appendix C). Because of different observational
approaches adopted, quantitative  comparison of this value to the binary frequencies returned by
other surveys is potentially misleading. Nevertheless, even before correcting for bias and
incompleteness our results appear qualitatively consistent with the hypothesis  that the high
binary fraction of very massive ($>50M_{\odot}$) stars continues the trend found for lower mass OB
stars  by other large scale surveys (Sect. 5.1.2).

Such a finding has important implications for a number of diverse astrophysical questions including implications for the birth of very massive stars in extreme environments, which has already been discussed, the inference of the (I)MF from an observational luminosity function (cf. Schneider et al. \cite{schneider}), and at larger scales, there is an increasing realisation that such extremely luminous stars play a substantive role in
(i) the provision  of ionising, radiative feedback (e.g. Doran et al. \cite{doran})
and (ii) the production of cosmic rays and subsequently high energy (TeV) $\gamma$-ray emission via particle acceleration in wind shocks.

In order to fully address these questions it is clear that we will have to expand and refine our current dataset to better understand the impact of binarity on the formation, evolution and fate of WNh stars. 
In terms of the Arches this will necessitate (i) Monte Carlo simulations to account
for biases introduced by differences in the cadence and number of observations of individual stars,
(ii) the population of a cluster HR diagram via model atmosphere analysis in order to place the
binaries into context, and (iii) an increased observational effort to derive orbital solutions for
the full cohort of RV variables. Expanding this undertaking to include Mercer 30 \& 81, Danks 1,
HM-1, Sco OB2 and W43  would raise the sample size exposed to multi-epoch spectroscopic observations
by 15, with a further 12 unstudied examples distributed  through the CMZ (Clark et al.
\cite{clark21}) and seven additional targets within the outskirts of Wd2, NGC3603 and 30 Dor.
Although observationally expensive, such an effort would increase the number of WNLh stars subject
to such  investigations  to 77, effectively doubling the sample size. Assuming an empirical  binary
fraction of $\gtrsim60$\%, this would allow the determination of the  full suite of orbital
parameters for a statistically robust sample of  $\gtrsim50$ binaries containing the most massive
stars currently forming within the local Universe. We note that confirmation  of the frequency of
occurrence would also be a central outcome of such a programme.

\section*{Acknowledgements}
We thank the referee, Alex de Koter, for providing a thorough review with helpful comments and interesting arguments.
F.N. acknowledges funding by grants PID2019-105552RB-C41 and MDM-2017-0737-19-3 Unidad de Excelencia "María de Maeztu".
LRP acknowledges the support of the Generalitat Valenciana through the grant APOSTD/2020/247.
This research is partially supported by the Spanish Government under grant PGC2018-093741-B-C21 (MICIU/AEI/FEDER, UE).

\section*{Data Availability}
Table 1 contains the results of the radial velocity analysis and is available via the CDS service. The individual radial velocity measurements for each epoch for all targets are available in Table~\ref{tab:appA1} and is available as supplementary material (online).
Appendix~\ref{app:Xray-data} contains the details of how the X-ray observations were compiled and is available as supplementary material (online).
The raw spectroscopic observational data can be freely obtained from the ESO Science Archive Facility. 
 




{}



\appendix

\section{Observing log and presentation of RV measurements}

Table~\ref{tab:appA1} contains the observing log and RV measurements for all epochs for all targets.

\begin{table*}
 \caption{Dates of epochs of observation for Arches sample and associated RV measurements}
 \label{tab:appA1}
  \begin{tabular}{ccl}
 \hline
 Star & Instrument & Epochs of observations (and associated RV measurements) \\
 \hline
      & & \\
    B1 & Sinfoni & 10/06/05 (0.21$\pm$3.05 km~s$^{-1}$), 08/08/13 (-2.95$\pm$2.97 km~s$^{-1}$), 21/04/17 (0.09$\pm$2.35 km~s$^{-1}$), \\ 
       &         & 26/04/17 (0.07$\pm$3.16 km~s$^{-1}$), 30/04/17 (2.58$\pm$3.30 km~s$^{-1}$) \\
       & KMOS    & 30/04/14 (-3.60$\pm$5.96 km~s$^{-1}$), 23/07/14 (2.86$\pm$9.78 km~s$^{-1}$), 04/08/14 (11.31$\pm$15.10 km~s$^{-1}$), \\          
       &         & 05/08/14 (-7.02$\pm$6.55 km~s$^{-1}$), 11/08/14 (1.03$\pm$5.97 km~s$^{-1}$), 12/08/14 (-10.32$\pm$5.89 km~s$^{-1}$), \\        
       &         & 13/08/14 (5.75$\pm$7.25 km~s$^{-1}$) \\
      
       &         & \\

    F1 & Sinfoni & 10/06/05 (-6.43$\pm$6.53 km~s$^{-1}$), 29/06/11 (4.70$\pm$5.57 km~s$^{-1}$), 27/08/11 (-9.50$\pm$5.81 km~s$^{-1}$), \\ 
       &         & 28/08/11 (-1.67$\pm$5.01 km~s$^{-1}$), 17/07/13 (4.46$\pm$5.38 km~s$^{-1}$, 08/08/13 (4.70$\pm$5.86 km~s$^{-1}$), \\
       &         & 26/04/17 (2.97$\pm$5.04 km~s$^{-1}$), 30/04/17 (-5.45$\pm$5.55 km~s$^{-1}$), 26/07/17 (-6.90$\pm$5.49 km~s$^{-1}$), \\
       &         & 07/05/18 (-9.93$\pm$7.01 km~s$^{-1}$), 07/08/18 (4.85$\pm$6.66 km~s$^{-1}$, 10/08/18 (4.73$\pm$9.26 km~s$^{-1}$), \\
       &         & 12/08/18 (11.38$\pm$7.61 km~s$^{-1}$), 17/08/18 (2.07$\pm$4.94 km~s$^{-1}$) \\
       & KMOS    & 12/08/14 (-2.95$\pm$3.13 km~s$^{-1}$), 13/08/14 (2.95$\pm$3.13 km~s$^{-1}$) \\
    
     & & \\
    
    F2 & Sinfoni & 10/06/05 (-16.05$\pm$8.22 km~s$^{-1}$), 29/06/11 (-230.15$\pm$8.26 km~s$^{-1}$), 27/08/11 (96.20$\pm$7.19 km~s$^{-1}$), \\
       &         & 28/08/11 (-85.25$\pm$11.19 km~s$^{-1}$), 17/07/13 (130.44$\pm$6.78 km~s$^{-1}$), 08/08/13 (104.81$\pm$7.10 km~s$^{-1}$) \\
       & KMOS    & 30/04/14 (-70.04$\pm$14.07 km~s$^{-1}$), 23/07/14 (-44.94$\pm$12.45 km~s$^{-1}$), 04/08/14 (-56.15$\pm$11.73 km~s$^{-1}$), \\
       &         & 05/08/14 (-0.05$\pm$12.62 km~s$^{-1}$), 11/08/14 (198.74$\pm$11.38 km~s$^{-1}$), 12/08/14 (54.41$\pm$16.62 km~s$^{-1}$), \\
       &         & 13/08/14 (-81.97$\pm$11.56 km~s$^{-1}$) \\
    
       & & \\

    F3 & Sinfoni & 10/06/05 (-4.81$\pm$4.39 km~s$^{-1}$), 08/08/13 (1.29$\pm$2.39 km~s$^{-1}$), 21/04/17 (4.98$\pm$2.33 km~s$^{-1}$), \\
       &         & 26/04/17 (-1.88$\pm$2.78 km~s$^{-1}$), 30/04/17 (-0.60$\pm$2.31 km~s$^{-1}$), 08/08/18 (2.42$\pm$3.11 km~s$^{-1}$), \\
       &         & 17/08/18 (-1.39$\pm$2.49 km~s$^{-1}$), 18/08/18 (low S/N) \\
    
       & & \\

    F4 & Sinfoni & 29/05/13 (0.54$\pm$2.19 km~s$^{-1}$), 08/06/13 (-0.17$\pm$2.16 km~s$^{-1}$), 17/07/13 (-6.78$\pm$2.50 km~s$^{-1}$), \\
       &         & 03/08/13 (-3.57$\pm$1.79 km~s$^{-1}$), 04/08/13 (-2.21$\pm$1.99 km~s$^{-1}$), 06/08/13 (-4.98$\pm$2.66 km~s$^{-1}$), \\
       &         &  08/08/13 (12.66$\pm$2.63 km~s$^{-1}$), 26/04/17 (4.75$\pm$2.18 km~s$^{-1}$), 30/04/17 (0.10$\pm$1.76 km~s$^{-1}$), \\
       &         &  27/07/17 (-0.33$\pm$1.95 km~s$^{-1}$) \\ 
    
      & & \\

    F5 & Sinfoni & 10/06/05 (0.93$\pm$3.14 km~s$^{-1}$), 08/08/13 (-7.49$\pm$1.92 km~s$^{-1}$), 21/04/17 (6.91$\pm$1.77 km~s$^{-1}$), \\
       &         & 26/04/17 (5.32$\pm$2.46 km~s$^{-1}$), 30/04/17 (4.52$\pm$1.92 km~s$^{-1}$), 08/08/18 (-1.70$\pm$1.98 km~s$^{-1}$), \\
       &         & 18/08/18 (-8.51$\pm$2.09 km~s$^{-1}$) \\ 
    
     & & \\

    F6 & Sinfoni & 14/04/11 (-17.17$\pm$6.31 km~s$^{-1}$), 17/07/13 (-4.28$\pm$4.19 km~s$^{-1}$), 26/04/17 (-9.37$\pm$3.92 km~s$^{-1}$), \\ 
       &         & 30/04/17 (15.87$\pm$3.99 km~s$^{-1}$), 26/07/17 (-8.94$\pm$5.00 km~s$^{-1}$), 07/05/18 (-14.99$\pm$5.11 km~s$^{-1}$), \\
       &         & 07/08/18 (-11.45$\pm$4.95 km~s$^{-1}$), 12/08/18 (50.44$\pm$6.40 km~s$^{-1}$), 17/08/18 (-1.19$\pm$3.96 km~s$^{-1}$), \\
       &         &  18/08/18 (1.08$\pm$3.99 km~s$^{-1}$) \\
       & KMOS    & 30/04/14 (-0.44$\pm$11.38 km~s$^{-1}$), 23/07/14 (7.78$\pm$12.98 km~s$^{-1}$), 04/08/14 (26.05$\pm$12.73 km~s$^{-1}$),\\      
       &         & 05/08/14 (27.41$\pm$15.40 km~s$^{-1}$), 11/08/14 (-28.18$\pm$12.95 km~s$^{-1}$), 12/08/14 (-16.96$\pm$10.10 km~s$^{-1}$), \\
       &         & 13/08/14 (-15.66$\pm$14.04 km~s$^{-1}$) \\
    
      & & \\

    F7 & Sinfoni & 29/05/13 (17.56$\pm$2.93 km~s$^{-1}$), 18/06/13 (17.01$\pm$3.86 km~s$^{-1}$), 17/07/13 (5.76$\pm$3.09 km~s$^{-1}$), \\
       &         & 03/08/13 (5.61$\pm$3.39 km~s$^{-1}$), 04/08/13 (22.99$\pm$3.03 km~s$^{-1}$), 06/08/13 (9.57$\pm$3.05 km~s$^{-1}$), \\
       &         & 08/08/13 (19.13$\pm$3.24 km~s$^{-1}$), 26/04/17 (-17.26$\pm$2.98 km~s$^{-1}$), 30/04/17 (-12.43$\pm$2.89 km~s$^{-1}$), \\
       &         & 27/07/17 (-20.84$\pm$3.10 km~s$^{-1}$), 07/05/18 (-6.31$\pm$4.31 km~s$^{-1}$), 07/08/18 (-12.14$\pm$2.77 km~s$^{-1}$), \\
       &         & 12/08/18 (-8.22$\pm$2.62 km~s$^{-1}$), 17/08/18 (-12.72$\pm$3.49 km~s$^{-1}$), 18/08/18 (-7.70$\pm$3.00 km~s$^{-1}$)\\
    
      & & \\

    F8 & Sinfoni & 14/04/11 (-12.36$\pm$3.78 km~s$^{-1}$), 17/07/13 (5.21$\pm$2.73 km~s$^{-1}$), 06/08/13 (8.15$\pm$2.67 km~s$^{-1}$), \\
       &         & 26/04/17 (9.34$\pm$3.41 km~s$^{-1}$), 30/04/17 (12.12$\pm$2.33 km~s$^{-1}$), 07/05/18 (-1.91$\pm$2.85 km~s$^{-1}$), \\
       &         & 07/08/18 (-0.95$\pm$2.30 km~s$^{-1}$), 12/08/18 (-8.83$\pm$2.56 km~s$^{-1}$), 17/08/18 (-3.95$\pm$2.66 km~s$^{-1}$), \\
       &         & 18/08/18 (-6.81$\pm$2.35 km~s$^{-1}$) \\
       & KMOS    & 30/04/14 (-3.60$\pm$5.96 km~s$^{-1}$), 23/07/14 (2.86$\pm$9.78 km~s$^{-1}$), 04/08/14 (11.31$\pm$15.10 km~s$^{-1}$), \\
       &         & 05/08/14 (-7.02$\pm$6.55 km~s$^{-1}$), 11/08/14 (1.03$\pm$5.97 km~s$^{-1}$), 12/08/14 (-10.32$\pm$5.89 km~s$^{-1}$), \\
       &         & 13/08/14 (5.75$\pm$7.25 km~s$^{-1}$) \\
    
       & & \\

    F9 & Sinfoni & 10/06/05 (-6.86$\pm$7.18 km~s$^{-1}$), 29/06/11 (6.97$\pm$4.09 km~s$^{-1}$), 27/08/11 (-16.60$\pm$5.52 km~s$^{-1}$), \\
       &         & 28/08/11 (3.03$\pm$4.53 km~s$^{-1}$), 17/07/13 (6.47$\pm$5.00 km~s$^{-1}$), 08/08/13 (11.26$\pm$5.92 km~s$^{-1}$), \\
       &         & 25/04/17 (-3.35$\pm$4.76 km~s$^{-1}$), 26/04/17 (2.63$\pm$4.73 km~s$^{-1}$), 30/04/17 (1.31$\pm$4.66 km~s$^{-1}$), \\
       &         & 23/06/17 (-7.50$\pm$4.62 km~s$^{-1}$), 07/05/18 (-1.43$\pm$4.54 km~s$^{-1}$), 07/08/18 (4.97$\pm$4.21 km~s$^{-1}$), \\
       &         & 10/08/18 (0.32$\pm$3.82 km~s$^{-1}$), 12/08/18 (3.70$\pm$4.30 km~s$^{-1}$), 17/08/18 (-4.91$\pm$7.48 km~s$^{-1}$)\\
       &  KMOS   & 30/04/14 (-9.74$\pm$10.15 km~s$^{-1}$), 23/07/14 (3.85$\pm$9.07 km~s$^{-1}$), 04/08/14 (0.17$\pm$8.78 km~s$^{-1}$), \\
       &         & 05/08/13 (1.24$\pm$9.85 km~s$^{-1}$), 12/08/14 (1.38$\pm$7.83 km~s$^{-1}$), 13/08/14 (3.09$\pm$11.35 km~s$^{-1}$) \\
    
      & & \\

\hline
\end{tabular}
\end{table*}

\begin{table*}
 \contcaption{Dates of epochs of observation for Arches sample and associated RV measurements}
  \begin{tabular}{ccl}
 \hline
 Star & Instrument & Epochs of observations (and associated RV measurements) \\
 \hline
      & & \\

    F10 & Sinfoni & 10/06/05 (-0.31$\pm$5.43 km~s$^{-1}$), 25/04/17 (4.97$\pm$3.36 km~s$^{-1}$), 26/04/17 (0.55$\pm$3.41 km~s$^{-1}$), \\
        &         & 30/04/17 (-4.84$\pm$3.01 km~s$^{-1}$), 23/06/17 (-0.61$\pm$6.03 km~s$^{-1}$), 07/05/18 (6.40$\pm$4.27 km~s$^{-1}$), \\
        &         & 07/08/18 (2.99$\pm$3.32 km~s$^{-1}$), 10/08/18 (-2.97$\pm$3.78 km~s$^{-1}$), 12/08/18 (-1.93$\pm$3.29 km~s$^{-1}$), \\
        &         &   17/08/18 (-4.26$\pm$4.88 km~s$^{-1}$) \\

     & & \\
    
    F12 & Sinfoni & 10/06/05 (-8.42$\pm$6.20 km~s$^{-1}$), 29/06/11 (5.47$\pm$6.26 km~s$^{-1}$), 27/08/11 (-0.28$\pm$5.79 km~s$^{-1}$), \\
        &         & 28/08/11 (10.62$\pm$5.31 km~s$^{-1}$), 17/07/13 (-0.14$\pm$6.85 km~s$^{-1}$), 03/08/13 (-0.03$\pm$6.65 km~s$^{-1}$), \\
        &         & 04/08/13 (low S/N), (06/08/13), 08/08/13 (-1.02$\pm$6.04 km~s$^{-1}$), 26/04/17 (1.72$\pm$4.41 km~s$^{-1}$), \\
        &         & 30/04/17 (-5.44$\pm$4.66 km~s$^{-1}$), 27/07/17 (-2.48$\pm$4.89 km~s$^{-1}$) \\
    
     & & \\

    F13 & Sinfoni & 10/06/05 (-7.19$\pm$11.31 km~s$^{-1}$), 25/04/17 (3.43$\pm$6.63 km~s$^{-1}$), 26/04/17 (16.34$\pm$7.68 km~s$^{-1}$), \\
        &         & 30/04/17 (-0.93$\pm$7.78 km~s$^{-1}$), 23/06/17 (0.10$\pm$6.97 km~s$^{-1}$), 07/05/18 (-13.19$\pm$6.80 km~s$^{-1}$), \\
        &         & 07/08/18 (-2.96$\pm$6.84 km~s$^{-1}$), 10/08/18 (-4.21$\pm$6.72 km~s$^{-1}$), 12/08/18 (8.61$\pm$6.24 km~s$^{-1}$) \\
    
    & & \\

    F14 & Sinfoni & 14/04/11 (-11.98$\pm$3.20 km~s$^{-1}$), 17/07/13 (-8.51$\pm$2.68 km~s$^{-1}$), 26/04/17 (15.39$\pm$2.75 km~s$^{-1}$), \\
        &         & 30/04/17 (20.40$\pm$2.96 km~s$^{-1}$), 26/07/17 (-6.61$\pm$3.51 km~s$^{-1}$), 07/05/18 (6.49$\pm$3.97 km~s$^{-1}$), \\
        &         & 07/08/18 (3.44$\pm$3.25 km~s$^{-1}$), 08/08/18 (0.60$\pm$3.03 km~s$^{-1}$), 12/08/18 (-3.77$\pm$2.68 km~s$^{-1}$),  \\
        &         & 17/08/18 (-12.41$\pm$2.55 km~s$^{-1}$), 18/08/18 (-3.04$\pm$2.85 km~s$^{-1}$) \\
    
    &   &  \\

    F15 & Sinfoni & 29/05/13 (28.79$\pm$4.21 km~s$^{-1}$), 18/06/13 (-1.63$\pm$6.15 km~s$^{-1}$), 17/07/13 (-39.28$\pm$4.47 km~s$^{-1}$), \\
        &         & 03/08/13 (13.99$\pm$4.17 km~s$^{-1}$), 04/08/13 (16.52$\pm$3.74 km~s$^{-1}$), 06/08/13 (15.26$\pm$4.20 km~s$^{-1}$),\\
        &         & 08/08/13 (13.31$\pm$5.41 km~s$^{-1}$), 26/04/17 (35.95$\pm$4.93 km~s$^{-1}$), 30/04/17 (25.80$\pm$3.41 km~s$^{-1}$), \\
        &         & 23/06/17 (-8.32$\pm$3.80 km~s$^{-1}$), 27/07/17 (24.97$\pm$3.78 km~s$^{-1}$), 07/05/18 (-42.62$\pm$3.66 km~s$^{-1}$), \\
        &         & 07/08/18 (-38.63$\pm$4.07 km~s$^{-1}$), 12/08/18 (-26.95$\pm$3.80 km~s$^{-1}$), 17/08/18 (-5.37$\pm$5.57 km~s$^{-1}$), \\
        &         & 18/08/18 (-11.78$\pm$3.68 km~s$^{-1}$)\\
    
   &   &   \\

    F16 & Sinfoni & 14/04/11 (0.01$\pm$3.81 km~s$^{-1}$), 17/07/13 (-0.06$\pm$3.83 km~s$^{-1}$), 06/08/13 (-5.03$\pm$3.45 km~s$^{-1}$), \\
        &         & 26/04/17 (-10.82$\pm$2.92 km~s$^{-1}$), 30/04/17 (1.68$\pm$3.57 km~s$^{-1}$), 07/05/18 (2.66$\pm$3.73 km~s$^{-1}$), \\
        &         & 07/08/18 (-1.00$\pm$2.97 km~s$^{-1}$), 12/08/18 (15.14$\pm$3.30 km~s$^{-1}$), 17/08/18 (-5.03$\pm$2.98 km~s$^{-1}$), \\
        &         &  18/08/18 (2.45$\pm$3.24 km~s$^{-1}$)\\
        & KMOS    & 30/04/14 \\
     
      & & \\

    F17 & Sinfoni & 10/06/05 (2.23$\pm$6.08 km~s$^{-1}$), 25/04/17 (-1.93$\pm$5.02 km~s$^{-1}$), 26/04/17 (2.40$\pm$4.31 km~s$^{-1}$), \\
        &         & 30/04/17 (10.94$\pm$6.02 km~s$^{-1}$), 23/06/17 (-9.53$\pm$4.55 km~s$^{-1}$), 07/05/18 (-8.14$\pm$6.21 km~s$^{-1}$),\\
        &         & 07/08/18 (1.92$\pm$4.01 km~s$^{-1}$), 10/08/18 (-9.03$\pm$4.27 km~s$^{-1}$), 12/08/18 (1.93$\pm$4.58 km~s$^{-1}$), \\
        &         & 17/08/18 (9.21$\pm$5.19 km~s$^{-1}$) \\
    
     & & \\

    B4  & Sinfoni & 29/05/13 (-0.65$\pm$9.68 km~s$^{-1}$), 18/06/13 (-5.08$\pm$5.43 km~s$^{-1}$), 17/07/13 (-7.77$\pm$6.16 km~s$^{-1}$), \\
        &         & 03/08/13 (13.27$\pm$4.95 km~s$^{-1}$), 04/08/13 (5.50$\pm$4.82 km~s$^{-1}$), 06/08/13 (-10.01$\pm$5.40 km~s$^{-1}$), \\
        &         & 08/08/13 (-6.80$\pm$6.59 km~s$^{-1}$), 26/04/17 (7.74$\pm$6.32 km~s$^{-1}$), 30/04/17 (2.76$\pm$4.82 km~s$^{-1}$), \\
        &         & 27/07/17 (2.48$\pm$6.25 km~s$^{-1}$), 17/08/18 (-1.44$\pm$5.56 km~s$^{-1}$) \\
    
     & & \\

    F18 & Sinfoni & 29/05/13 (14.15$\pm$7.14 km~s$^{-1}$), 18/06/13 (13.40$\pm$7.66 km~s$^{-1}$), 17/07/13 (1.61$\pm$5.14 km~s$^{-1}$), \\
        &         & 03/08/13 (-0.02$\pm$5.67 km~s$^{-1}$), 04/08/13 (-9.13$\pm$5.09 km~s$^{-1}$), 06/08/13 (-7.85$\pm$5.05 km~s$^{-1}$), \\
        &         & 08/08/13 (-18.23$\pm$4.96 km~s$^{-1}$), 26/04/17 (13.94$\pm$5.01 km~s$^{-1}$), 30/04/17 (8.56$\pm$5.11 km~s$^{-1}$), \\
        &         & 27/07/17 (-16.42$\pm$5.72 km~s$^{-1}$) \\
    
    & & \\

    F19 & Sinfoni & 10/06/05 (low S/N), 29/06/11 (low S/N), 27/08/11 (low S/N), 17/07/13 (low S/N), 08/08/13 (low S/N) \\
    
     & & \\
     & & \\

    F20 & Sinfoni & 29/05/13 (0.77$\pm$8.91 km~s$^{-1}$), 18/06/13 (0.62$\pm$14.07 km~s$^{-1}$), 17/07/13 (7.06$\pm$8.18 km~s$^{-1}$), \\
        &         & 03/08/13 (-10.50$\pm$7.35 km~s$^{-1}$), 04/08/13 (-14.15$\pm$8.00 km~s$^{-1}$), 06/08/13 (-15.23$\pm$7.31 km~s$^{-1}$), \\
        &         & 08/08/13 (-1.34$\pm$8.69 km~s$^{-1}$), 26/04/17 (15.40$\pm$7.17 km~s$^{-1}$), 30/04/17 (-4.29$\pm$6.67 km~s$^{-1}$), \\
        &         & 27/07/17 (low S/N), 07/05/18 (-0.13$\pm$7.67 km~s$^{-1}$), 07/08/18 (1.10$\pm$8.29 km~s$^{-1}$), \\
        &         & 12/08/18 (5.18$\pm$9.45 km~s$^{-1}$), 17/08/18 (4.79$\pm$9.39 km~s$^{-1}$), 18/08/18 (10.73$\pm$6.66 km~s$^{-1}$)\\
    
     &  &  \\

\hline
\end{tabular}
\end{table*}

\begin{table*}
 \contcaption{Dates of epochs of observation for Arches sample and associated RV measurements}
  \begin{tabular}{ccl}
 \hline
 Star & Instrument & Epochs of observations (and associated RV measurements) \\
 \hline
    F21 & Sinfoni & 29/05/13 (5.77$\pm$6.65 km~s$^{-1}$), 18/06/13 (7.08$\pm$10.76 km~s$^{-1}$), 17/07/13 (-10.74$\pm$6.74 km~s$^{-1}$), \\
        &         & 03/08/13 (-6.53$\pm$9.81 km~s$^{-1}$), 04/08/13 (-7.02$\pm$5.52 km~s$^{-1}$), 06/08/13 (-12.92$\pm$6.37 km~s$^{-1}$),\\
        &         & 08/08/13 (-9.74$\pm$7.97 km~s$^{-1}$), 26/04/17 (7.19$\pm$5.54 km~s$^{-1}$), 30/04/17 (-0.20$\pm$6.25 km~s$^{-1}$), \\
        &         & 27/07/17 (4.62$\pm$7.15 km~s$^{-1}$), 07/05/18 (-11.57$\pm$8.70 km~s$^{-1}$), 07/08/18 (8.02$\pm$6.95 km~s$^{-1}$), \\
        &         & 12/08/18 (7.55$\pm$7.85 km~s$^{-1}$), 17/08/18 (-0.34$\pm$6.35 km~s$^{-1}$), 18/08/18 (18.82$\pm$7.81 km~s$^{-1}$) \\
    
     & & \\

    F22 & Sinfoni & 10/06/05 (-5.10$\pm$5.35 km~s$^{-1}$), 29/06/11 (-19.16$\pm$5.82 km~s$^{-1}$), 27/08/11 (8.71$\pm$6.44 km~s$^{-1}$), \\
        &         & 28/08/11 (2.90$\pm$5.41 km~s$^{-1}$), 17/07/13 (4.54$\pm$6.80 km~s$^{-1}$), 08/08/13 (-3.38$\pm$5.19 km~s$^{-1}$), \\
        &         & 26/04/17 (4.30$\pm$5.59 km~s$^{-1}$), 30/04/17 (8.03$\pm$5.85 km~s$^{-1}$), 27/07/17 (-0.85$\pm$4.93 km~s$^{-1}$) \\
    
      & & \\

    F23 & Sinfoni & 10/06/05 (10.16$\pm$7.14 km~s$^{-1}$), 07/05/18 (-13.09$\pm$5.62 km~s$^{-1}$), 07/08/18 (-3.62$\pm$4.92 km~s$^{-1}$), \\
        &         & 12/08/18 (-0.03$\pm$4.79 km~s$^{-1}$), 17/08/18 (-1.13$\pm$7.85 km~s$^{-1}$), 18/08/18 (7.71$\pm$5.76 km~s$^{-1}$)\\
    
      &  & \\

    F24 & Sinfoni & 10/06/05 (low S/N), 29/06/11 (low S/N), 27/08/11 (-12.15$\pm$10.03 km~s$^{-1}$), 28/08/11 (low S/N), \\
        &     & 17/07/13 (low S/N), 08/08/13 (low S/N), 07/08/18 (13.23$\pm$8.42 km~s$^{-1}$), 10/08/18 (8.28$\pm$7.69 km~s$^{-1}$), \\        
        &         & 12/08/18 (3.17$\pm$6.79 km~s$^{-1}$), 17/08/18 (-12.54$\pm$5.73 km~s$^{-1}$) \\
    
      &  &  \\

    F25 & Sinfoni & 10/06/05 (95.71$\pm$13.49 km~s$^{-1}$), 29/06/11 (-72.56$\pm$11.84 km~s$^{-1}$), 27/08/11 (low S/N),  \\
        &         & 17/07/13 (-20.43$\pm$11.87 km~s$^{-1}$), 08/08/13 (-43.61$\pm$13.97 km~s$^{-1}$), 25/04/17 (37.19$\pm$12.57 km~s$^{-1}$),\\
        &         & 26/04/17 (98.02$\pm$8.60 km~s$^{-1}$), 30/04/17 (-54.12$\pm$9.52 km~s$^{-1}$), 23/06/17 (-54.47$\pm$10.97 km~s$^{-1}$), \\
        &         & 07/05/18 (-59.07$\pm$9.15 km~s$^{-1}$), 07/08/18 (-28.04$\pm$10.28 km~s$^{-1}$), 10/08/18 (54.49$\pm$9.93 km~s$^{-1}$), \\
        &         & 12/08/18 (46.90$\pm$8.21 km~s$^{-1}$) \\
    
       &  &  \\

    F26 & Sinfoni & 10/06/05 (-12.93$\pm$15.09 km~s$^{-1}$), 14/04/11 (21.25$\pm$15.09 km~s$^{-1}$), 17/07/13 (29.97$\pm$14.37 km~s$^{-1}$), \\
        &         & 26/04/17 (5.45$\pm$10.90 km~s$^{-1}$), 26/07/17 (0.30$\pm$13.88 km~s$^{-1}$), 08/08/18 (-10.26$\pm$11.38 km~s$^{-1}$), \\
        &         & 17/08/18 (-23.37$\pm$12.76 km~s$^{-1}$), 18/08/18 (-10.41$\pm$13.26 km~s$^{-1}$) \\
    
      &  & \\

    F27 & Sinfoni & 29/05/13 (10.97$\pm$6.74 km~s$^{-1}$), 18/06/13 (-0.35$\pm$6.96 km~s$^{-1}$), 17/07/13 (-5.64$\pm$4.99 km~s$^{-1}$), \\
        &         & 03/08/13 (0.73$\pm$6.38 km~s$^{-1}$), 04/08/13 (7.96$\pm$5.69 km~s$^{-1}$), 06/08/13 (-6.79$\pm$4.84 km~s$^{-1}$), \\
        &         & 08/08/13 (-10.68$\pm$6.11 km~s$^{-1}$), 26/04/17 (2.16$\pm$5.13 km~s$^{-1}$), 30/04/17 (-3.23$\pm$4.77 km~s$^{-1}$), \\
        &         & 27/07/17 (-21.47$\pm$8.01 km~s$^{-1}$), 07/05/18 (9.45$\pm$5.64 km~s$^{-1}$), 07/08/18 (5.94$\pm$6.26 km~s$^{-1}$), \\
        &         & 12/08/18 (5.91$\pm$7.38 km~s$^{-1}$), 17/08/18 (5.03$\pm$6.73 km~s$^{-1}$), 18/08/18 (low S/N) \\
    
     &  &  \\

    F28 & Sinfoni & 14/04/11 (-21.58$\pm$5.79 km~s$^{-1}$), 17/07/13 (10.94$\pm$5.29 km~s$^{-1}$), 26/04/17 (7.48$\pm$5.36 km~s$^{-1}$), \\
        &         & 30/04/17 (2.50$\pm$8.12 km~s$^{-1}$), 26/07/17 (-0.07$\pm$6.33 km~s$^{-1}$), 07/05/18 (-13.96$\pm$4.76 km~s$^{-1}$), \\
        &         & 07/08/18 (5.76$\pm$7.18 km~s$^{-1}$), 12/08/18 (6.00$\pm$5.53 km~s$^{-1}$), 17/08/18 (-4.67$\pm$5.74 km~s$^{-1}$), \\
        &         & 18/08/18 (7.61$\pm$6.10 km~s$^{-1}$) \\
    
     &  &  \\

    F29 & Sinfoni & 29/05/13 (-1.21$\pm$6.51 km~s$^{-1}$), 18/06/13 (3.89$\pm$8.28 km~s$^{-1}$), 17/07/13 (-7.61$\pm$6.59 km~s$^{-1}$), \\
        &         & 03/08/13 (6.66$\pm$6.11 km~s$^{-1}$), 04/08/13 (-1.40$\pm$6.04 km~s$^{-1}$), 06/08/13 (-20.30$\pm$5.88 km~s$^{-1}$), \\
        &         & 08/08/13 (-3.31$\pm$7.56 km~s$^{-1}$), 26/04/17 (-1.28$\pm$5.29 km~s$^{-1}$), 30/04/17 (1.04$\pm$4.63 km~s$^{-1}$), \\
        &         & 23/06/17 (8.67$\pm$7.54 km~s$^{-1}$), 27/07/17 (-10.32$\pm$6.71 km~s$^{-1}$), 07/05/18 (-3.33$\pm$5.45 km~s$^{-1}$), \\
        &         & 07/08/18 (0.10$\pm$7.26 km~s$^{-1}$), 12/08/18 (14.70$\pm$5.68 km~s$^{-1}$), 17/08/18 (-2.47$\pm$8.23 km~s$^{-1}$), \\
        &         &  18/08/18 (16.16$\pm$5.62 km~s$^{-1}$)\\
    
     &  &   \\

    F30 & Sinfoni & 29/06/11 (-6.89$\pm$7.76 km~s$^{-1}$), 27/08/11 (0.69$\pm$14.61 km~s$^{-1}$), 28/08/11 (0.36$\pm$8.24 km~s$^{-1}$), \\
        &         & 17/07/13 (7.20$\pm$7.63 km~s$^{-1}$), 08/08/13 (-1.68$\pm$7.84 km~s$^{-1}$), 26/04/17 (4.76$\pm$9.07 km~s$^{-1}$), \\
        &         & 30/04/17 (-3.81$\pm$8.60 km~s$^{-1}$), 27/07/17 (-0.64$\pm$9.33 km~s$^{-1}$) \\
    
     &  &    \\
 & & \\
  & & \\
 & & \\

    F32 & Sinfoni & 14/04/11 (-2.03$\pm$5.84 km~s$^{-1}$), 17/07/13 (-3.07$\pm$5.15 km~s$^{-1}$), 06/08/13 (-1.79$\pm$6.95 km~s$^{-1}$), \\
        &         & 26/04/17 (9.54$\pm$7.00 km~s$^{-1}$), 30/04/17 (-2.80$\pm$5.45 km~s$^{-1}$), 07/05/18 (-9.88$\pm$7.06 km~s$^{-1}$),\\
        &         & 07/08/18 (-17.89$\pm$5.39 km~s$^{-1}$), 12/08/18 (3.53$\pm$6.78 km~s$^{-1}$), 17/08/18 (-1.58$\pm$7.42 km~s$^{-1}$), \\
        &         & 18/08/18 (25.97$\pm$6.40 km~s$^{-1}$)\\
    
     &   &  \\

    F33 & Sinfoni & 14/04/11 (-17.07$\pm$12.70 km~s$^{-1}$), 17/07/13 (-10.13$\pm$8.76 km~s$^{-1}$), 06/08/13 (-9.45$\pm$7.39 km~s$^{-1}$), \\
        &         & 26/04/17 (5.78$\pm$7.06 km~s$^{-1}$), 30/04/17 (13.18$\pm$6.56 km~s$^{-1}$), 07/05/18 (-15.09$\pm$8.57 km~s$^{-1}$), \\
        &         & 07/08/18 (4.80$\pm$9.40 km~s$^{-1}$), 12/08/18 (6.61$\pm$7.39 km~s$^{-1}$), 17/08/18 (0.57$\pm$8.34 km~s$^{-1}$), \\
        &         & 18/08/18 (20.80$\pm$7.59 km~s$^{-1}$) \\
    
& & \\  
\hline
\end{tabular}
\end{table*}

\begin{table*}
 \contcaption{Dates of epochs of observation for Arches sample and associated RV measurements}
  \begin{tabular}{ccl}
 \hline
 Star & Instrument & Epochs of observations (and associated RV measurements) \\
 \hline
    F34 & Sinfoni & 14/04/11 (-11.79$\pm$6.68 km~s$^{-1}$), 17/07/13 (2.46$\pm$5.90 km~s$^{-1}$), 06/08/13 (-17.65$\pm$5.79 km~s$^{-1}$), \\
        &         & 26/04/17 (3.30$\pm$4.74 km~s$^{-1}$), 30/04/17 (3.44$\pm$6.79 km~s$^{-1}$), 07/05/18 (-2.51$\pm$6.28 km~s$^{-1}$), \\
        &         & 07/08/18 (5.68$\pm$8.82 km~s$^{-1}$), 12/08/18 (1.54$\pm$6.28 km~s$^{-1}$), 17/08/18 (1.40$\pm$5.21 km~s$^{-1}$), \\
        &         &  18/08/18 (14.12$\pm$7.93 km~s$^{-1}$)\\
    
    &   &    \\

    F35 & Sinfoni & 29/05/13 (-9.87$\pm$13.87 km~s$^{-1}$), 18/06/13 (-85.16$\pm$14.84 km~s$^{-1}$), 17/07/13 (-76.45$\pm$14.20 km~s$^{-1}$), \\
        &         & 03/08/13 (142.45$\pm$13.46 km~s$^{-1}$), 04/08/13 (-55.16$\pm$11.97 km~s$^{-1}$), 06/08/13 (-60.85$\pm$11.18 km~s$^{-1}$), \\
        &         & 08/08/13 (55.56$\pm$14.31 km~s$^{-1}$), 26/04/17 (-60.49$\pm$12.60 km~s$^{-1}$), 30/04/17 (8.06$\pm$12.00 km~s$^{-1}$), \\
        &         & 23/06/17 (134.17$\pm$12.05 km~s$^{-1}$), 27/07/17 (-41.69$\pm$19.00 km~s$^{-1}$), 07/05/18 (4.08$\pm$10.95 km~s$^{-1}$), \\
        &         & 07/08/18 (97.23$\pm$13.10 km~s$^{-1}$), 12/08/18 (14.03$\pm$11.51 km~s$^{-1}$), 17/08/18 (-73.13$\pm$10.93 km~s$^{-1}$), \\
        &         & 18/08/18 (7.22$\pm$14.25 km~s$^{-1}$)\\
    
         &  &   \\

    F38 & Sinfoni & 14/04/11 (-3.69$\pm$5.60 km~s$^{-1}$), 17/07/13 (-14.65$\pm$6.12 km~s$^{-1}$), 06/08/13 (-14.28$\pm$5.89 km~s$^{-1}$), \\
        &         & 26/04/17 (2.17$\pm$5.99 km~s$^{-1}$), 30/04/17 (-7.13$\pm$5.11 km~s$^{-1}$), 07/05/18 (5.92$\pm$5.13 km~s$^{-1}$), \\
        &         & 07/08/18 (6.23$\pm$5.40 km~s$^{-1}$), 12/08/18 (10.94$\pm$5.99 km~s$^{-1}$), 17/08/18 (low S/N), \\
        &         & 18/08/18 (14.49$\pm$6.09 km~s$^{-1}$) \\
    
        &   &  \\

    F40 & Sinfoni & 29/05/13 (-9.57$\pm$9.51 km~s$^{-1}$), 18/06/13 (-3.30$\pm$18.00 km~s$^{-1}$), 17/07/13 (-20.06$\pm$11.19 km~s$^{-1}$), \\
        &         & 03/08/13 (-2.71$\pm$10.88 km~s$^{-1}$), 04/08/13 (-0.87$\pm$10.08 km~s$^{-1}$), 06/08/13 (0.90$\pm$11.02 km~s$^{-1}$), \\
        &         & 08/08/13 (-18.25$\pm$14.85 km~s$^{-1}$), 26/04/17 (18.17$\pm$11.69 km~s$^{-1}$), 30/04/17 (19.44$\pm$9.69 km~s$^{-1}$), \\
        &         & 23/06/17 (-28.02$\pm$10.18 km~s$^{-1}$), 27/07/17 (19.52$\pm$9.07 km~s$^{-1}$), 07/05/18 (-5.06$\pm$10.54 km~s$^{-1}$), \\
        &         & 07/08/18 (-4.57$\pm$14.84 km~s$^{-1}$), 12/08/18 (8.52$\pm$12.17 km~s$^{-1}$), 18/08/18 (25.86$\pm$11.06 km~s$^{-1}$) \\
& & \\  
\hline
\end{tabular}
\end{table*}

\section{The LBV phenomenon and secular variability within the Arches} \label{AppB}

As massive stars transition away  from the Main Sequence they appear to encounter an instability of unknown origin: the so-called luminous 
blue variable (LBV) phase. Historically it has been thought  that the extreme mass loss that characterises LBVs helps  facilitate the formation of H-depleted WRs, although observations of Romano's star (Clark et al. \cite{clark12}, Polcaro et al. \cite{polcaro}) and the erupting component of the eclipsing hierarchical system HD5980 indicate this behaviour may also be present in  WN-subtype WRs.
 The LBV phenomenon appears to extend to very high masses, with the LBV component of  HD5980 
having a dynamical mass of  M$_{\rm current}\sim60M_{\odot}$, implying  M$_{\rm initial}>80M_{\odot}$ (Koenigsberger et al. \cite{K}; Hillier et al. \cite{hillier19}). 
However, this extreme is exceeded by that of the primary in the binary  $\eta$ Carinae, which is generally supposed to be $>100M_{\odot}$ (Hillier et al.
\cite{hillier01}),  with a mass of $\sim40-60M_{\odot}$ suggested for the unseen secondary (Mehner et al. \cite{mehner}).  $\eta$ Carinae has been the subject of intensive observation due to both its current exceptional properties and its unprecedented eruption in the $19^{\rm th}$ Century, during which a luminosity of $\gtrsim10^7L_{\odot}$ and a mass loss rate of $\gtrsim0.5M_{\odot}$ yr$^{-1}$ have been inferred (Smith et al. \cite{smith03}). 

$\eta$ Carinae is located within the Trumpler 16 + Collinder 228 cluster aggregate, for which Smith (\cite{smith06}) infers an age of 2-3Myr. As such -  although of significantly lower integrated mass - it appears directly comparable to the Arches; indeed it hosts three WNLha stars, a single O4 supergiant and a number of mid-O giants and main sequence stars (cf. papers I and III).  Given this similarity one might ask whether the Arches hosts any star exhibiting the LBV phenomenon? Trivially, despite the presence of similarly  massive binaries within the Arches, there appears to be  no direct analogue of $\eta$ Carinae; its K-band morphology is very different from any cluster member (Hamann et al. \cite{hamann}), while any nebula comparable to the Homunculus would be clearly visible (cf. the Pistol nebula; Figer et al. \cite{figer99}). 

Surprisingly,  none of the stars within our sample showed any evidence of long term changes in their  spectral morphologies during this period. 
 Simulations of $60M_{\odot}$ stars by Groh et al. (\cite{groh}) predict  that the LBV phenomenon occurs directly after a late-O/early-B hypergiant phase around $\gtrsim3$Myr, implying  that the cohort of mid-O supergiants and, potentially, the mid- to late-hypergiants have yet to reach this evolutionary juncture. Unfortunately no predictions exist for when, or even if, more massive stars such as the Arches WNLha cohort  become LBVs, nor how long this episode endures. Given the limited cadence of the observations, one could imagine that if the outbursts/eruption duty cycle were long, or outbursts rapid, they could have been missed. Indeed, while $\eta$ Carinae was highly (photometrically) active throughout the 19$^{\rm th}$ and 20$^{\rm th}$ Centuries (Smith \& Frew \cite{smith11}) and the characteristic variability of LBVs occurs over months to years (e.g. Humphreys \& Davidson \cite{HD}), observations of  P Cygni reveal that LBVs may reside in a quiescent state for centuries (cf Lamers \& de Groot \cite{lamers}). Alternatively one might suppose that WNLha stars have yet to reach the LBV phase,  or that such stars circumvent it entirely. 
However such a hypothesis would require the 19$^{\rm th}$ Century eruption of $\eta$ Carina to be driven by a mechanism other than the nuclear evolution of the star, such as binary interaction/merger (Portegies Zwart \& van den Heuvel \cite{PZ}).

\section{The X-ray and binary properties of  WNLh stars}

In order to better understand  the Arches binaries it is instructive to contrast their properties with those of 
other known binary  populations. Young massive clusters provide the perfect laboratory to undertake such a comparison,
 since their ensemble properties allow for constraints to be placed on their distances, extinction  and ages, in turn aiding
 the determination of properties such as luminosity for  individual members. Moreover, selection effects may be more 
easily understood and controlled in such an environment. When assessing the binary fraction of clusters, 
multi-epoch spectroscopic surveys clearly represent the most powerful 
methodology available, especially if combined with photometric observations. However not all the clusters that host
WNLha stars have been subject to such an approach, and in such cases X-ray observations - 
which are sensitive to the signature of shocked material generated by wind collisions -
are an excellent substitute. 

In order to undertake such an assessment, the X-ray properties of both single and binary WNLh stars must first be quantified.
 Clark et al. (\cite{clark09b}) presented such a summary; however a reappraisal is timely,  and we provide that here. 
For the reasons described above, stars located within stellar aggregates form a `gold standard'; moreover  many more such regions have now been subject
to targeted X-ray  investigation (Sect. C.1.1 and refs. therein). Nevertheless, we supplement the resultant sample with the remaining, 
isolated examples for which suitable data exist, an inclusion driven by the possibility of better understanding the intrinsic
 multi-wavelength properties of single stars (Sect. C.1.2). For the avoidance of bias, we emphasise that the isolated stellar cohort  are not included in the
subsequent determination of binary fractions. We also highlight that a large number of WNLha stars have been identified within the the central molecular
zone of the Galaxy (Clark et al. \cite{clark21}). Unlike the cohorts listed above these were in large part identified via X-ray 
observations; as a consequence of the obvious resultant  bias towards detecting colliding wind binaries we explicitly exclude them from 
the analysis presented here.

\subsection{Data collation and presentation} \label{app:Xray-data}

In the following section we present the results of a literature review undertaken to determine both the X-ray and binary properties of WNLha stars, encompassing both
isolated galactic examples and the larger population found within star clusters and associations. In order to produce the mostly complete dataset possible, so as to interpret their X-ray behaviour, we also present the preliminary results of tailored model-atmosphere analysis (cf. Sect. 4) of the X-ray bright WNLha stars found within the Galactic clusters Danks 1 and Mercer 81 
(Najarro et al. in prep.).

\subsubsection{WNLh stars within stellar aggregates} 

Given their rich stellar populations  the  young massive cluster R136 and its host star forming region 30 Doradus are obvious targets for 
such an approach. Unfortunately, their location within the  Large Magellanic Cloud leads to significant blending in the crowded core regions 
of R136, and limits spectral analysis to the brightest of the X-ray sources. As a consequence we also review the contents of Galactic clusters,
following the compilation of  Rate et al. (\cite{rate}). From an initial sample of 19 candidate clusters, a number were rejected due to spectral 
re-classification of relevant cluster members on the basis of published data\footnote{ Cl 1813-178 (Messineo et al. \cite{messineo11}) and 
VVV CL009 (Chen\'{e} et al. \cite{chene13}).} and/or the cluster population suggesting an age at which, by analogy with the Quintuplet and Westerlund 1 (Clark et al. \cite{clark18b},
\cite{clark20a}), core-H burning WNLha stars should  not be present\footnote{The Quartet (Messineo et al. \cite{messineo09}),
VVV CL036, VVV CL074, and  VVV CL099 (Chen\'{e} et al. \cite{chene13}).}. Four  further clusters host WNLha stars but lack the multi-epoch spectroscopic, X-ray and radio observations 
required to identify binaries\footnote{ [DBS2003]-179 (WR84-1, WR84-6, and WR84-7; Borissova et al. \cite{borissova}),  
VVV CL041 (WR62-2; Chen\'{e} et al. \cite{chene15}), 
VVV CL073 (WR75-26; Chen\'{e} et al. \cite{chene13}), and  
Mercer 23 (WR125-3; Hanson et al. \cite{hanson10})}. This left ten clusters hosting WNLha stars for which the requisite observational data were available,  in addition to R136 and 30 Dor\footnote{While 13 WNLha stars, including the binary CXOU J1745-28,  have currently been identified within the Central Molecular Zone of the Galaxy (Mikles et al. \cite{mikles06}, Clark et al. \cite{clark09b}) we do not consider them here since a large number have been selected on the basis of their X-ray properties, potentially biasing the sample towards colliding wind binaries.}.
The binary and X-ray properties of the WNLh (and related) stars in these clusters are summarised in Tables C.1 and C.2, while the thumbnail sketches below present a synopsis of
 the physical properties in each aggregate, including relevant references and conclusions derived from the reported observations.
 
\begin{table*}
{\small
\caption{The X-ray properties of WNha stars in 30 Doradus and galactic clusters}
\label{tab:appC1}
\begin{center}
\begin{tabular}{l l l c c c c c}
\hline
\hline 
Cluster & Dist. & ID & log   & $L_{\rm x}$         & $kT_1$ & $kT_2$  & log \\ 
        & (kpc) &    & $(L_{\rm bol}/L_{\odot})$   &($10^{32}$ergs$^{-1}$) & (keV) & (keV) & $(L_{\rm X}/L_{\rm bol})$ \\
\hline
 & & & & & &  & \\
R136        & $\sim50$ & R136a1 & 6.94$^a$ & 100 & -  & 1.2 & -\\
            &          &    +a2 & 6.63$^a$ &  &   &  & \\
            &          & R136a3 & 6.58$^a$ & 70 & - &4.2 & - \\
            &          &  +a6   & 6.52$^a$ & & & & \\
 & & R136c$^*$ & 6.75$^a$ & 700 & - & 3.0 & -5.5 \\

30 Dor & $\sim50$ & Mk30$^*$ & 6.16$^a$ & $<10$ & - & - & $<-6.8$\\
 & & Mk34$^*$ & 6.70$^b$ & {\em 1000} & 1.2 & 4.5 & -5.3  \\ 
 & & Mk35 & 6.3$^a$ &  10 & - & - & -6.9 \\
 & & Mk37$^*$ & 6.48$^a$ & 100 & - & - & -6.1 \\
 & & Mk37Wb & 6.21$^a$ & $<10$ & - & - & $<-6.8$ \\
 & & Mk39 & 6.4$^a$ & 130 & - & 2.0 & -5.9 \\
 & & Mk51 & 6.2$^a$ & $<10$ & - & - &$<-6.8$ \\
 & & RMC134 &6.2$^a$ &  7 & - & - & -7.0 \\
 & & RMC135$^*$ & 6.2$^a$ & $<6.3$ & - & - & $<-7.0$ \\
 & & RMC140b$^*$ & 6.4$^a$ & 30 & - & 2.5 & -6.5 \\

 & & RMC144$^*$ & 6.72$^c$ & {\em 25} & - & 2.1 & -6.9 \\

 & & RMC145$^*$ & 6.5$^a$ &19 & - & 1.6 &  $-6.8$ \\
 & & RMC146     & 6.29$^a$ & $<10$ & - & - & $<-6.9$ \\
 & & RMC147     & 6.36$^a$ & $<10$ & - & - & $<-7.0$ \\
 & & VFTS682    & 6.51$^a$ & $<10$ & - & - & $<-7.1$ \\    
Arches & $\sim8$ & F6$^*$ & 6.32$^d$ & 110 & - & 2.2 & -5.9 \\
      &         & F7$^*$ & 6.27$^d$ & 72  & - & 1.8 & -6.0 \\
      &         & F9 & 6.15$^e$ & 46  & - & 2.5 & -6.1 \\    
      &  &  &  &  & & \\
Bochum 7 & $\sim$6.0 & WR12$^*$  & 5.98$^f$ & $<0.4$ & - & - & $<-8.0$ \\
Westerlund 2 & $\sim4.2$ & WR20a$^{*}$ & 6.3$^g$ & {\em 57} & 0.4 & 1.6 & -6.1 \\
             &           & WR20b & - & 120   & 0.4 & 5.4 & - \\      
& &  [WR20aa & - & 24 & 0.9 & - & -]\\             
& & [WR21a* & - & {\em 194} & 0.8  & 3.0 & -] \\
Carina  &  $\sim2.6$     & WR22$^{*}$ & 6.35$^f$  & {\em 4.2} & 0.6 & 2.0 & -7.3 \\
        &                & WR24 &  6.19$^f$   & 6.4    & 0.7 & 3.3 & -7.0 \\
        &                & WR25$^{*}$ & 6.61$^f$   & {\em 84} & 0.6 & 2.7 & -6.3 \\
NGC 3603 & $\sim7.6$ & WR43A$^{*}$  & 6.39$^h$ &  360 & 0.8  & 3.1 & - \\
         &           & +WR43B   & 6.46$^h$  &  & & & \\
         &           & WR43C$^{*}$ & 6.35$^h$ & {\em 260} & 1.0  & 7.9 & -5.5 \\
         &           & [WR42e     &  - &  2.3  &  -   & - & - ] \\
         &           & [WR43-2$^*$ &  -   & 2.9 & - & - & - ] \\
Danks 1  & $\sim3.8$   & WR48-8 & 5.44$^e$ & {\em 17} & - & - & -5.8 \\
         &           & WR48-9 & 5.99$^e$ & {\em 43} & - & - & -6.0 \\
         &           & WR48-10 & 5.70$^e$ & 14   & - & -  & -6.2 \\
Mercer 81 & $\sim10$ & WR76-3 & 6.17$^e$ &14 & -& 2.0 & -6.6 \\
         &     & WR76-7 & 6.39$^e$ & 35 &-& 2.6 & -6.4 \\
Sco OB1  & $\sim$2 & WR78$^*$  & 5.8$^f$ &   6.9 & 0.6 & 2.3 & -6.6 \\    
         &         & WR79a$^*$ & 5.78$^i$  & 1.4 & 0.6 & 2.7 & -7.2 \\ 
HM-1     & $\sim3.3$ & WR87 & 6.21$^f$ & 1.6 & - & 2 & -7.6 \\
                &           & WR89$^*$ & 6.33$^f$ &{\em 154} & 0.6 & 2 & -5.7 \\
W43     &  $\sim6$   & WR121a$^*$  & - & {\em 154} & 1.0 & 3.6 & -\\
\hline
\end{tabular}
\end{center}
Sources in the vicinity of the clusters but for which a physical association
has yet to be proved are given in square brackets.  
Identifiers marked with an asterisk are those which are identified
as binaries via optical or IR observations, non-thermal radio emission (WR78 and WR89) or periodically modulated X-ray activity (WR121a).   Variable X-ray sources have their 
representative fluxes given in italics. Parameters for R136a1+a2, R136a3+a6 and WR43A+WR43B represent a
blend of both sources (see text for further details). References for the X-ray properties are given in the body of the text, while those for the bolometric luminosities of the system primaries are $^a$Crowther et al. (\cite{crowther16}), $^b$Tehrani et al. (\cite{tehrani}), $^c$Shenar et al. (\cite{shenar21}), $^d$this work, $^e$Najarro et al. (in prep.), $^f$Hamann et al. (\cite{hamann19}), $^g$Rauw et al. (\cite{rauw05}), $^h$Crowther et al. (\cite{crowther10}), and $^i$Crowther \& Bohannan (\cite{CB}). We highlight that Mk34, RMC144, and WR20a are composed of two `twins' - as with other systems the bolometric luminosities presented are those of the combined binary; individual  components in these systems will be 0.3dex fainter.
Finally we have scaled the bolometric luminosities for WR24, WR24, and WR25 (Hamann et al. (\cite{hamann19}) to a common distance of 2.6kpc for the Carina complex.
}
\end{table*}

\begin{table}
\caption{The X-ray properties of isolated WNha stars}
\label{tab:appC2}
\begin{center}
\begin{tabular}{l c c c c}
\hline
\hline 
ID  & Dist.  & log                         & $L_{\rm x}$           & log \\ 
    & (kpc)  & $(L_{\rm bol}/L_{\odot})$   &($10^{32}$ergs$^{-1}$) & $(L_{\rm X}/L_{\rm bol})$ \\
\hline
WR16 & 2.6 & 5.72$^a$ & 0.13 & -8.2  \\
WR29$^b*$ & 6.0 & 5.73$^b$ & 2 - 3 & -7.1 - -6.9 \\
WR40 & 4.0& 5.91$^a$ & $<0.7$ &  $<-7.7$  \\
WR105$^*$ & 1.8 & 5.89$^a$ & 4 - 12 & -6.9 - -6.4 \\
WR148$^*$ & 8.3 & 6.3$^{a,c}$ & 2.1 & -7.6 \\
\hline
\end{tabular}
\end{center}
Identifiers marked with an asterisk are those which are identified
as binaries via optical (WR29 and WR148) and radio (WR105) observations. References for the X-ray properties are given in the body of the text, while those for
the bolometric luminosities of the system primaries are $^a$Hamann et al. (\cite{hamann19}), $^b$Naze et al. (\cite{naze20}) and $^c$Zhekov (\cite{zhekov}). Note the X-ray luminosities listed for WR29 and WR105 represent the range allowed under the modelling assumptions of Naze et al. (\cite{naze20}).
\end{table}

\begin{itemize}

\item {\bf R136:} Schnurr et al. (\cite{schnurr09b}) undertook  the most comprehensive RV survey of the six WN5h stars within R136 to date, 
 but report only a single potential binary - R136c  - with a tentative $\sim8.4$d period (noting the survey was only sensitive to periods $\lesssim40$d).
X-ray observations reveal that it is one of the brightest objects within the LMC and demonstrates a hard spectrum (Table~\ref{tab:appC1}); strongly 
suggestive of a colliding wind binary (see below). Of the remaining WN5h stars within R136, Townsley et al. (\cite{townsley06}) report  a blend of emission from R136a1+a2 and
 R136a3+a6 (the  latter being O2 If; Crowther et al. \cite{crowther16}). Both composite sources are luminous and hard, noting that 
only single temperature models were employed due to data quality (Table~\ref{tab:appC1}). Finally,  R136b is heavily blended with other core stars and so is not discussed further, while no mention is made of R136a5 by these authors.

\item {\bf 30 Doradus:} A further 15 WNLh and related stars  are found within the wider confines of 30 Dor (Crowther et al. \cite{crowther16}). Of these, seven have been confirmed as  binaries via spectroscopic observations\footnote{Mk30 (O2 If/WN5), Mk34 (WN6(h)), Mk37 (O3.5 If/WN7),  RMC135 (WN6h),  RMC140b (WN5h), RMC144 (WN5-6h), and RMC145 (WN6h).}; one is an RV variable but lacks an orbital solution\footnote{Mk35 (O2 If/WN5).}; three appear to lack RV variability\footnote{Mk51 (O3.5 If/WN7), RMC134 (WN6(h)), and VFTS682 (WN5h).}; and appropriate datasets are lacking for four\footnote{Mk37Wb (O2 If/WN5), Mk39 (02.5 If/WN6), RMC146 (WN5ha), and RMC147 (WN5h).} (Table~\ref{pubin}, Schnurr et al. \cite{schnurr08b}, Bestenlehner et al. \cite{best}, Tehrani et al. \cite{tehrani}). X-ray data have been presented for 
a number of these stars, resulting in eight detections. Of these six arise from confirmed binaries/RV variables (Table~\ref{tab:appC1}). The  most comprehensive analyses are of Mk34 and RMC144. Pollock et al. (\cite{pollock18}) show that the former is  an extremely luminous source (median  $L_{\rm X} \sim10^{35}$ergs$^{-1}$) that is modulated on the orbital period, with a spectrum  comprising soft and hard components. Despite having similarly luminous components RMC demonstrates a much more modest X-ray flux, although it too displays orbital modulation and an intrinsically hard spectrum   (Townsley et al. \cite{townsley06},
Shenar et al. \cite{shenar21}). Two further stars - RMC140b and RMC145 - are also luminous and  hard sources 
($L_{\rm X}\sim10^{33}-10^{34}$ergs$^{-1}$; $kT\gtrsim1.6$keV;  noting that only single temperature models were employed by Townsley et al. \cite{townsley06}).
 Mk37, the remaining binary detection, and the RV variable  Mk35  are similarly luminous, while RMC134 - which appears not to be RV variable - is comparatively faint; unfortunately no  spectral information is available for any of these three stars (Table~\ref{tab:appC1}; Schnurr et al. (\cite{schnurr08b}). Finally, Mk39 has yet to be subject to RV monitoring, but has X-ray properties directly  comparable to confirmed binaries (Table~\ref{tab:appC1}; Townsley et al. \cite{townsley06}); 
Schnurr et al. (\cite{schnurr08b}) returned individual upper limits for the last two binaries (Mk30 and RMC135; Table~\ref{tab:appC1}). This leaves a total of five stars for which we adopt the generic detection  threshold of $10^{33}$ergs$^{-1}$ reported by Townsley et al. (\cite{townsley06}) as upper limits, noting that in practice deviations from this are expected depending on the local environment of individual stars.

\item {\bf Bochum 7:} Bo7 hosts a single WN8h star, WR12, for which Fahed \& Moffat (\cite{fahed}) present a spectroscopic orbital solution.  Pollock et al. (\cite{pollock95}) report an X-ray  detection, albeit of low detection significance ($(10\pm5)\times10^{32}$ergs$^{-1}$ for a distance of 6kpc, subsequently quoted by Ignace et al. \cite{ignace}). However more recent and sensitive  observations by Naz\'{e} et al. (\cite{naze20}) failed to detect it (Table~\ref{tab:appC1}), the authors noting that it was observed at a phase at which the  WNLha primary and its dense stellar wind  would have (partially) obscured the secondary and any  putative colliding wind region (see also 
WR22 below). Clearly future observations will be required to ascertain the nature of its (variable?) emission although we adopt the upper limit of 
Naz\'{e} et al. (\cite{naze20}) in this work. 

\item {\bf Westerlund 2:} Two WNLha stars have traditionally been associated with Wd2; WR20a (WN6ha+WN6ha) and WR20b (WN6h). The former is a well known massive eclipsing system (Rauw et al. \cite{rauw05}; Table~\ref{pubin}) and is also a luminous, hard  X-ray source that varies on the orbital period (Naz\'{e} et al. \cite{naze08}); in Table~\ref{tab:appC1} we list the properties for the brightest of the epochs 
presented by the authors. They further demonstrate that WR20b is a fainter, harder source (representative $L_{\rm X}\sim2\times10^{33}$ergs$^{-1}$, $kT\sim3.6$keV) although Skinner et al. (\cite{skinner}) prefer a two-temperature fit resulting in a higher luminosity (see Table~\ref{tab:appC1}). As a consequence they propose that WR20b is also a colliding wind system;
however spectral monitoring by Rauw et al. (\cite{rauw11}) apparently excludes
short binary periods ($<50$d - subject to the usual {\em caveats} regarding unfavourable orbital inclination).
Three further WNLha stars are found in the periphery of Wd2, for which a physical association is currently uncertain: WR20aa (O2If$^*$/WN6), WR20c (O2IF$^*$/WN6ha), and WR21a (O3If$^*$/WN6ha+O3V). 
The very massive short period binary WR21a  (Table~\ref{pubin}) is, as may be anticipated,  a highly luminous, hard and periodically variable X-ray source (Gosset \& Naze \cite{gosset16}). WR20aa and WR20c are possible runaways originating from  Wd2 (Roman-Lopes et al. \cite{RL11}).  Townsley et al. (\cite{townsley}) report that the former appears to be a moderately bright and soft X-ray source (Table~\ref{tab:appC1}), while no X-ray data are available for the latter. In neither case has RV spectral monitoring been attempted.

\item {\bf Carina:} The Carina star-forming complex, encompassing the young massive clusters Trumpler 16 and Collinder 229, is host to three WNLha stars - 
 WR22 (=HD92740A; WN7+abs +09), WR24 (=HD93131; WN6ha), and WR25 (=HD93162; O2.5 If$^*$/WN6 + O). X-ray observations have been made of all three systems (Gosset et al. \cite{gosset}, Skinner et al. \cite{skinner}, Raassen et al. \cite{raassen}, Pollock \& Corcoran \cite{pollock06}) and are summarised in Table~\ref{tab:appC1}. 
WR22 and WR25 have already been confirmed as binaries via RV studies, and both sources also appear to show orbitally modulated X-ray fluxes  (Gosset et al. \cite{gosset}, Arora et al. \cite{arora}). WR22 is of particularly interest since it is observed to be particularly faint at the point at which the WNLha and associated wind might be expected to eclipse both the secondary and any associated wind interaction zone ($L_{\rm X}\sim6.3\times10^{31}$erg$^{-1}$).
Finally, while spectroscopic evidence for binarity  is lacking for WR24, we highlight the close similarity of its X-ray properties to those of WR22.

\item {\bf NGC 3603:} Three WNLha stars are found within the core of NGC3603 - WR43A (= cl* NGC 3603 BLW A1; WN6ha+WN6ha), WR43B (= cl* NGC 3603 MDS B; WN6ha) 
and WR43C (= cl* NGC3603 MDS C; WN6ha + ?), of 
which WR43A and WR43C are known short period binaries (Table~\ref{pubin}). X-ray observations are complicated by blending due to the compact nature of NGC3603 but Moffat et al. 
(\cite{moffat02}; their Fig. 4) suggest fluxes of $L_{\rm x} \sim 10^{33}$ergs$^{-1}$ for WR43B, $L_{\rm x} \gtrsim 4\times10^{34}$ergs$^{-1}$ for WR43C and
$L_{\rm x} \gtrsim 10^{34}$ergs$^{-1}$ for the blend comprising WR43A, cl* NGC3603 BLW A2 (O3 V) and cl* NGC3603 BLW A3 (O3 III(f$^*$)); unfortunately no 
spectral information was presented for any source. Subsequently, Huenemoerder et al. (\cite{huen}) analysed the emission spectra deriving 
from a blend of WR43A+WR43B and from WR43C, finding  both to be highly luminous and demonstrating  hard spectral components (cf. Table~\ref{tab:appC1}). Of these Pollock et al. (\cite{pollock18}) report WR43C as being variable.
Two further WNLha stars are located in the surrounds of NGC3603. Displaced $\sim6$pc from the cluster core, Roman-Lopes (\cite{RL12}) suggests that the O2 If$^*$/WN6h stars WR42e could be a runaway (and hence single); nevertheless they report a  X-ray luminosity that is comparable with the binary WR22 (Table~\ref{tab:appC1}) but no spectral information.  Finally WR43-2 (= cl* NGC3603 MTT 58;  O2-3.5 If$^*$/WN5-6ha + ?)) is 
located within a compact H\,{\sc ii} region on the periphery of the wind-blown bubble encompassing NGC3603, some $\sim$2pc distant from the cluster core. Again lacking spectral information, it is of comparable luminosity to WR42e (Roman-Lopes \cite{RL13}), but is unambiguously a short-period binary on the basis of periodic 
photometric modulation (Jaque Arancibia et al. \cite{JA}).  

\item {\bf Mercer 30:} de la Fuente et al. (\cite{dlF}) report two WNLh stars within the cluster Mercer 30  - WN46-5 (= [DNB2016] Mc30-7; WN6h) and 
WN46-6 (= [DNB2016] Mc30-8; WN7h), which modelling suggests are highly luminous, and hence massive (log$L/L_{\odot}\gtrsim6.0$ and
 $M_{\rm current} \gtrsim 50M_{\odot}$). Despite a limited dataset comprising three epochs of spectroscopic observations, the former is clearly a binary 
by virtue of $\Delta$RV$\sim191$km s$^{-1}$ between observations in 2009 and 2011; unfortunately no X-ray observations exist for this cluster.

\item {\bf Mercer 81:}  Davies et al. (\cite{davies12a}) and de la Fuente et al. (\cite{dlF13}) report five WN7-8ha stars within the distant and highly reddened cluster Mercer 81; 
  WR76-9 (=[DDN2012] 3), WN76-6 (=[DDN2012] 7), WR76-2 (=[DDN2012] 5), WR76-7 (=[DDN2012] 2), and 
WR76-3 (=[DDN2012] 8). No radial velocity surveys have been undertaken but the X-ray survey of Norma Arm by
Rahoui et al. (\cite{rahoui}) report detections  WR76-3 and WR76-7 (their sources \#1278 and \#1279 respectively), finding them to be hard, bright sources (Table~\ref{tab:appC1}; assuming a single temperature fit). Bolometric luminosities for the two X-ray detections derive from quantitative modelling to be presented in 
Najarro et al. (in prep.); we find both stars to be intrinsically highly luminous (Table~\ref{tab:appC1}).

\item {\bf Danks 1:} In the absence of RV observations,  Zhekov et al. (\cite{zhekov}) presents a preliminary analysis of  X-ray data, quoting fluxes for the three WNLha stars WR48-8 
(= [DCT2012] D1-5; WN8-9ha), WR48-9 (= [DCT2012] D1-1; WN8-9ha), and WR48-10 (= [DCT2012] D1-2; WN8-9ha), with Townsley et al. (\cite{townsley}) subsequently reporting that the first two sources are variable, with fluxes decreasing by factors of $\sim5$ and $\sim10$ respectively, between two epochs of observations separated by $\sim4$yr. While neither author presents spectral information for these detections both conclude that they are bona fide binaries based on their X-ray properties (a claim we revisit below). As with Mercer 81 we provide bolometric luminosities for the three X-ray detections derived from modelling in Table~\ref{tab:appC1} (Najarro et al. in prep.). We highlight that while WR48-9 and 48-10 are broadly comparable to other Galactic examples, WR48-8 is significantly fainter, while the emission lines in the K-band spectrum appear anomalously broad and weak (Davies et al. \cite{davies12b}). As a consequence we suspect its nature and/or evolutionary pathway differs from other examples; we return to this topic in Najarro et al. (in prep.).

\item {\bf Havlen Moffat 1;} As with Danks 1,  HM-1 as a whole has yet to be subject to a systematic,  modern  RV survey, and so we are forced to fall back on observations at other wavelengths; WR89 (WN7h) appears to be a non-thermal radio source and hence colliding wind binary (Cappa et al. \cite{cappa}), although 
inspection of archival spectra fails to reveal RV shifts (Naz\'{e} et al. \cite{naze13})\footnote{The lack of spectral variability is potentially explicable by unfavourable inclination and/or a  long or highly eccentric orbit.}.  
Nevertheless, these authors reveal it to have an extreme  (variable) X-ray luminosity and a spectrum comprising both soft and hard thermal components. Unfortunately, only X-ray observations are available to determine the nature of WR87 (WN7h); while it is comparatively faint with respect to WR79a, it too has a hard spectral component at least consistent with emission from a wind collision zone (Table~\ref{tab:appC1} and discussion below).   

\item {\bf Sco OB1:} Hosting the young massive cluster NGC6321 at its core, the Sco OB1 association contains two WNLh stars - WR 78 (= NGC 6231 305; WN7ha) and WR79a (= NGC6231 327; WN9ha) - that, in contrast to many other examples considered, demonstrate relatively modest bolometric luminosities Table~\ref{tab:appC1}). Both stars are detected at X-ray energies, and share similar properties, being rather faint and  hard sources (Skinner et al. \cite{skinner}, \cite{skinner12}).
WR78 appears to demonstrate aperiodic optical variability (Skinner et al. \cite{skinner12} and refs. therein) while  Chini et al. (\cite{chini}) lists  WR79a as an SB2 system but provides no links to an appropriate reference; as such, we refrain from assuming binarity for either on the basis of these works.
However de Becker \& Ruacq (\cite{deBecker}) and Cappa et al. (\cite{cappa}) report that both stars  are non-thermal radio sources and hence  colliding-wind binaries: a conclusion we adopt for the remainder of this study.

\item{\bf W43:} The giant H\,{\sc ii} region hosts a young, heavily embedded cluster at its core, which in turn contains a single WN7ha star: WR121a. Arora \& Pandey (\cite{arora}) provide an analysis of time-resolved X-ray observations of this star, finding it to be a hard and highly luminous source, which demonstrates an apparent $\sim4.1$d periodicity that they attribute to orbital modulation in 
 a massive compact binary. Examining near-IR and radio data, Luque-Escamilla et al. (\cite{LE}) further show that WR121a maybe resolved into two components separated by a projected angular distance of $\sim0.6$", suggesting the possible presence of a further stellar companion. 

\item {\bf Quintuplet:} Finally we turn to the Quintuplet, which Clark et al. (\cite{clark18b}) report contains two WN8-9ha  (qF256 and qF274) and an hybrid O7-8 Ia$^+$/WN8-9ha star (LHO001). Given the remaining cluster population their presence is a surprise, with the authors suggesting they could be the result of binary interaction. Unfortunately only LHO001 has been the subject of multi-epoch spectroscopy, which reveal it to be highly variable, although insufficient data exists to attempt to identify periodic modulation. Wang et al. (\cite{wang}) present X-ray observations of the Quintuplet; none of these sources are detected and, as with the Arches (Sect. 3.2.1), a combination of potential blending and the presence of  spatial variable  diffuse emission prevent the derivation of upper limits. Given the limited observational data for these stars as well as their uncertain provenance, we refrain from discussing them further.

\end{itemize}

\subsubsection{Isolated WNLh stars}

We may supplement the above dataset with five  isolated single and binary stars. Of the former, Skinner et al. (\cite{skinner12}) report a detection of the WN8h star WR16; unfortunately  it is too faint for spectral analysis (Table~\ref{tab:appC2}, with the flux scaled to the distance adopted by Hamann et al. \cite{hamann19}).
 Although optically variable it currently cannot be unambiguously identified as a  binary
(Antokhin et al. \cite{antokhin}). Similar conclusions are also reported for WR40 (WN8h) by these authors; unfortunately Gosset et al. (\cite{gosset}) are only able to provide a (conservative) upper limit to the X-ray flux which, suitably scaled,
is repeated in Table~\ref{tab:appC2}.
Turning to the binaries,  Naz\'{e} et al. (\cite{naze20}) find WR29 (WN7h+O; Table~\ref{pubin}) to be a  faint X-ray source; assuming a canonical 
$kT\sim0.6+6.9$keV  thermal spectrum it  appears moderately  luminous ($\sim2-3\times10^{32}$ergs$^{-1}$; Table~\ref{tab:appC2}). 
WR105 (WN9h) is a non-thermal radio source (Montes et al. \cite{montes09}) and hence  a strong colliding wind binary candidate. Although a weak X-ray  detection
(Naz\'{e} et al. \cite{naze20}), under the same modelling assumptions adopted for WR29 it appears to be intrinsically luminous  ($\sim4-12\times10^{32}$ergs$^{-1}$; Table~\ref{tab:appC2}).  

Finally we turn to the runaway binary WR148 (WN8h), the nature of which  has been the subject of some debate e.g. Drissen et al. \cite{drissen}. Recently Munoz et al. (\cite{munoz}) have suggested an O5V classification for the secondary; however
the canonical mass of such an object - they quote $\sim37M_{\odot}$ -  is in tension with the inclination derived by Drissen et al. (\cite{drissen}). Upon reappraisal of the relevant  photometric and polarimetric data the former authors suggest
that the results are `inconclusive'. Allowing for a lower inclination, the canonical mass of the O5V secondary implies a current mass of $\sim33M_{\odot}$ for the WN7ha primary - unexpectedly low in comparison to other examples (Table~\ref{pubin}) and for a luminosity of $\sim10^{6.3}L_{\odot}$\footnote{Obtained by scaling the luminosity of Hamann et al. (\cite{hamann19}) to the consensus distance of 8.3kpc (Zhekov \cite{zhekov12}), noting that  Rate \& Crowther (\cite{rate20}) suggest  $\sim9.47^{+1.77}_{-1.49}$kpc.}. Nevertheless, Zhekov (\cite{zhekov12}) report a rather moderate X-ray luminosity ($\sim2.1\times10^{32}$ergs$^{-1}$) and a thermal spectrum ($kT\sim0.4+1.0$keV) for the system.

For completeness  we briefly discuss the Galactic centre source CXO J174536.1-285638 (WN9ha). Although excluded from the sample for the reasons given above, we note that 
it is a long period binary ($\sim189$~d; Mikles et al. \cite{mikles}). Adopting  a two-component spectral fit ($kT\sim0.7 + 4.6$keV) implies an exceptional X-ray luminosity ($L_{\rm X}\sim10^{35}$ergs$^{-1}$) that is directly comparable to that of Mk34 (Table~\ref{tab:appC1}). This indicates that the latter is not a pathological system; however, the lower luminosity of the primary ($L_{\rm bol}\sim10^{6.24}$; Clark et al. \cite{clark09b}) leads to an even more extreme log$(L_{\rm X}/L_{\rm bol})\sim-4.8$ for CXO J174536.1-285638 than was found for Mk34 (Table~\ref{tab:appC1}). 
\subsection{Data interpretation}

\subsubsection{The X-ray properties of WNLh stars}

Historically, and by analogy to O stars,  WR binaries have been expected to be hard, overluminous  X-ray sources. However recent studies suggest that such a picture may be overly simplistic. In particular, while a number of confirmed O star and WR binaries within Wd1 conform to this expectation, others are found to have comparatively soft spectra consistent with emission from single stars (Clark et al. \cite{clark19c}; see also Oskinova \cite{oskinova15}, \cite{oskinova16}). Considering a dataset exclusively comprising  WR stars  allowed Skinner et al. (\cite{skinner}, \cite{skinner12}) to report the ubiquity of a hard ($kT\gtrsim2$keV) component in the X-ray spectra of all WN stars with spectra sufficient to accommodate  modelling. Conversely, Oskinova (\cite{oskinova16}) state that, on average,  single WN stars display a softer ($kT\sim0.4$keV) spectrum. 
Clearly, reconciling these potentially conflicting findings requires an accurate determination of the binary properties of the stars studied. 
Nevertheless, studies focusing  on individual  binary systems via multi-epoch, phase-resolved observations indicate that where present  - and as with O stars -  the  high-temperature spectral component originates in the wind collision zone (e.g. Gosset et al. \cite{gosset}, Montes et al. \cite{montes}, Parkin \& Gosset \cite{parkin}). 

Mindful of the fact that the cohorts considered by Skinner et al. (\cite{skinner12}) and Oskinova (\cite{oskinova15}, \cite{oskinova16})  comprise in part H-free WNE stars - and so are not directly comparable to our sample -  we highlight that the spectral properties of the WNLh stars considered here are uniformly characterised by the presence of a hard thermal component (Table~\ref{tab:appC1}), potentially indicative of binarity.  However, while the possibility remains that single WR stars may also demonstrate similarly hard  X-ray spectra (Skinner et al. \cite{skinner12}) it appears premature to utilise this observational feature to unambiguously identify WR binaries.

 This leaves X-ray luminosity as a prospective binary diagnostic. Oskinova (\cite{oskinova15}) and  Gagn\'{e} et al. (\cite{gagne}) provide compilations of 
 X-ray properties for  single and binary WRs, finding that the luminosities of both show an exceptional variance, spanning
  $\sim$four orders of magnitude. Nevertheless, they identify {\em potential}  trends whereby (i) binaries appear to be more prevalent amongst the 
 brighter X-ray sources, apparently obeying the canonical $L_{\rm X}/L_{\rm bol}\sim10^{-7}$ relationship delineated by hot, massive stars
 and (ii) the most luminous systems  are biased towards the binaries with the longest periods. However, interpretation is complicated  by the heterogenous nature of the samples appraised by these authors, comprising both WNLh  and the more evolved H-free `classical' WNE stars 
(with the latter being of systematically lower luminosity).

Motivated by their evolutionary uniformity we  plot $L_{\rm bol}$ versus $L_{\rm X}$ for our sample of  WNLh  stars in Fig. C.1 (c.f. Tables C.1 and C.2; note that due to uncertainty regarding  individual contributions, the blended R136 and NGC3603 sources are not included). For consistency, in all instances we plot the integrated bolometric luminosity of the stellar system in question (with that of F9 obtained via an identical methodology to that described in Sect. 4; Najarro et al. in prep.). It is expected that for the majority of objects considered this will be equivalent to the output from the WNLh star that dominates the emergent spectrum. However at least three systems - WR20a, MK34 and RMC144 - are known binaries comprising twin WNLh stars; in such cases the luminosity of individual components is $\sim0.3$~dex lower than represented in Fig. C.1.
We find that X-ray luminosities of the stars range over four orders of magnitude: from $\sim10^{31}$ergs$^{-1}$ (for the apparently single WR16) to $\gtrsim10^{35}$ergs$^{-1}$ (the binaries Mk34 and CXO J174536.1-285638), the latter stars being amongst the most luminous colliding-wind systems known (Gagne et al. \cite{gagne}). Despite considerable scatter, a  positive trend  between X-ray and bolometric luminosities is apparent. With 23/39 of the WNLh stars considered exceeding the  empirical  $L_{\rm X}/L_{\rm bol}\sim10^{-7}$ relationship found for O stars, it appears that the former have a  systematically greater X-ray luminosity than the latter. Moreover there are hints that the correlation is also steeper for WNLh stars, with 12/15 examples  with log$(L_{\rm bol}/L_{\odot})>6.3$ exceeding this threshold (Fig. C.1).

Of the remaining stars,  the constant RV sources WR24 and R134  essentially straddle this division\footnote{Based on the results presented in Fig. 4 of  Moffat et al. (\cite{moffat02}), WR43B also lies on this threshold which, based on the values in Table~\ref{tab:appC1} would imply that WR43A significantly exceeds it.}, while the upper limits to the X-ray flux of five stars - the binary Mk30 and  apparently single star Mk51 as well as  MK37Wb, RMC146, and RMC147 (which are currently of indeterminate nature) -  render their placement uncertain. Indeed,  only nine stars are unambiguously found  below 
this  threshold; these comprise the {\em apparently} single stars  WR16, WR40 and VFTS682;  the binaries WR12, WR22, WR79a, WR148 and RMC135;  and WR87, the nature of which is currently uncertain.

Of these the binaries WR12 and WR22 are of particular interest given the stringent upper  limits obtained for their X-ray fluxes ($L_{\rm X}\lesssim 10^{31.8}$ergs$^{-1}$; Fig. C.1). These derive from  observations of both systems obtained at orbital  phases when their WNLha primaries  potentially obscure their companions and  any putative wind collision zones. Following the arguments of Gosset et al. (\cite{gosset}) and Naz\'{e} et al. (\cite{naze20}), one might expect the dense, metal-rich winds of the WNLh stars to efficiently absorb high energy emission from the latter components, leaving only their own intrinsically weak emission visible. Indeed, the  observations of both  WR16 and WR40 support the hypothesis  that single WNLh stars are  particularly faint X-ray sources.

A corollary of this hypothesis would be that all of  the X-ray  luminous WNLh stars are colliding wind binaries. Indeed 16 of the 21 confirmed WNLh binaries\footnote{WR20a, WR25, WR29, WR43C, WR78, WR89, WR105, F6, F7, Mk34, Mk35, Mk37, RMC136, RMC140b, RMC144, and RMC145} are found above the $L_{\rm X}/L_{\rm bol}\sim10^{-7}$  threshold; of the remaining seven co-located stars 
 six currently lack diagnostic observations\footnote{WR48-8, WR48-9, WR48-10, WR76-3, WR76-7, and Mk39} and only Arches F9 shows no evidence for binary-induced RV variability at this time (although we note that its radio properties are indicative of binarity; Sect. 3.2).  Thus we might infer a  binary fraction of 15/16 ($\sim94$\%) amongst stars exceeding this limit at this time (and a minimum of $\sim70$\% if the remainder are found to be single). 

Consequently,  we adopt $L_{\rm X}/L_{\rm bol}\sim10^{-7}$ as a conservative and  empirical lower limit for the identification of colliding wind binaries  with WNLh primaries via their X-ray emission. In doing so we reiterate that five known binaries fall below this threshold; however lowering the limit to encompass all these system at all times would potentially permit the inclusion of genuinely single stars. This is evident upon consideration of the variable X-ray emission from the binaries  WR12 and WR24, which appear to manifest as single stars at certain orbital phases. In addition to their intrinsic variability, the large  variance in the X-ray properties of binaries likely results from  the diversity of their physical  properties - in particular  the nature of the secondary, orbital separation and eccentricity (Gagn\'{e} et al. \cite{gagne} and Table~\ref{pubin}) - which may allow some systems to remain inherently faint.  That the absolute X-ray luminosity of binaries with WNLh primaries  appears greater than that of those containing  O stars  likely reflects the difference in wind properties, with the former supporting more powerful, optically thick winds in comparison to the optically thin outflows of the latter (e.g. Gr\"{a}fener (\cite{grafener21}).

\begin{figure}
\includegraphics[width=\columnwidth, trim=0 5cm 0 5cm, clip]{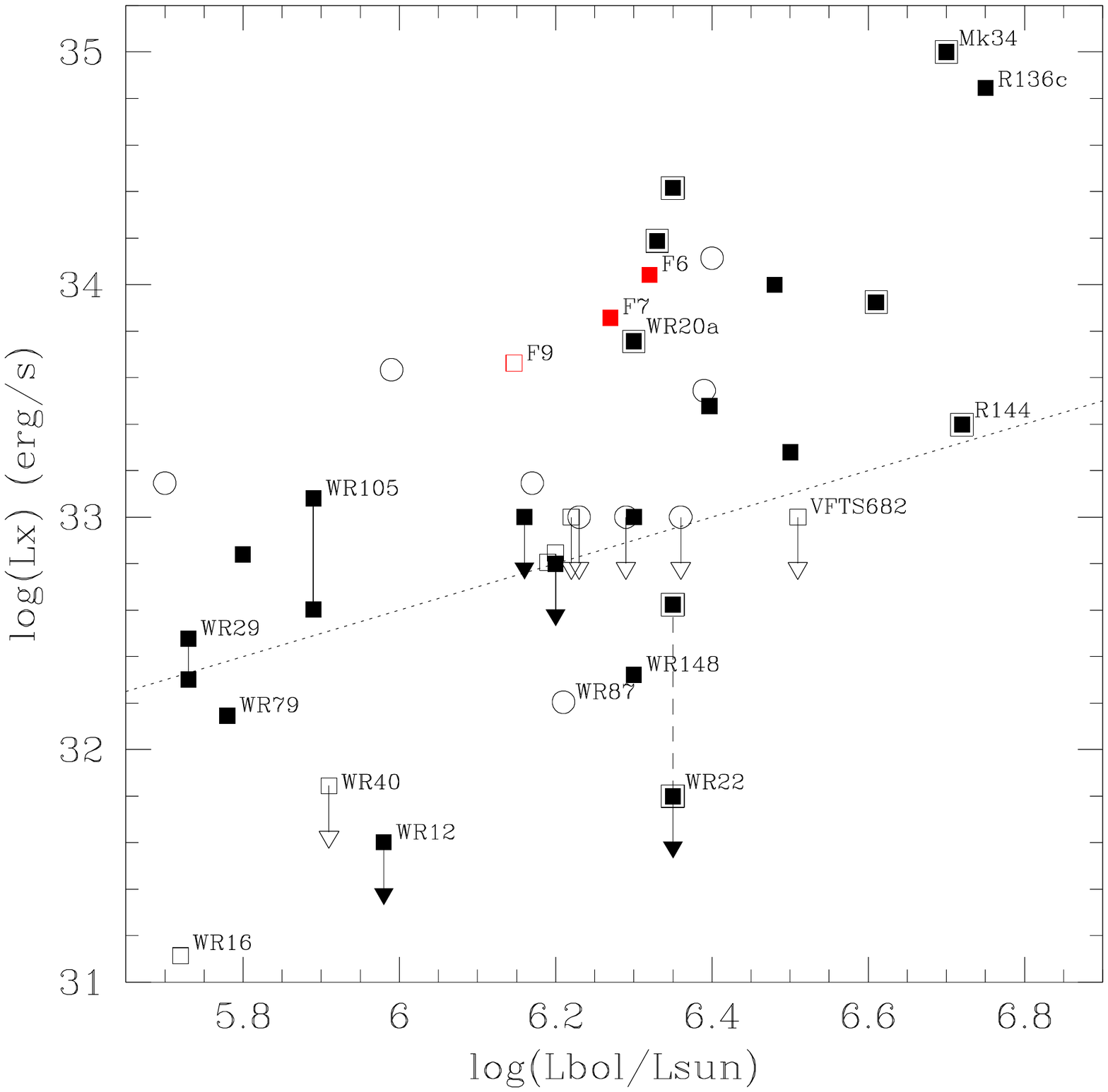}
\caption{Plot of bolometric and X-ray luminosities for the WNLh stars given in Tables C.1 and C.2 (black), with the addition of the Arches detections F6, F7, 
and F9 (red symbols). Confirmed binaries (including the RV variable Mk35 for  simplicity)  are given by filled
squares, stars with no indication of binarity are depicted by open squares, and the subset for which the requisite observations have yet to be undertaken 
represented by open circles. Stars known to be X-ray variable are indicated by the nested symbols. The second,  open symbol for W22 represents the 
upper limit to the intrinsic emission from the WN7h primary according to Gosset et al. (\protect\cite{gosset}). The solid lines joining the two points for WR29 and WR105 indicate the range of possible luminosities for both stars (Naz\'{e} et al. \protect\cite{naze20}).
Luminosities are shifted slightly for RMC140b (-0.003~dex),  RMC144 (+0.003~dex), Mk51 (+0.02~dex) and Mk37Wb (+0.02~dex) to aid clarity. 
The canonical 
$L_{\rm X} = 10^{-7} L_{\rm bol}$ relationship that 
Oskinova (\protect\cite{oskinova15}) suggests is followed by O and WN stars is given by the dotted line;
stars falling below this limit are individually labeled, as are certain  additional stars of interest (subject to figure clarity). Finally the low luminosity star WR48-8 is not plotted, but lies significantly above the threshold.}
\end{figure}

\subsubsection{The binary fraction of WNLh stars within clusters}

How may we interpret these data in terms of the incidence of binarity and the X-ray properties of such systems?
Regarding the frequency of occurrence, as a first step we may simply consider those stars that appear unambiguously binary via spectroscopic determination of an orbital period. Mindful of the limitations of extant RV and X-ray surveys,  we find 9/17 ($\sim53$\%) confirmed binaries within R136/30 Dor (including the RV variable Mk35).
Likewise, recognising  that clusters such as Danks 1, HM-1 and Mercer 81 have yet to be subject to systematic RV surveys (and the limited nature of the Mercer 30 study) we arrive at a total of 7/11 in five galactic aggregates (Table~\ref{tab:appC2}; Bochum 7, Westerlund 2, Carina, NGC3603 and Mercer 30).
To this tally  we may add the periodic X-ray source WR121 (W43) and the non-thermal radio sources WR78 (Sco OB1) and WR89 (HM-1) where emission is though to arise in  shocked material in the wind collision zone, increasing the Galactic count to 10/16.
This implies a minimum binary fraction of 19/33 ($\gtrsim58$\%) before allowance is made for the insensitivity of the extant RV surveys to longer period systems and those seen at unfavourable orbital inclinations.

This brings us to interpretation of the X-ray properties of these stars. Orbitally modulated  X-ray emission is apparent in  the spectroscopically confirmed binaries Mk34, RMC144, WR20a, WR21a, WR22, and  WR25; on this basis the  periodically variable X-ray source WR121a is also a compelling binary candidate (see also CXO 
J1745-28; Table~\ref{pubin}). However, many of the remaining candidate and confirmed WNLh binary systems have an insufficient timeline  of X-ray observations to permit a search for periodicities. For such systems we adopt  $L_{\rm X}/L_{\rm bol}\gtrsim10^{-7}$ 
 as a conservative threshold for binarity; in doing so we are mindful of (i) the relative paucity of observational constraints for apparently {\em bona fide} single stars, (ii) that genuine binaries may fall below this threshold (cf. Fig. C.1 and Sect. 3.2.2) and (iii) that we may not assess binarity for a number of X-ray detected WNLh stars  that currently lack bolometric luminosity determinations\footnote{WR20aa, WR20b, WR21a, WR42e and WR43-2.}.

Nevertheless, adopting this discriminant means we may assert binarity for Arches F9 (see Sect. 3.3.2) as well as Mk39 (within 30 Dor), WR48-8, WR48-9, WR48-10 (within Danks 1), and  WR76-3 and WR76-7 (Mercer 81). Critically, this would  mean recognising the possibility of  a binary fraction approaching unity for the most luminous subset of stars (log$(L_{\rm bol}/L_{\odot})>6.3$; Sect. C.2.1). Such a conclusion, however,  is subject to the {\em caveat} that the nature of some members of the  very luminous cohorts within R136 (a1, a2, a3, and a5) and NGC3603 (WR43A and WR43B) remain difficult to ascertain due to blending, although the properties of the respective  composite X-ray sources  are strongly suggestive of binarity amongst one or more of the component  stars (Table~\ref{tab:appC1}).  More provocatively, if we were to take the observations summarised in Fig. C.1 at face value, one would infer  that (apparently) single WNLha stars are intrinsically X-ray faint ($L_{\rm X}<10^{32}$ergs$^{-1}$). This would lead to the conclusion that all but one  of the stars reported as  X-ray detections in Table~\ref{tab:appC1} - which, with the exception of WR12,  uniformly exceed this 
value at one or more epochs -  are binaries: a verdict consistent with their uniformly hard X-ray spectra.

Nevertheless, under the former, more conservative X-ray cut, the binary count for R136/30 Dor rises to 10/21 ($\sim48$\%) with the inclusion of Mk39 (noting that the sample size has risen since the full cohort of WNLh stars has been subject to X-ray observations). 
Similarly, consideration of the additional X-ray diagnostic  leads to the identification of 15/24 ($\sim63$\%) binaries within the ten galactic aggregates considered above. Given the foregoing discussion and the diverse and incomplete nature of the multi-wavelength observations  employed in the preceding assessment, we consider it highly likely that the simple number counts presented above  are lower limits to the true binary fractions,  but the heterogenous  nature of the extant datasets means we are unable to assess the degree of incompleteness at this time. We discuss the Arches cluster in the context of these observations -  and the resultant conclusions - further in Sect. 4 and 5. 

\bsp	
\label{lastpage}
\end{document}